\documentclass[fleqn,usenatbib]{mnras}
\UseRawInputEncoding
\usepackage{xcolor}
\usepackage{newtxtext,newtxmath}
 
\usepackage[T1]{fontenc}
\DeclareRobustCommand{\VAN}[3]{#2}
\let\VANthebibliography\thebibliography
\def\thebibliography{\DeclareRobustCommand{\VAN}[3]{##3}\VANthebibliography}

\usepackage{graphicx}	
\usepackage{amssymb}
\usepackage{amsmath}	
\usepackage{threeparttable}

\definecolor{address}{rgb}{0.36,0.54,0.66}
\definecolor{red}{rgb}{0.8,0.,0.}

\newcommand{\Ha}{\mathrm{H}\,\alpha}
\newcommand{\HI}{\mathrm{H}\,\textsc{\large{i}}}
\newcommand{\HIs}{\mathrm{H}\,\textsc{\textmd{i}}}
\newcommand{\HIb}{\mathbf{H}\,\textsc{\large{i}}}
\newcommand{\Hmolb}{\mathbf{H_2}}
\newcommand{\Hmol}{\mathrm{H_2}}
\newcommand{\HIl}{\mathrm{H}\,\textsc{\Large{i}}}

\newcommand{\eagle}{\textsc{\large eagle}{ }}
\newcommand{\eagleNS}{\textsc{\large eagle}}

\usepackage{scalerel,tikz}
\usetikzlibrary{svg.path}
\definecolor{orcidlogocol}{HTML}{A6CE39}
\tikzset{orcidlogo/.pic={
 \fill[orcidlogocol] svg{M256,128c0,70.7-57.3,128-128,128C57.3,256,0,198.7,0,128C0,57.3,57.3,0,128,0C198.7,0,256,57.3,256,128z};
 \fill[white] svg{M86.3,186.2H70.9V79.1h15.4v48.4V186.2z}
 svg{M108.9,79.1h41.6c39.6,0,57,28.3,57,53.6c0,27.5-21.5,53.6-56.8,53.6h-41.8V79.1z M124.3,172.4h24.5c34.9,0,42.9-26.5,42.9-39.7c0-21.5-13.7-39.7-43.7-39.7h-23.7V172.4z}
 svg{M88.7,56.8c0,5.5-4.5,10.1-10.1,10.1c-5.6,0-10.1-4.6-10.1-10.1c0-5.6,4.5-10.1,10.1-10.1C84.2,46.7,88.7,51.3,88.7,56.8z};
}}
\newcommand\orcidicon[1]{\href{https://orcid.org/#1}{\mbox{\scalerel*{
\begin{tikzpicture}[yscale=-1,transform shape]
\pic{orcidlogo};
\end{tikzpicture}
}{|}}}}

\graphicspath{{Figures/}}

\title[Red misfits in \textsc{\huge eagle}]{Emergence of red, star-forming galaxies (red misfits) in a $\Lambda$CDM universe}

\author[A. Manuwal]{
Aditya Manuwal$^{\orcidicon{0000-0003-2893-2793}\,1}$\thanks{E-mail: adi.manuwal@astro.unam.mx}
\\
$^{1}$Instituto de Astronom\'{\i}a, Universidad Nacional Aut\'onoma de M\'exico, A.P. 70-264, 04510 CDMX, M\'exico \\
}

\date{Accepted 2024 June 12. Received 2024 June 03; in original form 2023 December 19}

\pubyear{2024}

\begin{document}
\label{firstpage}
\pagerange{\pageref{firstpage}--\pageref{lastpage}}
\maketitle

\begin{abstract}
We investigate the formation of red misfits (RM) using a cosmological, hydrodynamical simulation from the \textsc{\Large eagle} project. Similar to observations, the RM possess less dust, higher stellar metallicities, and older stellar populations compared to blue, star-forming galaxies (BA) at the same $M_\star$. Lagrangian particle-tracking reveals that the older ages of RM have resulted from a combined effect of higher star formation efficiency (SFE), and the earlier onset and faster \textit{net} depletion of their interstellar medium (ISM). For the centrals, the latter was partially due to higher efficiency of escape from ISM, driven by stronger stellar and/or AGN feedback (depending on the mass). There was an additional contribution to this escape from gas stripping for satellite RM, as suggested by the higher group masses ($\gtrsim 0.5$~dex) and $\Hmol/\HIl$ ratios ($\gtrsim 0.3$~dex). Moreover, accretion of circumgalactic gas (CGM) onto the galaxy has been less efficient for the satellites. On the metallicity front, the offsets are largely due to the disparity in SFE, causing varying degrees of enrichment through the mass-transfers associated with stellar winds and supernovae. We ascribe this SFE disparity to the lower specific angular momentum ($j$) of freshly accreted CGM for RM, which ultimately manifested in the ISM kinematics due to 
interactions with cooling flows. The impact on $j_{\rm ism}$ was further intensified by poorer alignment with the flow's $\vec{j}$, particularly for the satellites. Our results illuminate potential origins of RM, and motivate further exploration of this peculiar class through a synergy between observations and simulations.
\end{abstract}




\section{Introduction}\label{intro}
In its essence, the global colour of a galaxy represents the shape of its spectral energy distribution (SED). If corrected for dust extinction, it is mainly a function of ages and metallicities of the constituent stellar populations: for a given age, metal-rich stars are redder due to heightened opacities
at higher frequencies; and ageing stars are redder due to lower temperatures compared to younger counterparts with equivalent metallicity. Since newly born stars inherit the properties of their parent clouds in the interstellar medium (ISM), colour can be a valuable tool for shedding light on the intricate interplay between the physical mechanisms involved in galaxy formation. For example, blue colouration could be attributed to active star formation, which can 
be induced by a multitude of factors -- like ongoing gas accretion \citep{Young2014,Lazar2023}, mergers \citep{Kaviraj2011,Stark2013,Oh2019}, or tidal interactions \citep{Stark2013,Mesa2014,George2015}. Conversely, redness could be indicative of a low star formation rate (SFR) due to poor gas supply \citep{Marinacci2010,Forbes2012}, expulsion of gas via feedbacks \citep{Springel2005d,Trayford2016,Nelson2018}, or environmental gas-stripping \citep{Font2008,Jaffe2015,Brown2017}. 

Observationally, galaxy colour is quantified as a difference of magnitudes for a reasonably-chosen pair of photometric bands. Bimodality in such colours is one of the key discoveries from voluminous surveys, both in the local Universe \citep{Strateva2001,Hogg2002,Blanton2003,Balogh2004,Baldry2004,Baldry2006,Driver2006,Mateus2006}, and at higher redshifts \citep{Cassata2008,Brammer2009,Williams2009,Whitaker2011,Moresco2013,Garcia2019}. The trend is apparent even after the colours are corrected for dust extinction, which happens to make the
two populations further disparate \citep[e.g.][]{Schawinski2014,Taylor2015}. Hence, it is thought to reflect two genuinely distinct (yet overlapping) populations, rather than a mere artefact of dust obscuration or analysis techniques. A similar separation has also been claimed in the star formation activity of galaxies, generally parameterised as the specific star formation rate (sSFR, i.e. SFR divided by stellar mass) \citep[e.g.][but also see \citealt{Feldmann2017}]{Kauffmann2003,Zamojski2007,Santini2009,McGee2011,Evans2018}. Given that star formation activity is correlated to the amount of cold gas and galaxy morphology, altogether, these bimodalities are envisaged as a blue population that is predominantly star-forming (hereafter SF), disky and gas-rich (known as the `blue cloud'), and a red population largely composed of quiescent, elliptical, gas-poor galaxies (the `red sequence').

As there is no a priori reason to expect such a distinction in galaxy colour, a vast amount of effort on the theoretical side has been directed towards understanding the physics that gives rise to these populations, and governs their evolution throughout cosmic time \citep{Kang2005,Dekel2006,Cattaneo2007,Gabor2012,Trayford2015,Trayford2016,Trayford2017,Nelson2018,Cui2021,Bravo2022}. The current concordance model predicts that all galaxies start off as blue but evolve to get redder and/or end up in the red sequence, subject to a pathway that is a complex function of internal mechanisms and environment. The exact details about this transition, though, have not been ascertained. 

A substantial body of work purports that the dominant mode of transition varies with mass. For massive galaxies (with stellar mass $M_\star\gtrsim 10^{10.5}\,{\rm M}_\odot$), this is attributed to quenching (or an ongoing cessation of star formation) via black-hole/AGN feedback, which ejects gas out of the ISM, and prevents further accretion by injecting energy into the ambient medium
\citep[e.g.][]{Trayford2016,Nelson2018}. On the other hand, low-mass galaxies are believed to have become red as a result of environmental mechanisms that they are exposed to when they are accreted as satellites by more massive galaxies \citep{Trayford2016}. This, however, is not the only explanation that has been put forward. 

\citet{Cui2021} used the \textsc{\large simba} simulation \citep{Dave2019} to propose that the colour of a central galaxy is strongly driven by the assembly history of its dark matter halo. They showed that, at a given stellar mass, blue galaxies tend to reside in haloes that were assembled early (or faster), while the red ones prefer recently-assembled haloes. Considering that haloes grow hierarchically, this also implies that massive haloes contain redder galaxies. The reason is, early-formed haloes were able to accumulate copious amounts of cold gas due to high accretion rates, which aided in sustaining star formation for a long duration. Contrarily, haloes that formed late could not accrete enough cold gas to
prolong their star formation activity till present. The feedback from AGN activity just exacerbates this: low cold-gas fractions in late-formed haloes result in low black-hole accretion rates, which is conducive for AGN feedback via jets (termed as the `radio mode') and an expedited quenching. As a consequence, although all galaxies approach quiescence, the ones in late-formed haloes are expected to transition earlier.

Note that all the information presented above circumambulates around the implicit assumption that galaxies fall in two camps. While this is indeed a fairly accurate description of the majority, there are notable exceptions. Numerous studies have revealed the existence of blue, SF ellipticals \citep{Driver2006,Schawinski2009,Company2010,Ilbert2010}, blue, passive disks \citep{Goto2003,Mahajan2009,Wild2009}, passive, red spirals \citep{Poggianti1999,Goto2003b,Poggianti2004,Wolf2009,Masters2010,Fraser2016,Lopes2016},
and galaxies with colours between red and blue, called the `green valley' \citep[hereafter GV;][]{Martin2007,Salim2007,Fang2012,Schawinski2014}. These intriguing galaxies might represent transitions between the two main groups, and therefore be the key to uncovering the origin of colour bimodality. Alternatively, they could be entirely different classes that represent undiscovered stages of galaxy evolution. Either way, there is considerable merit in exploring these cases,
as doing so will facilitate the formulation of a complete theory of galaxy formation.

There is another peculiar class that is red and SF, the so-called `red misfits' (hereafter RM). 
These galaxies have garnered significant interest over time owing to the fact that, even though quenching is generally
thought to precede or accompany the transition to redness, RM have been rendered red \textit{before} getting quenched. It is this class that forms the focus of this paper. 

RM have been studied observationally for more than a decade \citep[e.g.][]{Wolf2005,Davoodi2006,Weinmann2006,Haines2008,Brand2009}, but it is only recently that dedicated studies spanning a diverse range of environments have been conducted \citep{Evans2018,Chown2022}. \citet{Evans2018} used the Data Release 7 of the Sloan Digital Sky Survey \citep[SDSS;][]{York2000} to show that, contrary to the popular opinion, RM are not dusty, SF galaxies, and their colour is primarily due to old stellar populations. They showed that RM are more likely to host an AGN than either blue, star-forming (or `blue actives'; hereafter BA) or red, quiescent galaxies (or `red passives';
hereafter RP); indicating that AGN feedback might play a role in facilitating the formation of RM. They also present morphologies that are intermediate between spirals and ellipticals, meaning that RM are distinct from red spirals. 

The last characteristic also applies to GV \citep{Mendez2011,Smethurst2015,Coenda2018,Smith2022}, but \citet{Evans2018} highlight various factors that distinguish them from RM. For one, RM do not seem to have any environmental preference, whereas the abundance of GV varies with halo mass \citep[e.g.][]{Schawinski2014,Jian2020}. They interpret this as a support in favour of the role of internal mechanisms (like stellar and AGN feedback) in the formation of RM. Secondly, RM do not occupy the same region as GV in the colour--mass plane (see figure 12 in \citealt{Evans2018}). Likewise, the authors further note that RM share myriad properties with lenticular galaxies (S0s), including the red colours, overlap with GV, higher sSFRs and dust masses than ellipticals. However, unlike RM, S0s do not have a constant abundance across haloes \citep{Wilman2012}, and are instead concentrated in dense regions \citep[e.g.][]{Houghton2015,Mishra2019}.

More recently, \citet{Chown2022} compared molecular gas fractions, total (cold) gas fractions, and dust fractions of RM in SDSS against that of the BA. They found that RM possess lower dust fractions than BA but similar dust-to-gas ratios, reaffirming that the redness of RM cannot be attributed to dust-reddening. These galaxies also
exhibit lower molecular gas and total gas fractions, where the difference is more pronounced for the latter, suggesting significant depletion of $\HI$ in RM. Based on the depletion times for molecular and total gas (computed by dividing the gas masses by 
the SFR), the authors concluded that the lower gas fractions of RM are owed to poor gas supply, and not differing SFEs. 

Observational works akin to those mentioned above provide a wealth of useful information about the formation of galaxies, but are -- by nature -- limited in their scope due to the lack of \textit{accurate} temporal information for each gas cloud, stellar population, and black hole (BH). These are crucial for proper disentanglement of the complex tapestry of the involved physical mechanisms, and are readily available in hydrodynamical simulations. In addition, by modelling dark matter explicitly, the simulations enable rigorous examination of the link between halo assembly and galaxy formation in a self-consistent framework. 

Despite the resolution and volume limitations, the last decade has been a testament to the success of such simulations in recovering the observed properties of galaxies and the surrounding baryons, and their utility in revealing the underlying physics. Yet, there has not been any numerical investigation of the formation of RM. In this work, we uncover, for the first time, the main formation trajectories for central and satellite RM using the statistics in the flagship `Reference' run from the \eagle suite of cosmological, hydrodynamical simulations \citep{Crain2015,Schaye2015}.

This paper is organised as follows. The simulation used in this work and its post-processing for identifying haloes/galaxies (Section~\ref{haloes}), obtaining galaxy colours and dust masses (Section~\ref{galcol}), $\HI$ and $\Hmol$ masses (Section~\ref{hydromass}), and constructing galaxy histories (Section~\ref{tree}), are described in Section~\ref{sim}. We explain the strategy behind selecting the RM in \eagle in Section~\ref{samp}, and present the insights gleaned by comparing various properties of RM and BA, and their host (group) haloes, in Section~\ref{juxt}. In Section~\ref{feedhist}, we assess the impact of stellar and AGN feedback by comparing the black hole accretion rates (BHARs) and SFRs of RM against that of BA. We utilise the particle histories to understand the role of mass-transfers/transformations across time in Section~\ref{galhist}. In Section~\ref{accnsfe}, we extend this analysis to examine the connection between star formation efficiency, gas angular momentum, and accretion geometry. Finally, we present a concise summary and main conclusions from this work in Section~\ref{conc}. 

The readers already familiar with \eagle are recommended to only peruse Section~\ref{galcol} and Section~\ref{hydromass} while reading Section~\ref{sim}. The quantities are in proper units throughout the paper, unless stated otherwise.

\section{The Simulated Data}\label{sim}
\eagleNS\footnote{\url{https://eagle.strw.leidenuniv.nl/}} \citep{Schaye2015,Crain2015} is a suite of hydrodynamical simulations that were run using a modified version of the \textsc{\large gadget}-3 code [a successor to \textsc{\large gadget}-2 (\citealt{Springel2005})], with the general aim of understanding the physics that governs the formation and evolution of structures in the Universe. The simulations cover a broad range in box size and physics. For this study, we use the fiducial `Reference' simulation with $100^3~{\rm Mpc}^3$ co-moving volume (i.e. Ref-L0100N1504; see Table 1 of \citealt{Mcalpine2016}). The run assumes a flat $\Lambda$--cold-dark-matter ($\Lambda$CDM) cosmology from \citet{Planck2014}, and initialised\footnote{As the universe
evolves, the number of dark matter particles remains conserved, but gas particles can be converted to stars or accreted by BHs.} with $1504^3$ gas and dark matter particles each; the dark matter and (initial) gas particle masses are $m_{\rm dm}=9.70\times10^{6}\,{\rm M}_\odot$ and $m_{\rm g}=1.81\times10^{6}\,{\rm M}_\odot$, respectively. 

The limited mass-resolution of gas particles is a compromise between volume and computational cost. This inevitably precludes detailed modelling of the intricate physics of the ISM and BHs, which includes, among others, star formation, stellar mass loss due to winds and supernovae, and feedbacks from AGN activity. These are implemented as `subgrid'\footnote{The term denotes that the process occurs at a scale below the spatial resolution of the simulation.} models that are approximations of the small-scale physics; see \citet{Crain2015} for an elaborate description. Given
the interdependent nature of various processes, the parameters in these models cannot be estimated a priori, and are chosen or calibrated to reproduce some key observations: gas to SFR surface density relation, thermal history of the intergalactic medium, galaxy stellar mass function, size--mass relation, supernova rate density, BH to stellar mass relation, and stellar to halo mass relation. While this limits its predictive power, the simulation is nevertheless broadly consistent with a wide variety of observational data that were not used for calibration. 
This also includes optical colours \citep{Trayford2015,Trayford2017}, $\Ha$ luminosity \citep{Trayford2017} and D4000 spectral index \citep{Trayford2017}, which makes the simulation particularly well-suited for the science in this work.

\subsection{Haloes and galaxies}\label{haloes}
Dark matter haloes are identified in a two-step approach. First, the `main haloes' are identified using the Friends-of-Friends \citep[FoF; ][]{Davis1985} algorithm by taking the linking length
as 0.2 times the mean interparticle separation. Then, \textsc{\large subfind} \citep{Springel2001,Dolag2009} is used to search for self-bound substructures (or subhaloes) within the main/group haloes. A baryonic particle is linked to a subhalo only if the nearest dark matter particle to the baryonic particle belongs to the subhalo. The baryons associated with a subhalo constitute the galaxy and its circumgalactic material. The centre of the subhalo (and the galaxy) is taken as the coordinates of the
particle that has the least gravitational potential energy as per \textsc{\large subfind} among all the particles bound to the subhalo. 

The stellar and dust measurements of $z=0$ \eagle galaxies in this paper are based on the baryons enclosed within 30 kpc of the galactic centres. This generally provides good agreement
with the observations\footnote{Deviations are apparent only for the most-massive galaxies at $M_\star\gtrsim 10^{11}\,{\rm M}_\odot$ \citep[e.g.][]{Schaye2015,deGraaff2022}.} \citep{Schaye2015,Camps2016,Trayford2017,Baes2020,deGraaff2022}.

\subsection{Galaxy colours and dust masses}\label{galcol}
The colours are based on mock luminosities for the \textit{ugrizYJHK} broadbands \citep{Fukugita1996,Hewett2006,Doi2010}, where the magnitudes are absolute and rest-frame in the AB-system \citep{Oke1974}. The SEDs of stellar populations older than 100 Myr are modelled using \textsc{\large galaxev} \citep{Bruzual2003}, taking into account initial masses, ages and smoothed metallicities. Young stellar populations in both gas and star particles are re-sampled at an increased resolution to effectively replace them with an ensemble of SF molecular clouds.
This is done to avoid stochasticity in colours due to limited mass-resolution.
For the populations with ages $\in$[10, 100] Myr post re-sampling, the spectra are modelled with
\textsc{\large galaxev}, and those with younger ages are modelled using \textsc{\large mappings-iii} \citep{Groves2008}, which accounts for birth-cloud dust absorption and nebular emission. The physics of dust--radiation interaction in the diffuse ISM is calculated using \textsc{\large skirt} \citep{Baes2003,Baes2011,Camps2015}, where the dust mass is obtained by discretising the gas density on an adaptive-mesh-refinement (AMR) grid, and assuming a constant dust-to-metal mass ratio.

We note that there is no direct comparison in the literature between SDSS and the \eagle magnitudes used in this study\footnote{These are different from the ones in \citet{Trayford2015}.}. Nonetheless, \citet{Trayford2017} demonstrated that these are consistent with Sersic magnitudes from the Galaxy And Mass Assembly (GAMA) survey \citep{Driver2011,Liske2015}, which are based on SDSS imaging. Since these magnitudes also agree well with (Petrosian) SDSS magnitudes \citep[cf.][]{Kelvin2012}, we infer -- albeit indirectly -- that the optical colours in \eagle are consistent with SDSS.

\subsection{$\HIb$ and $\Hmolb$ masses}\label{hydromass}
The atomic and molecular hydrogen ($\Hmol$) masses are post-processed for each gas particle according to the approach in \citet{Manuwal2023}. First, the mass in all the neutral hydrogen ($\HI+\Hmol$) is obtained using the prescription of \citet{Rahmati2013}. Here, the UV ionising background is
adopted as \citet{Haardt2012}, which is weaker than that used in \eagle (i.e. \citealt{Haardt2001}), but results in negligible differences. The temperature of dense, SF gas particles is taken as $T=10^{4}\,{\rm K}$, to mimic the warm, diffuse ISM around young stellar populations \citep{Crain2017}. The neutral hydrogen is further broken down into atomic and molecular components
using the equations from \citet{GD14}. The choice of this particular method is predicated on its demonstrated success in producing sensible results for both $\HI$ and $\Hmol$ simultaneously \citep[cf.][]{Manuwal2022,Manuwal2023}. The final values are computed using the iterative convergence technique of \citet{Stevens2019}, where the interstellar far-UV flux at $1000\,$\AA~is estimated as per \citet{Diemer2018}, and dust-to-gas ratio is assumed to scale with metallicity \citep[see][]{Lagos2015}. 

\subsection{Merger trees}\label{tree}
The \eagle data is spread over various timestamps in the form of `snapshots', and one needs to keep track of a galaxy/halo across these snapshots for constructing its history. This is done by linking the systems\footnote{The term `system' here refers to all the particles (dark matter, BH, stellar and gas) bound to the host subhalo.} across snapshots according to the approach of \citet{Qu2017}, where the particles in each `progenitor' are forward-tracked to identify the `descendant' in the next snapshot. Given that a system can form via mergers of a multitude of smaller systems, there can be several progenitors for a given descendant, but not vice versa. The final merger tree represents the complete history of the system, with each branch pertaining to a specific progenitor. The systems constituting the branch with the greatest cumulative mass are referred to as the `main progenitors'.

\section{Selecting and classifying SF galaxies}\label{samp}
The most straightforward approach for selecting RM is to focus on a particular region in the sSFR vs colour plane. There are, however, a myriad of colours that one can opt for this purpose. For this work, we take \citet{Chown2022} as the observational reference and use it to inform the selection. We employ the $g-r$ colour\footnote{We note that the $u-r$ colour is generally considered to be more appropriate for separating SF galaxies from quiescent ones, as the $u$ band is more sensitive to emission lines that trace recent star formation. Notwithstanding, we use $g-r$ colours for the sake of consistency with \citet{Chown2022}.} corresponding to the face-on orientation,\footnote{These are calculated using the \textit{SDSS\_g\_f} and \textit{SDSS\_r\_f} parameters in the RefL0100N1504\_DustyMagnitudes table of the public database.} where the orientation is based on the galaxy's stellar angular momentum vector, and chosen because the observed colours are inclination-corrected. This colour is not corrected for attenuation by the dust \textit{in} the galaxy, which is congruent with the observed colours in \citet{Chown2022} that are corrected for dust in the Milky Way \citep{Evans2018} but not in the galaxy itself. In accordance with the observations, we only consider the galaxies with $M_\star>10^{9}\,{\rm M}_\odot$, which also happens to select galaxies with well-resolved stellar distribution ($\gtrsim$ 500 stellar particles; \citealt{Schaye2015}). Furthermore, following \citet{Katsianis2021}, we enhance the default sSFRs in \eagle by $0.2$~dex to account for the systematic offset in SFR with respect to observed values \citep{Furlong2015,Katsianis2017}.

Dust properties of a simulated galaxy are only meaningful if the dust distribution is resolved to an appreciable level. This can be achieved in \eagle by enforcing a threshold on the number of particles that represent dust in the galaxy (or $N_{\rm dust}$; see \citealt{Camps2018}). Though it is not feasible to determine a specific value for this threshold, we follow \citet{Camps2018} and adopt $N_{\rm dust}>250$, which roughly demarcates the number below which resolution effects impart significant contamination. This omits about 50 per cent of the galaxies above $M_\star=10^{9}\,{\rm M}_\odot$, which can introduce potential biases.

\begin{figure}
  \includegraphics[width=1\columnwidth]{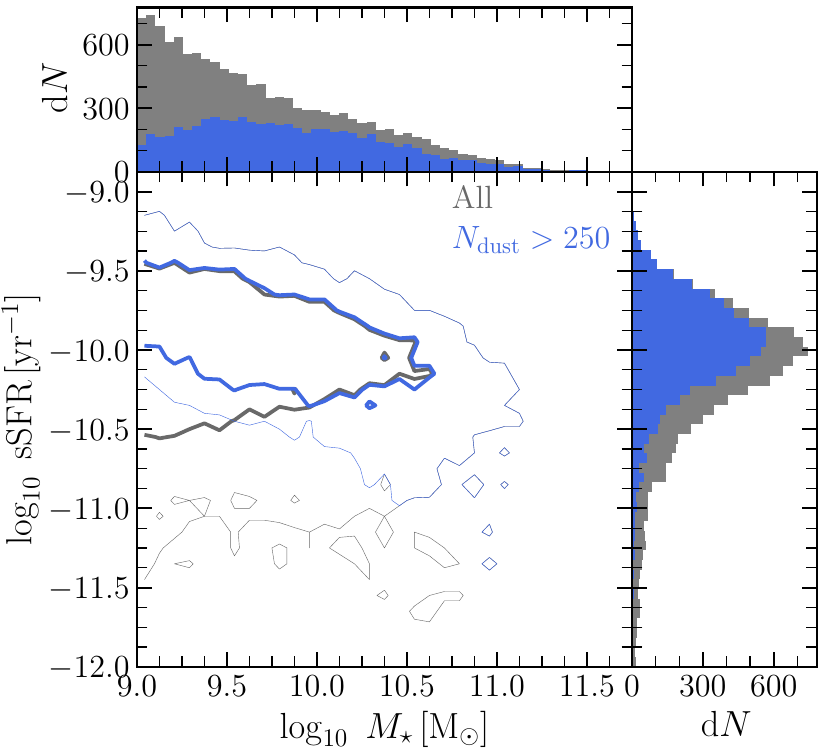}
   \caption{The specific star formation rate plotted against stellar mass for the \eagle galaxies with $M_\star>10^{9}\,{\rm M}_\odot$. The grey contours show the regions encompassing 68 and 95 per cent of all the galaxies, where the former is shown as a thicker contour. The blue contours correspond to the galaxies with $N_{\rm dust}>250$. The respective histograms for sSFR (right) and $M_\star$ (top) have the same colour correspondence.}
   \label{ssfrvsmstar}
\end{figure}

We examine this explicitly using the sSFR--$M_\star$ plane in Fig.~\ref{ssfrvsmstar}, with the full sample above $M_\star=10^9\,{\rm M}_\odot$ shown in grey, and that with $N_{\rm dust}>250$
shown in blue. The corresponding histograms for sSFR and $M_\star$ are shown on the right and the top,
respectively. The discrepancy between the contours in the bottom-left corner of the main panel clearly
indicates that the threshold primarily removes galaxies at low $M_\star$ and sSFR. This involves omission of $\gtrsim 50$ per cent of all the galaxies at any given mass below $M_\star\approx 10^{9.5}\,{\rm M}_\odot$,
or ${\rm sSFR}\lesssim 10^{-10.3}\,{\rm yr}^{-1}$. This means that, \textit{as a whole}, we are not sampling enough SF galaxies (with ${\rm sSFR}\gtrsim 10^{-11}\,{\rm yr}^{-1}$, for example) -- but a subsample limited to higher masses would still be representative. In fact, for the reasons stated in Section~\ref{dam}, the conclusions in this paper have only been derived for the SF galaxies above $10^{9.9}\,{\rm M}_\odot$.

\begin{figure}
  \includegraphics[width=1\columnwidth]{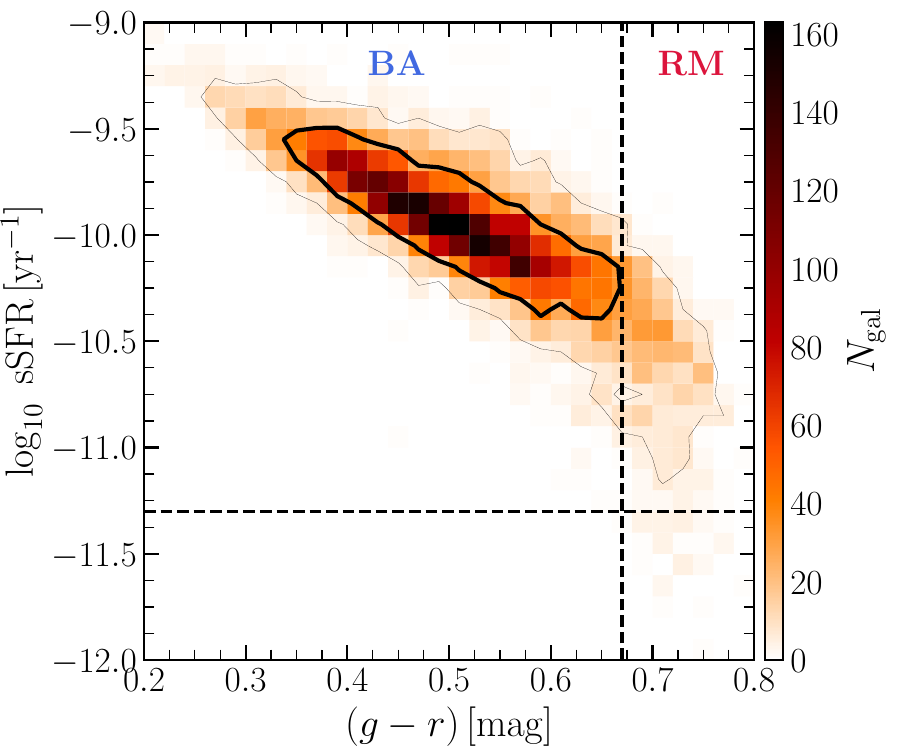}
   \caption{The specific star formation rate against the $g-r$ optical colour for the \eagle galaxies
   shown as a density plot. Only the galaxies with a well-resolved dust distribution (see the text) and $M_\star>10^{9}\,{\rm M}_\odot$ are considered here. The colour of a pixel shows the number of galaxies ($N_{\rm gal}$) in the parameter space occupied by the pixel, and the value can be gauged using the colour bar on the right. The dashed-black lines denote the thresholds used for selecting `red misfits' (RM) and `blue actives' (BA) in \eagleNS: ${\rm sSFR}=10^{-11.3}\,{\rm yr}^{-1}$ (horizontal line) and $g-r=0.67$ (vertical line).}
   \label{ssfrvscol}
\end{figure}

The sSFR vs $g-r$ plane is shown as a two-dimensional density plot in Fig.~\ref{ssfrvscol}, where the number of galaxies in each pixel is denoted according to the colour bar shown on the right. The black contours portray the region encompassing 68 and 95 per cent of the data; the former is shown using a thicker contour. As expected, most of the galaxies are SF and there is a dearth of quiescent galaxies in the lower-right corner. In addition to the aforementioned bias, the latter can be partially owed to the well-documented inability of \eagle to properly reproduce the quiescent population at $10^{-13}<{\rm sSFR}/{\rm yr}^{-1}<10^{-11}$ \citep{Baes2020,Trcka2020,Katsianis2020,Katsianis2021}. One may attribute this to the ${\rm SFR}=0\,{\rm M}_\odot\,{\rm yr}^{-1}$ galaxies that constitute $\approx 80$ per cent of \textit{all}
the $M_\star>10^{9}\,{\rm M}_\odot$ galaxies below ${\rm sSFR}=10^{-11}\,{\rm yr}^{-1}$. These 
are thought to arise due to poor sampling of SF particles \citep{Furlong2015,Trayford2017} and/or the feedback schemes which are sometimes too effective in quenching galaxies \citep{Katsianis2021}. However, accounting for them does not suffice to explain the discrepancy with SDSS (see \S3.2 of \citealt{Katsianis2021}), and further investigation is warranted.

We note that \citet{Chown2022} select RM and BA by imposing thresholds on sSFR and $g-r$ that roughly denote the saddle points (or minima) of their bimodal distributions. This corresponds to ${\rm sSFR}>10^{-11.3}\,{\rm yr}^{-1}$ and $g-r>0.67$ for RM, and ${\rm sSFR}>10^{-11.3}\,{\rm yr}^{-1}$ and $g-r<0.67$ for BA. Ideally, the thresholds for \eagle should be derived by following a similar approach for consistency with observations, but the scarcity of low ${\rm sSFR}$ galaxies implies that the sSFR bimodality is not recovered adequately, thus precluding identification of the saddle point. Therefore, we directly adopt the thresholds from \citet{Chown2022}, shown as the dashed-black lines in Fig.~\ref{ssfrvscol}. It can be seen that, although RP are virtually absent, the thresholds do result in appreciable statistics for both BA (5939) and RM (610)\footnote{Our sample contains a higher BA fraction among the SF systems than SDSS \citep{Evans2018}, because the proportion of galaxies rejected due to the $N_{\rm dust}$ threshold increases
at lower sSFR (see the histograms on the right in Fig.~\ref{ssfrvsmstar}).}. In what follows, we explore the physical origin of the RM thus defined.

\section{Insights from juxtapositions of galaxies and their host haloes}\label{juxt}
In any given group (containing a central along with its satellites), satellite galaxies are generally subjected to a higher degree of environmental perturbation than the central, owing to interactions with surrounding gas and galaxies while traversing their orbits. This mainly involves ram pressure exerted by the intragroup gas on satellites moving at high relative
velocities \citep[see the review by][]{Boselli2022}, and strong tidal forces engendered by steep gravitational gradients near the pericentres \citep[e.g.][]{Tollet2017,Carleton2019,Jackson2021,Wright2022}. These processes are particularly conducive for quenching \citep{Bahe2015,Steinhauser2016,Simpson2018,Cintio2021,Wright2022}, but also play a key role in transition to redder colours \citep[see e.g.][]{Steinhauser2016,Jiang2019,Borrow2023}. Thus, the physics responsible for the redness 
may not be the same for the centrals and the satellites. In the sections that follow, we present results for the two separately.

\subsection{Dust content, age, and metallicity}\label{dam}
\begin{figure*}
  \includegraphics[width=1\columnwidth]{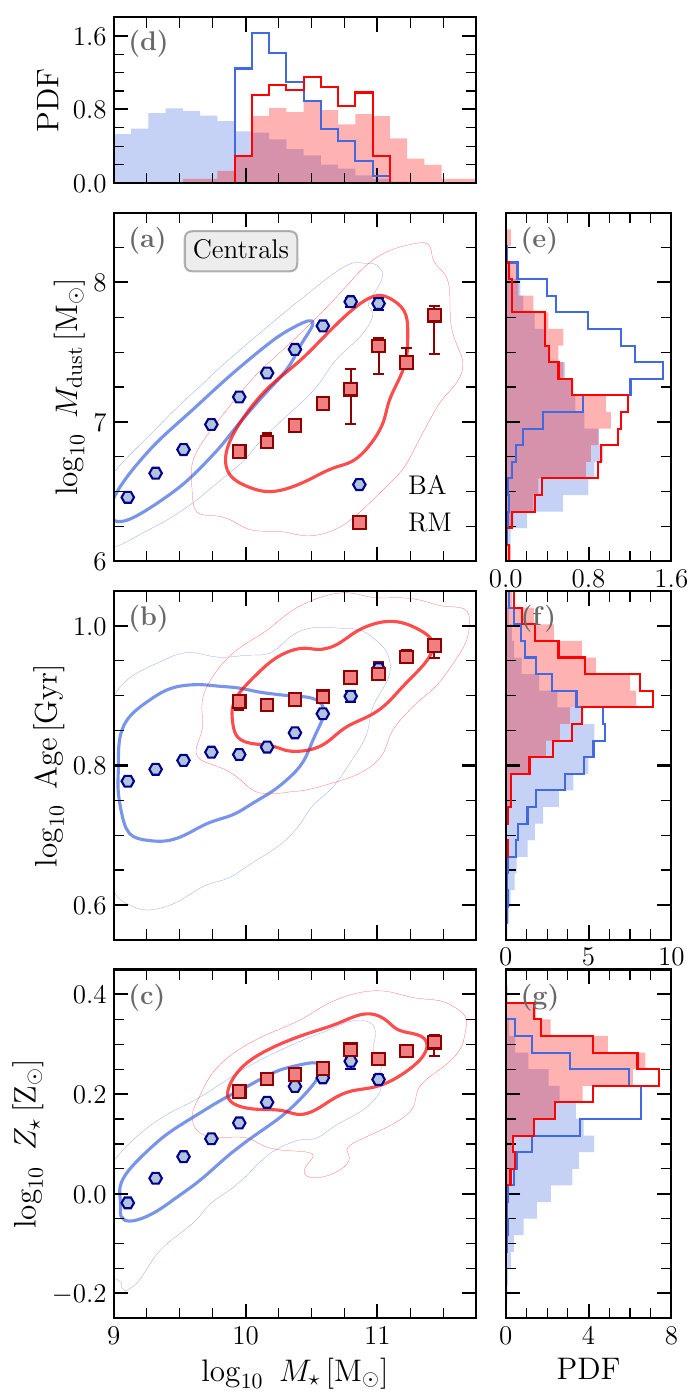}\quad
  \includegraphics[width=1\columnwidth]{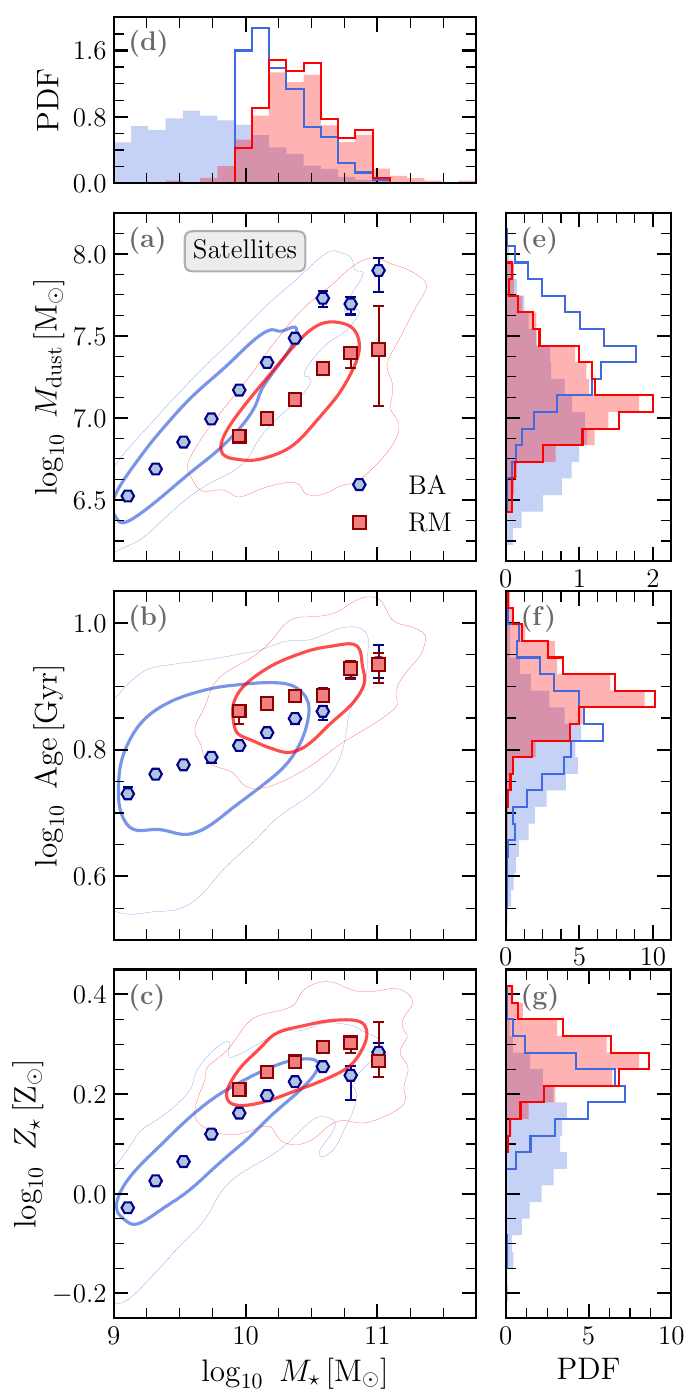}
   \caption{The relationship of dust mass, age and stellar metallicity (normalised by the solar metallicity, $Z_\odot=0.0127$) with stellar mass for SF galaxies classified as BA and RM. The left- and the right-halves show the results for the centrals and
   the satellites, respectively. In each half, the blue hexagons and the red squares show the medians for BA and RM, respectively, for the bins with at least 10 galaxies. The error bars on the medians are $1\sigma$ uncertainties based on bootstrapping. The blue (red) contours show the regions containing 68 and 95 per cent of BA (RM) for a Gaussian-smoothed density distribution; the former is thicker. Panel (d) shows the probability density functions (PDFs) for the stellar masses of the two samples with the same colour coding, where the filled histograms are for the full samples, and the step histograms are for the range where both RM and BA show at least 10 galaxies per bin. Each of the other rectangular panels [i.e. (e)--(g)] shows similar PDFs for the quantity plotted on the $y$-axis of the square panel on its left. RM show a preference
   for higher stellar masses compared to BA. For a given stellar mass, RM generally exhibit lower dust contents, higher
   metallicities, and older ages.}
   \label{damvsmstar}
\end{figure*}

As stated earlier (in Section~\ref{intro}), a galaxy's colour is dependent on its dust content, age and metallicity, but it is degenerate with respect to these parameters. It is, therefore, worthwhile to investigate which of these is responsible for the red colours of RM. We do so by comparing the parameters of RM against those of BA in Fig.~\ref{damvsmstar}, where the left- and the right-halves
of the figure show the results for centrals and satellites, respectively. In each half, panels (a), (b) and (c) show the dust content, mass-weighted stellar age, and mass-weighted stellar metallicity,\footnote{The metallicities have been smoothed using the SPH kernel.} respectively. The medians for BA (blue hexagons) and RM (red squares), along with their uncertainties (error-bars) derived from bootstrap resampling ($10^5$ samples),\footnote{These are evaluated using the {\tt bootstrap\_percentiles} module in the \textsc{\large python} code at \url{https://github.com/arhstevens/Dirty-AstroPy/blob/master/galprops/galcalc.py}.} are only shown for the stellar-mass bins that contain at least 10 galaxies. (The binning is to control for stellar mass, a quantity that is well-correlated with
each of the three parameters.) The coloured contours show the regions encompassing 68 and 95 per cent of the data; the former is shown using thicker contours. Panel (d) shows the probability density functions (PDFs) for stellar mass, and panels (e)--(g) show them for the parameters plotted on their left. The stellar metallicities are normalised by the solar value, $Z_\odot=0.0127$.

Both central and satellite RM tend to have higher stellar masses than BA, as is apparent both in the contours
and in the PDFs. This is reasonable because RM are biased to lower sSFRs than BA (Fig.~\ref{ssfrvscol}), and sSFR is negatively correlated\footnote{The Spearman-rank correlation coefficients for SF centrals and 
satellites are $-0.52$ and $-0.66$, respectively.} with $M_\star$. We note that nearly all the galaxies below $M_\star\approx 10^{9.9}\,{\rm M}_\odot$ are BA while those above $M_\star\approx 10^{11.1}\,{\rm M}_\odot$ are RM, where the former is primarily due to the selection bias towards high sSFRs (see Fig.~\ref{ssfrvsmstar}). Though these
extreme regimes are intriguing in their own right, a strong dominance of a single population precludes fixing the stellar mass, and hence, like-for-like comparisons. Such comparisons are nonetheless possible for the intermediate stellar masses with $\geq 10$ galaxies per bin for \textit{both} BA and RM, i.e. $M_\star/{\rm M}_\odot\in[10^{9.9},10^{11.1}]$. We refer to these as the `overlapping subsamples' (hereafter OS), and focus
the rest of the discussion only on this range. Note that, since the $N_{\rm dust}$ threshold (Section~\ref{samp}) only removes about 9 per cent of all the SF galaxies in this range, the OS are fairly representative. The corresponding PDFs are shown as step histograms throughout the paper.

For this range, the dust masses of both central and satellite RM are, on average, lower than those of BA at the same $M_\star$ by $\gtrsim 0.4$~dex, even after accounting for the uncertainties. Thus, the red colours of the simulated RM are evidently \textit{not} caused by dust-reddening, providing theoretical support for the conclusions of \citet{Evans2018} and \citet{Chown2022} for observed galaxies. Interestingly, the filled PDFs suggest that RM, as a whole, possess slightly \textit{more} dust than BA. A careful examination reveals that this apparent discrepancy occurs due to the paucity of RM at $M_\star\lesssim10^{9.9}\,{\rm M}_\odot$, which -- owing
to the $M_{\rm dust}$--$M_\star$ correlation -- in turn biases the distribution for BA to comparatively lower $M_{\rm dust}$. In fact, if one only considers the OS, the trend in the PDFs [(e) panels] is reversed, and consistent with what is suggested by the medians.

If dust is not responsible for rendering the RM red, it must be age and/or metallicity. As a matter of fact, panel (b), on either side, shows that RM contain stellar populations clearly older than those in BA by $\lesssim 0.1$~dex,
where the offset reduces with increasing $M_\star$ and vanishes at $\approx 10^{11}\,{\rm M}_\odot$; in agreement with the observed trend in \citet{Evans2018}. Metallicity, on the other hand, shows a systematic difference over a wider range of stellar mass, where RM are enriched more than BA by about $0.05$~dex [(c) panels]. Similar results have previously been reported for 
the gas-phase metallicities of observed RM, corresponding to differences of $\approx 0.03$~dex \citep{Evans2018}. 

The (f) and (g) panels show that the general trends in age and metallicity are also reflected in the PDFs, albeit at a lesser degree for the OS. The latter stems from the fact that the bins corresponding to the lowest masses of BA and the highest masses of RM mostly contain galaxies from the respective classes. While these points populate the extremes of the (filled) PDFs for the full samples, they are absent from those for the OS (step).

\subsection{\textit{In situ} vs \textit{ex situ} origin of stars}\label{invsex}
In principle, a galaxy can gain stars in two ways: a) star formation within its ISM (\textit{in situ}), or b) accretion of stars from other galaxies in its proximity (\textit{ex situ}). It is now believed that galaxies undergo a `two-phase growth' (coined by \citealt{Oser2010}), such that, initial star formation occurs \textit{in situ} due to cosmological accretion of cold gas or internal gas recycling, and \textit{ex situ} stellar mass is accumulated later through mergers and tidal stripping of other galaxies. This means that examining the contributions of \textit{in situ} and \textit{ex situ} stars to the overall stellar population can provide some insights into the galaxy's evolutionary history. 

We compute the fraction of stars formed \textit{ex situ} for each galaxy in our sample. Every gas particle in \eagle has a unique ID, which is retained when it converts to a star particle. We leverage this to identify the gas particles in a main progenitor existing as star particles in its descendant in the next snapshot. By carrying this out for all pairs of successive snapshots, we identify all the stars formed within the ISM of the main progenitors of a galaxy, and flag other stars in the galaxy as \textit{ex situ}.

\begin{figure}
  \includegraphics[width=1\columnwidth]{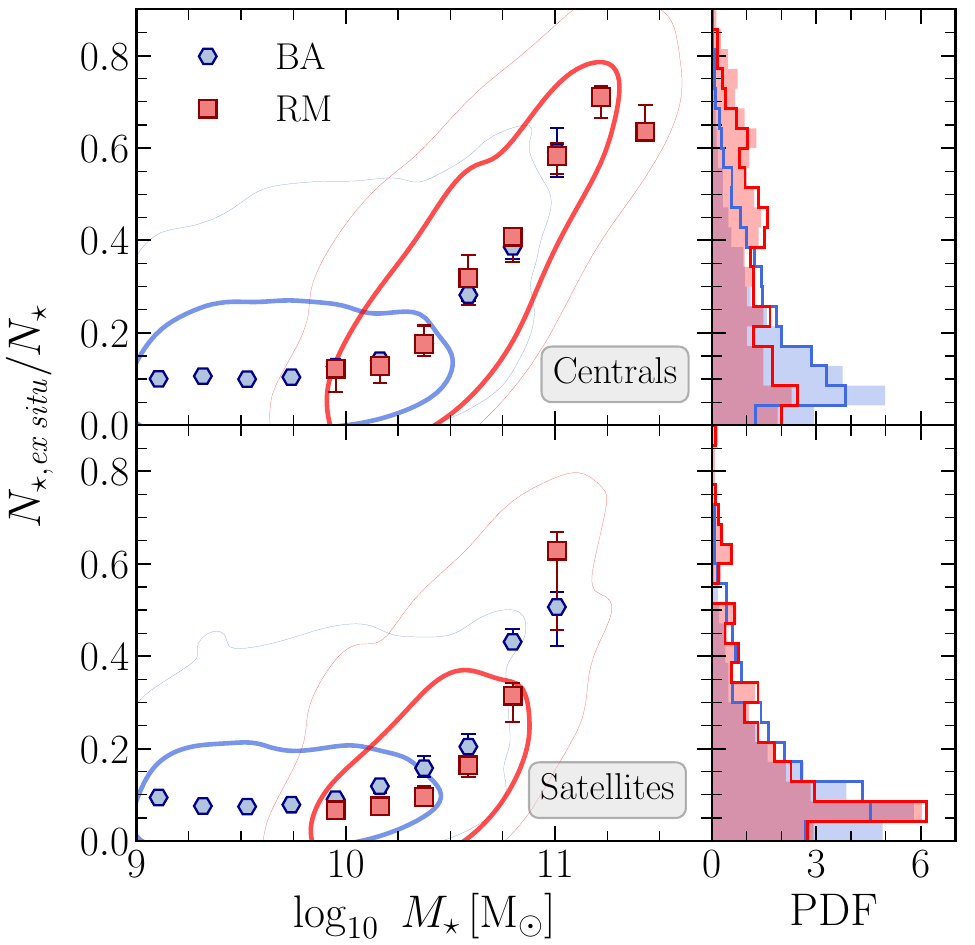}
   \caption{Fraction of stars formed \textit{ex situ} against stellar mass for RM and BA, where $N_{\star,{\it ex\,situ}}$ and $N_{\star}$ are the number of stars formed \textit{ex situ}, and the total number of stars in the galaxy, respectively. The
   top and bottom panels show the results for the centrals and the satellites, respectively. As a general trend, the fraction remains constant for $M_\star\lesssim 10^{10}\,{\rm M}_\odot$ and increases with stellar mass for higher $M_\star$, reaching $\gtrsim 0.5$ for $M_\star>10^{11}\,{\rm M}_\odot$. Almost all BA predominantly possess \textit{in situ} stars, whereas RM show a wide range in stellar origin. The fractions are broadly consistent between RM and BA, regardless
   of their central/satellite status.}
   \label{invsexfig}
\end{figure}

The resultant \textit{ex situ} fractions for RM and BA at different stellar masses are shown in Fig.~\ref{invsexfig}, where
the top and bottom panels show the values for centrals and satellites, respectively. It remains constant up until $M_\star \approx 10^{10}\,{\rm M}_\odot$ and then increases with mass, reaching $\gtrsim 0.5$ for $M_\star>10^{11}\,{\rm M}_\odot$. This is in congruence with previous studies on simulated and observed galaxies (see \citealt{Gomez2016}; \citealt{Clauwens2018}; \citealt{Tacchella2019}; \citealt{Davidson2020};
and the references therein). 

The contours show that almost all central BA possess stellar populations that are predominantly (more than 50 per cent) \textit{in situ}, whereas the RM span a wide diversity. The medians are consistent between the RM and the BA, but the distributions for the OS (step) do not exhibit similar shapes; the RM show a nearly uniform spread, while the BA are skewed towards low fractions. In fact, the median fraction for the RM is higher than that for the BA by $\approx 0.1$. This is a direct consequence of a similar discrepancy between the $M_\star$ distributions, where the RM have a uniform spread and the BA are biased towards lower masses [see panel (d), left-half, Fig.~\ref{damvsmstar}]. 

A similar $M_\star$ bias exists in satellite OS [panel (e), right-half, Fig.~\ref{damvsmstar}], but the RM and the BA in this sample are consistent in their \textit{ex situ} fractions as populations\footnote{The Kolmogorov-Smirnov test reveals a $p$-value of 0.24 ($<2\sigma$ significance)}. At a fixed mass, though, the fractions of satellite BA tend to be slightly higher. The typical offset is just $\lesssim 0.07$, except for an offset of $\approx 0.15$ at $M_\star\approx 10^{10.8}\,{\rm M}_\odot$, which also corresponds to the maximum difference in the median $Z_\star$s [panel (c), right-half, Fig.~\ref{damvsmstar}]. In the absence of these offsets, the PDF for the BA would have been biased towards lower fractions. However, the preference for higher fractions somewhat compensates for the bias towards lower $M_\star$s, thereby resulting in similar PDFs. Nevertheless, the inner contour shows that most of the RM possess $M_\star\lesssim 10^{10.8}\,{\rm M}_\odot$, where the differences between the fractions are rather minor. Thus, for an overwhelming majority of our galaxies, the difference in colours of BA and RM cannot be imputed to \textit{in situ} vs \textit{ex situ} nature of their stars.

\subsection{Group-scale environment}\label{grouphist}
The stellar origins of RM and BA (as inferred from Fig.~\ref{invsexfig}) suggest similar merger histories for these classes. Does this mean RM and BA have experienced the same environmental conditions?
We address this explicitly using halo and cold gas characteristics; one of the most reliable indicators of intrahalo environmental processing. 

\subsubsection{Host group}\label{groupmass}
\begin{figure}
  \includegraphics[width=1\columnwidth]{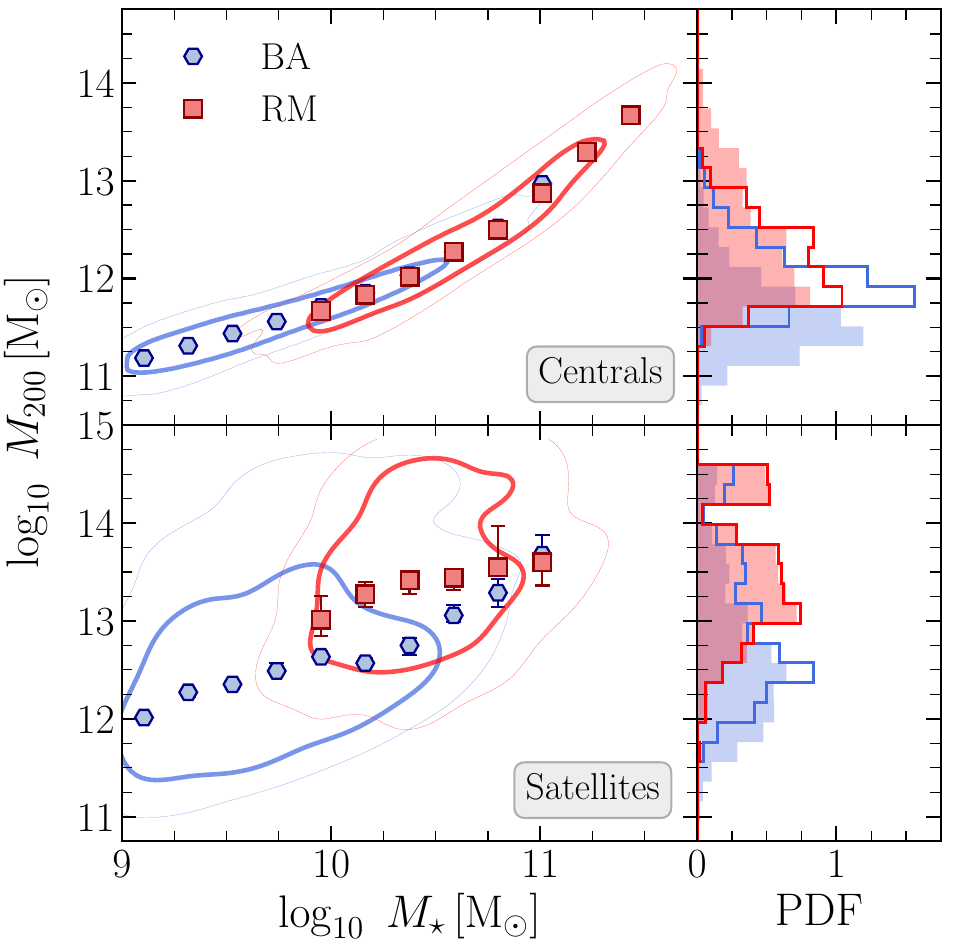}
   \caption{The mass of the host group plotted against the galaxy's stellar mass. The plotting style follows Fig.~\ref{invsexfig}. The group masses of the centrals (top) are agnostic to their class, but satellite RM (bottom) generally prefer haloes with masses greater than their BA counterparts by $\gtrsim 0.5$~dex.}
   \label{m200vsmstar}
\end{figure}

The mass of the host group, like $M_{200}$, can serve as a first-order metric to assess the cumulative impact of environmental forces that act on a satellite over its orbital history. This is owed to an overall increase in galaxy density \citep{Haas2012}, orbital eccentricity \citep{Wetzel2011} and velocity, and intragroup gas temperature \citep{Babyk2023}, with group mass. We show the $M_{200}$--$M_\star$ relations for our galaxies in Fig.~\ref{m200vsmstar}. The masses for central RM are near identical to those for the BA, essentially ruling out virial shock heating\footnote{This process is efficient only above a critical halo mass, which is $M_{200}\approx 10^{11.7}\,{\rm M}_\odot$ for \eagle \citep{Correa2018}, close to the lower bound for the OS (defined in Section~\ref{dam}).} as a probable cause for the colour difference. The satellites, however, show a clear separation between the two, where the RM inhabit haloes that are more massive than the hosts of BA by $\gtrsim 0.5$ dex. This happens to be one
of the strongest trends for the satellites in this paper, and suggests that satellite RM experience `harsher' environments than the BA. Since the difference is maximised for $M_\star/{\rm M}_\odot\in[10^{10.2},10^{10.5}]$, the impact of local environment on satellite RM must manifest to the most degree in this range. (This is certainly the case, as shown in the following sections.)

Note that the consistency in halo mass of the centrals does not necessarily translate to assembly history. At a fixed $M_{200}$, haloes that have assembled early tend to possess more gas due to higher accretion rates at those epochs. Consequently, galaxies residing
in such haloes would have enough gas available for extending star formation till present, leading to bluer colours \citep[see][]{Cui2021}. This effect is particularly relevant for centrals because they can accrete more efficiently than satellites. 

\begin{figure}
  \includegraphics[width=1\columnwidth]{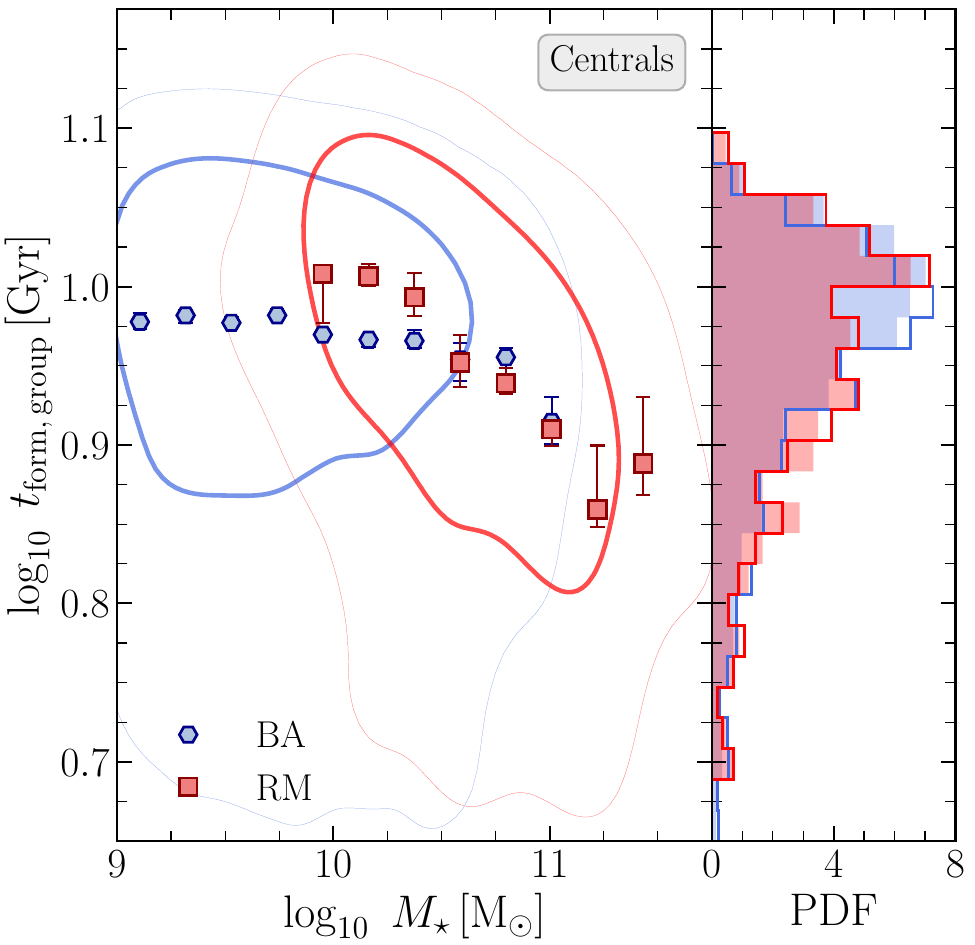}
   \caption{Formation time of the host group-halo against stellar mass for the centrals, where the former is defined as the look-back time when 50 per cent of the final halo mass had assembled. The group haloes of central RM and BA, on average, assembled at similar times.}
   \label{tformhalo}
\end{figure}

We compare the assembly history of the groups hosting our central RM and BA. One way to quantify the `pace' of halo assembly is to derive the instant in time when a large fraction ($\gtrsim 50$ per cent) of its final mass was accumulated. The mass history of a group halo in \eagle can be obtained by noting the mass of the groups hosting the main progenitors of its central. We interpolate these histories to determine the look-back time when 50 per cent of $M_{200}$ was accumulated ($t_{\rm form,\,group}$). The results are shown in Fig.~\ref{tformhalo}. It is evident that the group haloes of the RM and BA were, for the most part, formed at similar times. There is some bias ($\lesssim 12$ per cent) for RM at $M_\star/{\rm M}_\odot\in[10^{10.2},10^{10.5}]$, but this is towards higher $t_{\rm form,\,group}$ (or earlier formation). Hence, the red colours of central RM in \eagle do not seem to arise as a result of late assembly of their host haloes.

\subsubsection{Cold gas}\label{coldgas}
Next, we turn our attention to the cold gas reservoir, mostly composed of neutral hydrogen. This phase is susceptible to environmental perturbations, which reflects in the content, kinematics, and morphology of both $\HI$ and $\Hmol$ \citep[e.g.][]{Marasco2016,Stevens2019,Watts2020,Stevens2021,Manuwal2022,Manuwal2023}. Stripping of cold gas (caused by ram pressure and tidal forces) proceeds outside-in, such that the diffuse, more extended phases are the first ones to bear the brunt, and compact phases are affected later, leading to a greater loss of $\HI$ than that of $\Hmol$. If
the phenomenon has a preponderant bearing on the cold gas content, it can manifest as higher $\Hmol/\HI$ ratios of satellites in comparison to centrals with the same stellar mass \citep[e.g.][]{Stevens2021}. Unlike the centrals, satellite RM evidently prefer haloes more massive than their BA analogues (Fig.~\ref{m200vsmstar}). It is, therefore, of interest to also investigate their $\Hmol/\HI$ ratios.

\begin{figure}
  \includegraphics[width=1\columnwidth]{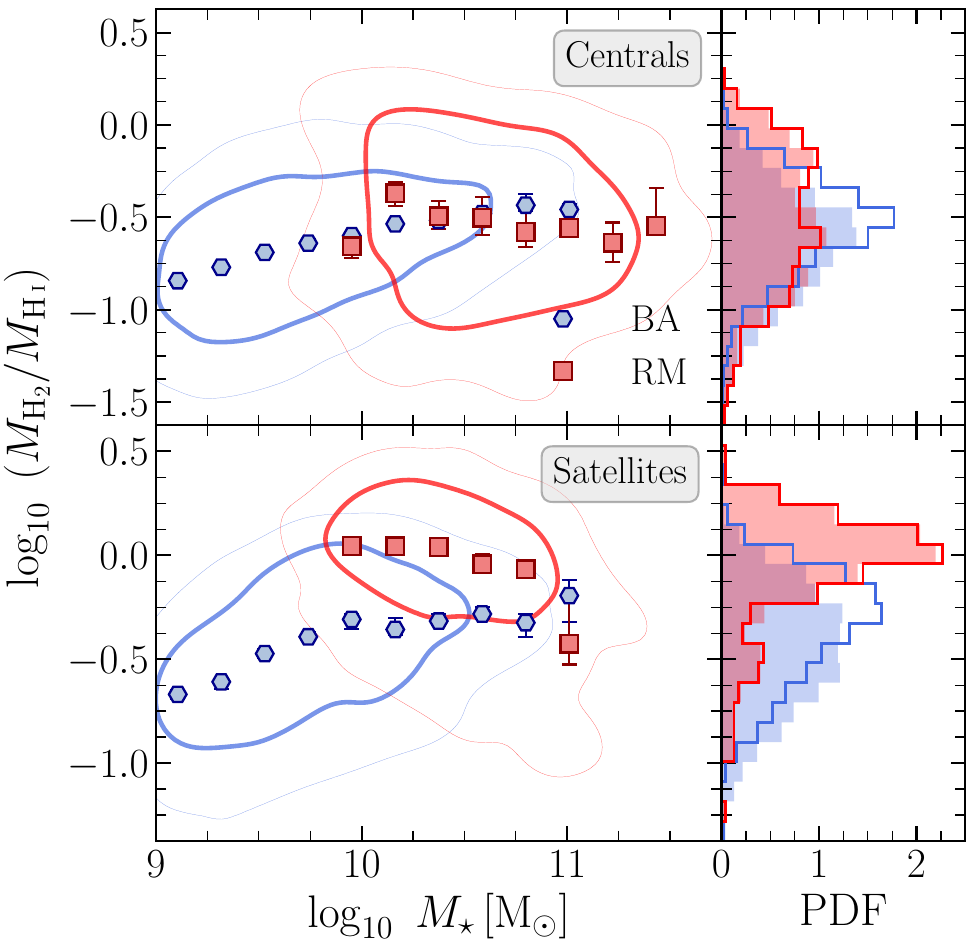}
   \caption{$\Hmol/\HI$ ratios for our galaxies. At a given stellar mass, RM generally exhibit ratios greater than those for the BA by $\gtrsim 0.3$~dex for the satellites (bottom panel), but show consistency with BA for the centrals (top panel).}
   \label{h2byhivsmstar}
\end{figure}

We derive the gas masses using all the gas particles bound to subhalo, and not just those enclosed within the 30 kpc aperture. This is because $\HI$ tends to extend further out than stars, and stars can sometimes migrate inwards after their birth. The ratios 
are presented in Fig.~\ref{h2byhivsmstar}. While the ratios for the centrals are consistent between RM and BA, satellite RM show a clear preference for higher $\Hmol/\HI$ ratios (by $\gtrsim 0.3$~dex). Unsurprisingly, this general dichotomy is also discernible in the PDFs, both for the complete sample (shaded) and the OS (step). Note that
the offset in $\Hmol/\HI$ ratio is maximised in the same $M_\star$ range as that in $M_{200}$ (i.e. $10^{10.2}\leq M_\star/{\rm M}_\odot\leq 10^{10.5}$; see Fig.~\ref{m200vsmstar}). Moreover, we have confirmed that the differences in $\Hmol/\HI$ ratios
of the satellites are primarily due to the lower $M_{\HIs}$ of RM, and not due to higher $M_\Hmol$. In fact, both satellite and central RM have lower $M_\Hmol$ than their BA counterparts, but unlike the centrals, the offsets between the satellites' $\Hmol$ contents are smaller than those in the $\HI$ contents. Together, these results provide substantial support for the scenario where the higher ratios of satellite RM are caused by environmental processing within groups.

\subsection{Black hole mass}\label{agn}
\begin{figure}
  \includegraphics[width=1\columnwidth]{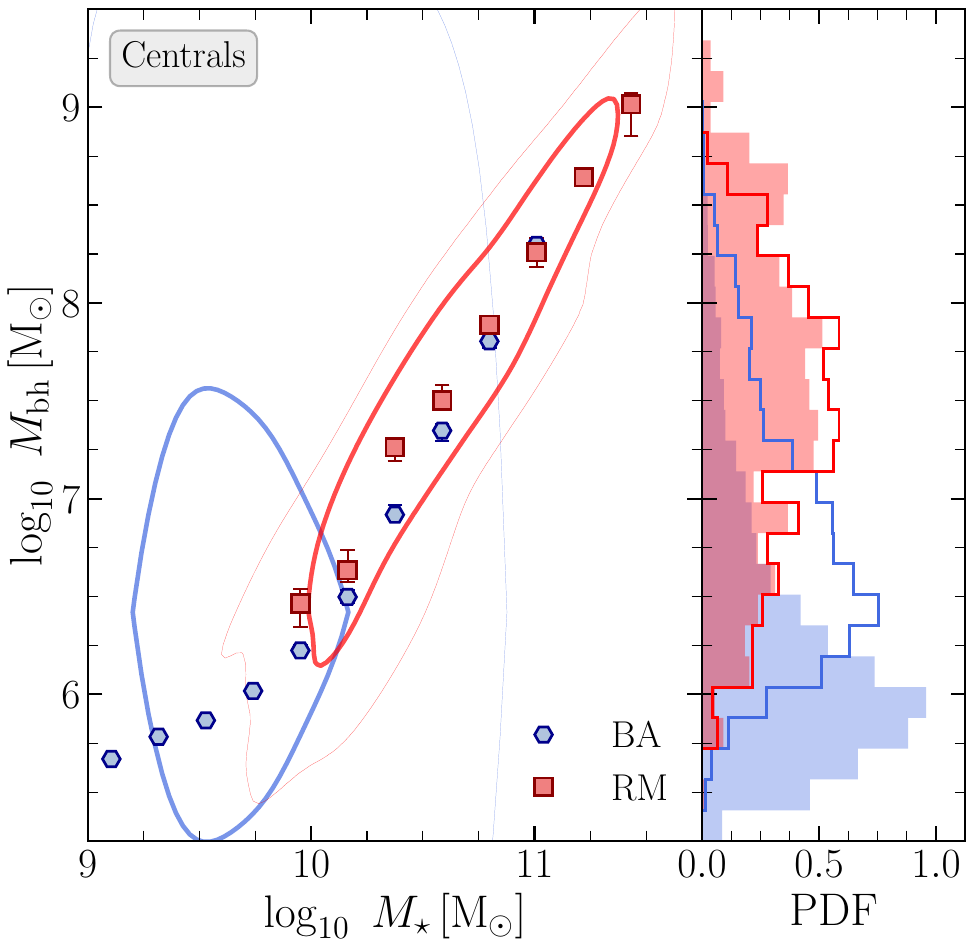}
   \caption{Comparison of BH masses of central RM and BA. For $M_\star\lesssim 10^{10.7}\,{\rm M}_\odot$, the RM 
   possess BHs systematically more massive than those in the BA by $\lesssim 0.45$~dex. This results in an explicit distinction between the respective PDFs.}
   \label{mbhvsmstar}
\end{figure}
We have established in Section~\ref{grouphist} that local environment is a salient contributor to the colour of SF satellites (Figs~\ref{m200vsmstar} and \ref{h2byhivsmstar}), but the physical reasons behind the RM--BA disparity in SF centrals are still uncertain. Besides external factors, feedback processes occurring inside the galaxy are also influential in quenching the galaxy, and rendering it optically red \citep[e.g. see][]{Taylor2015a,Nelson2018,Correa2019}. In fact, the feedback concomitant with AGN activity is particularly efficient in driving the gas out of a galaxy and injecting energy into its CGM \citep[e.g.][]{Zinger2020}. In order for us to understand how this relates to the current state of the galaxy, we can examine a characteristic that captures the cumulative effect of AGN feedbacks across time. Recent works using \eagle centrals have established that this is best achieved through the BH mass \citep{Piotrowska2022,Bluck2023}. This is interesting because, while the energy injected during feedback is a function of BHAR \citep[see][]{Schaye2015}, accretion is not the sole mode for a BH's growth.

Considering this, we compare the BH masses of central RM against that of the BA in Fig.~\ref{mbhvsmstar}. It shows that the RM -- on average -- possess BHs more massive than those in the BA, provided $M_\star\lesssim 10^{10.7}\,{\rm M}_\odot$. 
The offset is $\lesssim 0.45$~dex and exhibits an erratic variation with stellar mass, the largest being at $M_\star\approx 10^{10.5}\,{\rm M}_\odot$. The difference is also evident as a population-wide mismatch between the respective PDFs, even for the OS.

\subsection{Hot gas fraction}\label{hotgas}
Based on the results in Fig.~\ref{mbhvsmstar}, one might expect a tangible impact of AGN feedback on the ISM of central RM,
at least for $M_\star\lesssim 10^{10.7}\,{\rm M}_\odot$. Recall that we did mention (in Section~\ref{coldgas}) a dearth of $\Hmol$ in the RM, which is a phase that -- being centrally concentrated -- is expected to be affected more by such events. We also observe a quantitatively similar offset for the $\HI$ (resulting in the same $\Hmol/\HI$ ratios for the RM and the BA; Fig.~\ref{h2byhivsmstar}). This phase is 
partly replenished via cooling of the hot halo gas, and \citet{Correa2018} showed that AGN feedbacks in centrals typically result in smaller hot gas fractions; which means less gas available for cooling-driven replenishment of $\HI$. Stellar feedback,
on the other hand, tends to increase the hot gas fraction by driving out the ISM gas. This begs the question: have the 
past feedbacks left any signature in the hot gas around the centrals?

\begin{figure}
  \includegraphics[width=1\columnwidth]{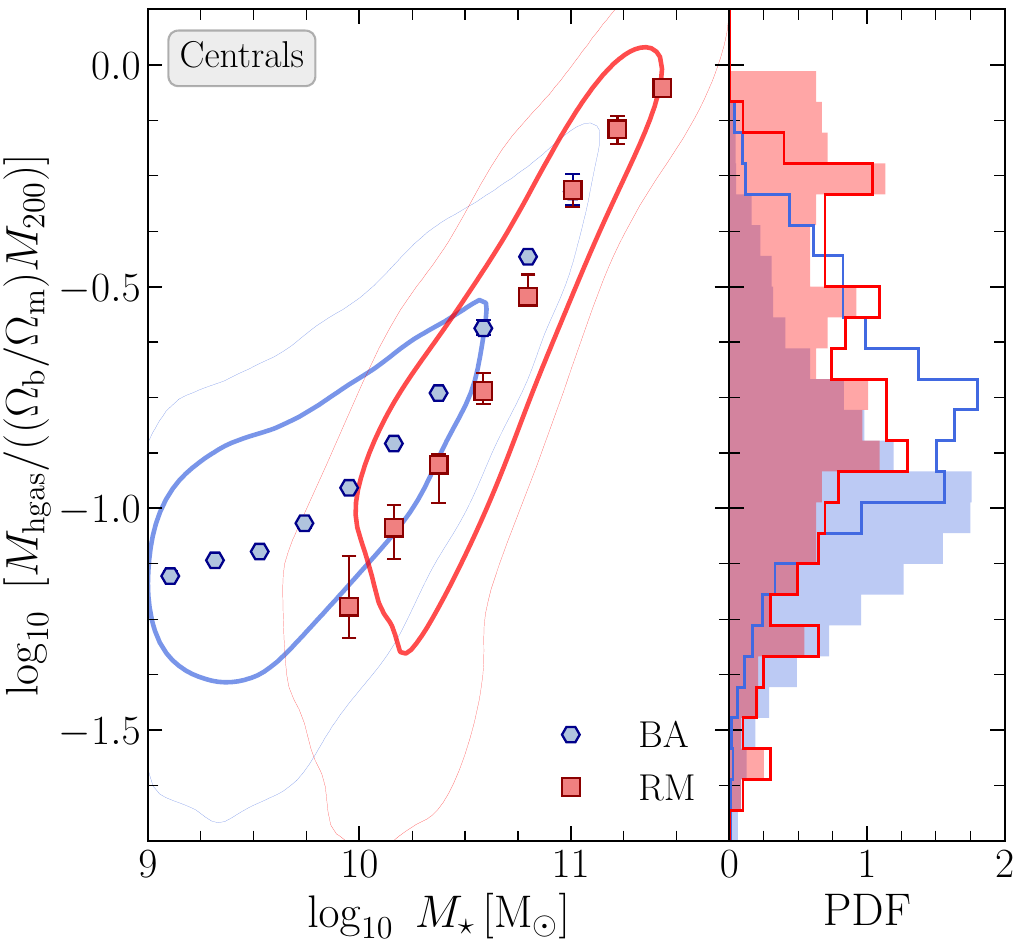}
   \caption{The fraction of halo mass in hot gas plotted against the stellar mass of the centrals. The values
   are normalised by the universal baryon fraction, $\Omega_{\rm b}/\Omega_{\rm m}=0.146$. The haloes of central RM at $M_\star\lesssim 10^{10.7}\,{\rm M}_\odot$ exhibit smaller hot gas fractions relative to the BA. This is the same
   range where the RM show higher BH masses in Fig.~\ref{mbhvsmstar}.}
   \label{hotgasvsmstar}
\end{figure}

We define hot gas as \textit{all} the gas in the host group with temperature greater than the virial temperature of the halo,
\begin{equation}\label{tvir}
T_{\rm vir} = \frac{G\mu m_{\rm H}}{3k}(M_{200})^{2/3}(200\rho_{\rm c})^{1/3},    
\end{equation}
where $m_{\rm H}$ is the mass of the hydrogen atom, $\mu$ is the mean particle mass in units of $m_{\rm H}$, $G$ is the
gravitational constant, and $\rho_{\rm c}$
is the critical density of the universe. We opt $\mu=0.59$ assuming primordial gas with fully-ionised hydrogen and helium \citep{Wijers2022}. For groups spanning $M_{200}/{\rm M}_\odot=[10^{11},10^{14}]$, this implies $T_{\rm vir}/{\rm K} = [10^{4.8},10^{6.8}]$. 

The hot gas fractions for the haloes hosting the centrals are shown in Fig.~\ref{hotgasvsmstar}. 
The RM with $M_\star\lesssim 10^{10.7}\,{\rm M}_\odot$, on average, exhibit fractions that are smaller than those for the BA by $\lesssim 0.4$~dex,
and the offset decreases monotonically with $M_\star$. The results hold even if we exclude the gas in satellites or -- like \citet{Correa2018} -- the gas within $0.15R_{200}$. These differences exist is the same mass range where the RM show higher BH masses (Fig.~\ref{mbhvsmstar}), providing further support for stronger impact of AGN feedbacks on the RM. However, we note that the offset in BH mass does not correlate with that in hot gas fraction. This is partially because BH accretion history can vary for a fixed \textit{final} BH mass, but one cannot definitively rule out contribution from stellar feedback, especially considering the results that follow.

\section{Black hole accretion and star formation history}\label{feedhist}
\begin{figure*}
  \includegraphics[width=1.5\columnwidth]{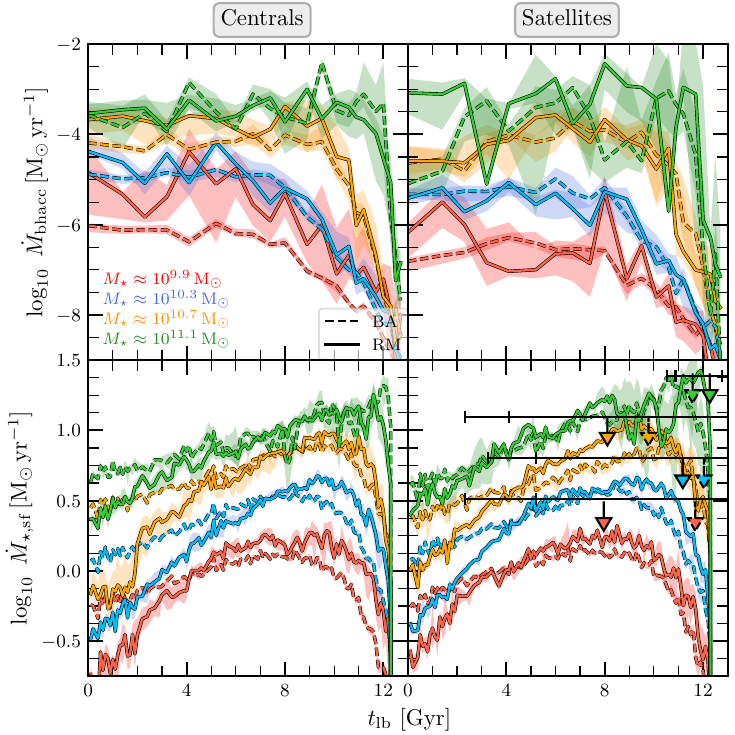}
   \caption{The BHAR (top row) and SFR (bottom row) across time for our centrals (left column) and satellites (right column). The medians for the BA and the RM are shown as dashed and solid curves, respectively. The shaded region around each curve shows the uncertainty derived using bootstrap resampling. The $x$-axis is the look-back time, implying that the cosmic time traverses from right to left. The colours denote the $z=0$ stellar masses mentioned in the bottom-left corner of the top-left panel. Each downward-pointing arrow in the bottom-right panel
   marks the first instance (or the greatest $t_{\rm lb}$) when the average galaxy became a satellite (see the text), and has the same colour and line-style as the curve it represents. The associated
   scatter is shown using the horizontal error-bar that spans 20th to 80th percentile.}
   \label{bharnsfr}
\end{figure*}
As mentioned earlier, BH accretion history can vary even for a fixed BH mass. In principle, this also applies to star formation history and stellar mass, and is strongly suggested for SF \eagle galaxies by the age- and metallicity-offsets in Fig.~\ref{damvsmstar}. This is important with regard to the impact of feedbacks because, for \eagleNS, the energy released due to AGN and star formation activity is proportional to the BHAR and SFR, respectively (see \S4.5 and \S4.6.4 of \citealt{Schaye2015}). Therefore, these characteristics can be leveraged to examine the strength of stellar\footnote{Though
the feedback efficiency has an additional dependence on $f_{\rm th}$ \citep{Schaye2015}, we have
verified that the rate of energy release follows the SFR, supporting the reliability of the latter for this purpose.} and AGN feedback over time.

We calculate the BHAR of each main progenitor as the cumulative, instantaneous accretion rate of all the BHs instead of just the central BH. This is to account for every feedback site
in the galaxy\footnote{Multiple, supermassive BHs are 
indeed expected after a recent galaxy merger, and have even been observed occasionally \citep[e.g.][]{Boroson2009,Deane2014}. However, the simulation likely underpredicts such instances due to overly-efficient coalescence caused by artificial repositioning of BHs to the location of the minimum potential \citep{Booth2009,Tremmel2015}.}. 
Likewise, since we are only interested
in stellar feedbacks within the main progenitors, the star formation history is derived by binning
\textit{in situ} stars according to their birth times in intervals of $\Delta t=100$~Myr, computing the total \textit{birth} mass in each bin, and
normalising by $\Delta t$. 

The median BH accretion and star formation histories for RM and BA are displayed
in Fig.~\ref{bharnsfr}. The figure contains four panels arranged in a $2\times2$ grid. The left and right columns show the results for the centrals and the satellites, respectively. The top and bottom rows show the BHAR and SFR, respectively. In each panel, the solid and dashed curves show the medians for the RM and the BA, respectively. As $\Delta t$ is smaller than the typical time span between
consecutive snapshots,\footnote{The time elapsed between successive snapshots only becomes comparable to $100$~Myr for redshifts beyond $z=5$ (or $t_{\rm lb}\gtrsim 12.66$~Gyr).} the star formation histories exhibit greater variability per Gyr than the accretion histories. The colour of a curve corresponds to a specific $z=0$ stellar mass mentioned in the bottom-left corner of the top-left panel. For the satellites, we
also mark the first instance when they were accreted by any halo using the downward-pointing arrows that follow the same colour and line-style as the curves. These are determined by identifying the earliest snapshot where the galaxy existed as a satellite, and taking the median value of the corresponding look-back times for all the satellites in the stellar mass bin. The horizontal error-bars span the 20th to 80th percentiles.

We first focus on the centrals. The red curves in the top-left panel show that the RM at 
$M_\star\approx 10^{9.9}\,{\rm M}_\odot$ always had BHARs higher than the BA by about 0.5 dex or more. 
In fact, this is also the scale corresponding to one of the largest offsets in BH mass (Fig.~\ref{mbhvsmstar}). 
The only other mass where we see a similar systematic is $M_\star\approx 10^{10.7}\,{\rm M}_\odot$, but mainly 
for the past 7 Gyr and at a smaller magnitude, consistent with the smaller $M_{\rm bh}$-offset. Hence, AGN 
feedback only played a notable role in producing red misfits at these mass scales.

The bottom-left panel shows that there was an additional influence from the stellar feedback at $M_\star\approx 10^{9.9}\,{\rm M}_\odot$, where the RM clearly exhibit higher SFRs until the look-back time of $t_{\rm lb}\approx 4$~Gyr, that is, about
70 per cent of their history. The combined influence of both feedback mechanisms has likely contributed to the maximum
difference in hot gas fraction at this mass ($\approx 0.4$~dex; Fig.~\ref{hotgasvsmstar}). Note that the SFRs
also declined faster for the RM, which explains the large age-offset seen in Fig.~\ref{damvsmstar}. 

Though AGN feedback does not seem to have been relevant for the RM at $M_\star\approx 10^{10.3}\,{\rm M}_\odot$, most of these galaxies did have higher SFRs than the BA around the peak of their star formation history (i.e. $t_{\rm lb}\gtrsim 8$~Gyr). 
The RM at higher masses do not exhibit any difference relative to the BA for most of their star formation history,
which suggests that stellar feedback was not an important factor at those scales. The star formation histories, particularly the early phases, appear more and more similar between the classes with increasing $M_\star$. Since
most of the stars were produced \textit{in situ} (Fig.~\ref{invsexfig}), this reflects as the decline in age-offset with mass (Fig.~\ref{damvsmstar}).

We now turn to the satellites in our sample. The top-right panel in Fig.~\ref{bharnsfr} shows that the BHAR of the satellites
was generally agnostic to RM/BA classification for most masses. Therefore, unlike the centrals, AGN feedback was generally not relevant in the formation of satellite RM. However, as expected from the trends in age (Fig.~\ref{damvsmstar}), 
the RM exhibit steeper star formation histories that peaked above that of the BA, implying stronger stellar feedbacks for all the satellites at $M_\star<10^{11.1}\,{\rm M}_\odot$. 

This does not mean stellar feedback and environmental processes acted together in the formation of satellite RM \textit{at all times}. For this to occur, the galaxies must exist as satellites for the time-span when the feedback was
stronger in the RM. While this did happen for $M_\star\approx 10^{11.1}\,{\rm M}_\odot$ (green arrows), the wide error-bars indicate that this was not always true for the satellites below this scale.

We note that there can be some instances of surge or drop in stellar/AGN activity that were overlooked due to the limited time resolution of the snapshots. This demands caution in interpretation of short-term behaviour of the associated rates. Our main conclusions, however, are likely to hold even after including finer time stamps, because they are based on the general behaviour of the rates over several Gyr.

Next, we extend our analysis on galaxy histories by leveraging the Lagrangian nature of mass elements to track their movement
and transformation with time.

\section{Mass-transfers associated with gas flows, star formation, and stellar migration}\label{galhist}
Here, we further contextualise the results in previous sections by quantifying the impact of specific astrophysical processes on the galaxies and their CGM. The idea is to derive the crucial insights into the formation of RM
that have not been gleaned from the results that came before. As earlier, we carry this out separately for the centrals and the satellites.

For this, we first need to define the ISM component. This is not a trivial exercise, as there is no clear distinction between ISM and CGM. Nonetheless, there are some methods that have served well
in the past in providing reasonable results for \eagle \citep[e.g.][]{van2017,Correa2018,Wijers2020,Smithson2021,Wright2022}. We follow \citet{Wright2022} and define the galaxy's extent, $R_{\rm bary}$, as the radius beyond which the cumulative baryonic mass profile -- based on stars, cool gas ($T<5\times 10^4~{\rm K}$), and SF gas -- resembles an isothermal sphere \citep[see][]{Stevens2014}. The ISM is assumed to be all the cool and SF gas within $R_{\rm bary}$, and the rest of the gas bound to the subhalo constitutes the CGM\footnote{The CGM of a central does not include the gas in its satellites.}. Similarly,
the galactic stars are all the stars within $R_{\rm bary}$, while the rest represent
its stellar halo. 

We begin by tracking the gas and star particles between snapshots to determine the gaseous and metallic mass that was lost or added as a result of:
\begin{enumerate}
    \item star formation,
    \item removal of ISM gas from the subhalo,
    \item removal of CGM gas from the subhalo,
    \item movement of gas from ISM to CGM,
    \item movement of gas from CGM to ISM,
    \item addition of fresh (previously unbound) gas into the ISM, and
    \item addition of fresh gas into the CGM.
    \newline We also quantify the stellar and metallic mass that was lost or added due to:
    \item loss of halo stars from the subhalo,
    \item removal of galactic stars from the subhalo, and
    \item introduction of fresh stars into the galaxy via accretion. 
\end{enumerate}
The last three represent the addition and loss of stars concomitant with mergers and tidal interactions.

\begin{figure*}
\centering
  \includegraphics[width=2.07\columnwidth]{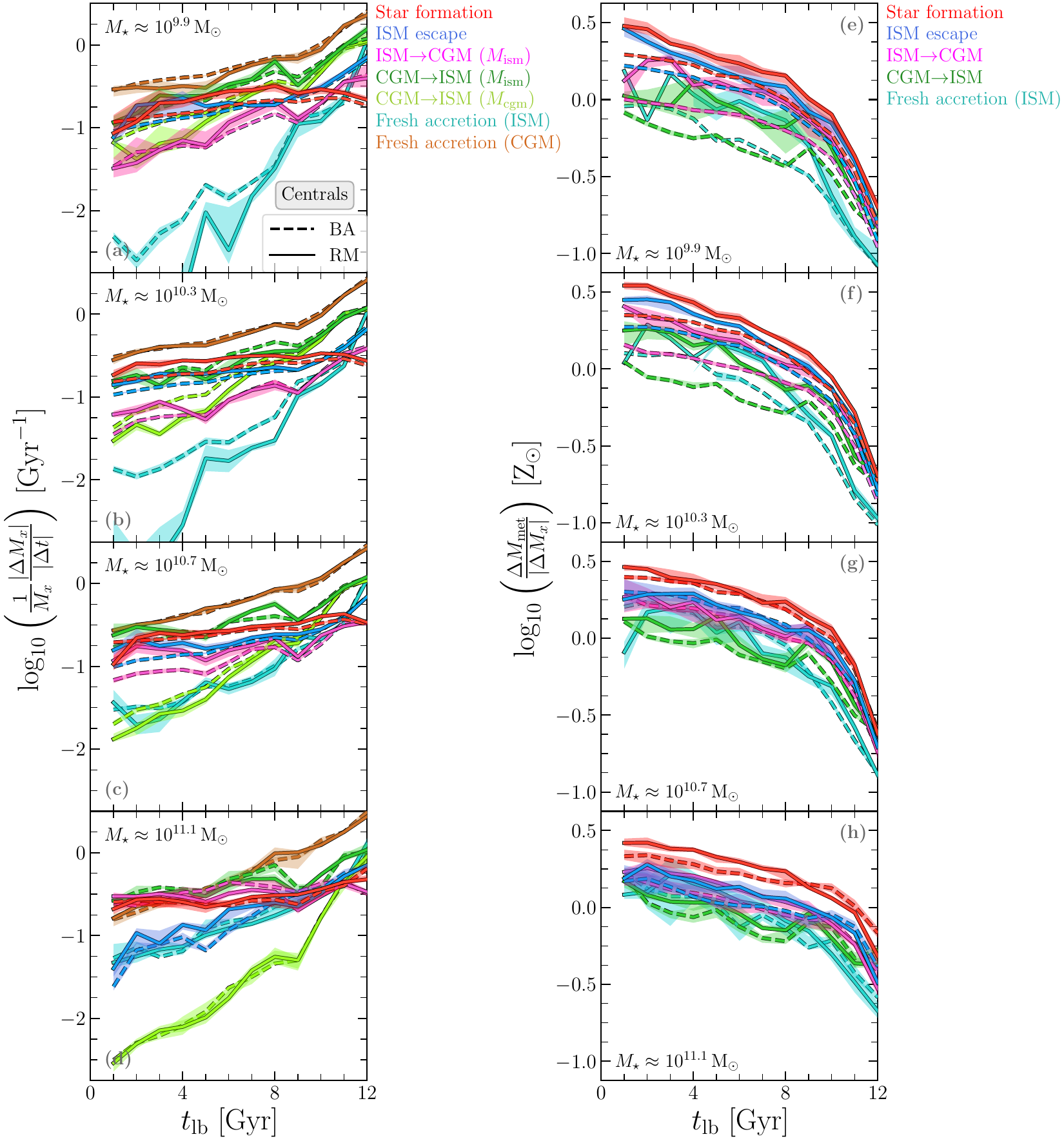}
   \caption{Specific gas-/stellar-mass transfer rates (left column) and metallicities (right column) across time for various processes involved in the evolution of the centrals. The choice of normalising mass ($M_x$) depends on the process (see the text). Each panel shows the results for the galaxies whose $z=0$ stellar mass lies within $0.1$ dex of that mentioned in its bottom-left corner; the $M_\star$ increases from top to bottom. The medians for the BA and the RM are shown as dashed and solid curves, respectively. The shaded region around each curve shows the uncertainty derived using bootstrap resampling. The colour pertains to the astrophysical process mentioned in the top-right. The dashed and solid curves correspond to the BA and the RM, respectively.}
   \label{cenex}
\end{figure*}

The results for the centrals and the satellites are displayed in Figs~\ref{cenex} and~\ref{satex}, respectively. The
figures include two vertical columns of panels, where each panel corresponds to the $z=0$ stellar mass mentioned in a corner. In the left column, each curve shows the mass-loss/gain rate normalised by the mass that is relevant \textit{for that process}, and thus, acts as a proxy for the efficiency of the process in modulating the overall gas (or stellar) content. In addition, the
normalisation controls for the diversity in progenitor mass at a fixed $z=0$ stellar mass\footnote{The gas mass, for example, can vary by $\approx 0.3$ dex.}. The dashed- and solid-curves are the medians for the BA and the RM, and the shaded regions are their bootstrapping uncertainties. The astrophysical process corresponding to a curve is denoted using its colour and can be gauged through the legend next to the top-most panel. 

The rates for star formation (red), ISM escape (blue), ISM$\rightarrow$CGM (magenta), CGM$\rightarrow$ISM (green), and fresh accretion (ISM; turquoise) are normalised by the ISM mass. This means that the rates portray the impact of these processes on the ISM. Note that the rate used for star formation is the fraction of ISM mass converted to stars per Gyr, otherwise known as the star formation efficiency (SFE). We also show CGM$\rightarrow$ISM rate normalised by the CGM mass (green-yellow) wherever required. The rates for CGM escape (orange) and fresh accretion (CGM; brown) are normalised by the CGM mass ($M_{\rm cgm}$). The rates for fresh accretion (stars) and stellar escape (grey) are normalised by the galactic stellar mass (i.e. the stellar mass within $R_{\rm bary}$), and that for stellar escape (halo; yellow) is normalised by the stellar
halo's mass. Panels [(e) to (h)] in the right column have a similar arrangement to the left column, but show the mass-weighted average metallicity of the gas/stellar particles involved in the process.
 
\subsection{Centrals}\label{cenflow}
\subsubsection{How did these processes shape the stellar growth?}\label{sfh}
Panel (a) shows that, before $t_{\rm lb}\approx 9$~Gyr, the ISM of both the RM and the BA underwent a net replenishment due to CGM$\rightarrow$ISM (green) and at the same efficiency. By comparing the temperatures of gas particles in successive snapshots, we find that CGM$\rightarrow$ISM was \textit{always} accompanied by cooling, and the gas had a temperature\footnote{This is the mean temperature of the gas particles.} $10^{4.4}\lesssim T/{\rm K}\lesssim 10^{5.1}$ in
the last snapshot before mixing with the ISM; characteristic of cooling flows originating from the hot halo \citep{Stevens2017}. The red curves show that star formation was more efficient for the RM; indeed, the RM exhibit higher SFRs over this
time span (bottom-left panel, Fig.~\ref{bharnsfr}). During this time, ISM escape was the primary depletion mode, but gradually waned in its efficiency. 

During $6\lesssim t_{\rm lb}/{\rm Gyr}\lesssim 9$, star formation overtook the escape for the RM, and became close to (but slightly less than) CGM$\rightarrow$ISM. Meanwhile, for the BA, the decline in the escape rate caused it to equate to star formation, such that the total depletion rate (star formation + ISM escape) remained slightly less than the total replenishment rate (i.e. green). Thus, the gas reservoir incremented at similar rates for both the classes, but the RM formed stars at a higher efficiency, sustaining the higher SFRs.

For $t_{\rm lb}\lesssim 6$~Gyr, the escape rate increased for the RM and progressively approached star formation, which commenced a net depletion of their ISM, and hence, a decline in their SFRs (Fig.~\ref{bharnsfr}). For the BA, CGM$\rightarrow$ISM rate reduced and got a bit closer to the \textit{combined} rate for star formation and ISM escape, stifling any \textit{net} gains in the ISM
mass. The SFRs nevertheless declined because of the reduction in SFEs, albeit at a weaker rate than the RM. This enabled the BA to compensate for the lower stellar content relative to the RM during initial stages, and end up with the same stellar mass at $z=0$.

We would like to point out that stellar escape typically removed $\lesssim 0.5$ per cent of the stellar mass at any given moment, and hence, was never an important factor in the evolution of the centrals. Similarly, fresh accretion of stars
added $\lesssim 10$ per cent to the stellar mass. Likewise, interactions with the stellar
halo were inconsequential, because the loss of stars balanced the accretion. Therefore, the corresponding curves have been omitted from the figure for the sake of clarity.

Comparative analyses of the above results with those for the other stellar mass bins provide important insights into the decreasing age-offset with mass (Fig.~\ref{damvsmstar}). Consider panel (b) of Fig.~\ref{cenex}, corresponding to $M_\star\approx 10^{10.3}\,{\rm M}_\odot$. For these galaxies, star formation was always the predominant mode of gas depletion, regardless of their class. The corresponding rates for the RM were again higher, but by a smaller magnitude than the previous bin for most instants. Since the replenishment rates (CGM$\rightarrow$ISM) were independent of class, the difference in star formation histories was not as drastic as that for $M_\star\approx 10^{9.9}\,{\rm M}_\odot$. Panels (c) and (d) show that the differences between the RM and the BA for these rates got progressively smaller with increasing $M_\star$, which manifested in the star formation histories (Fig.~\ref{bharnsfr}), and the final ages of the constituent stellar populations.

\subsubsection{The flow of metals and its implication for the stars}\label{metflow}
We know that star formation and ISM escape were the main depletion modes for the galaxies at $M_\star\approx 10^{9.9}\,{\rm M}_\odot$, and SFEs of the RM were usually higher than the BA. The gas particles in \eagle can gain metals from stars due to supernovae or stellar winds (for details, see \S4.4 of \citealt{Schaye2015}). This means that a rise in SFE can enhance gas metallicity, provided the affected gas is not removed or mixed with metal-poor clouds. We examine the rate of change/enhancement of metallicity for the gas \textit{that remained} in the ISM between two consecutive snapshots. The rates suggest that the ISM of RM was enriched more than BA, and the offset between their enhancement rates followed that in the SFEs (see Appendix~\ref{sfenrich}). 
Conversely, panel (e) indicates that the cooling flows carried lower metallicities than the ISM, and therefore served as a dilution mechanism. Though these flows had the same efficiency regardless of the class, they did not
carry the same metallicities. In fact, the metallicity-difference between ISM (red) and CGM$\rightarrow$ISM (green) was generally smaller for the RM, implying poorer dilution. This, along with the mass-transfers from stars, was conducive for the higher $Z_\star$ of the RM. Similar trends exist at higher masses, albeit with progressively smaller differences between the classes with increasing mass, leading to smaller metallicity-offsets in Fig.~\ref{damvsmstar}.

\subsubsection{Why was the metallicity of CGM$\rightarrow$ISM gas historically higher for the RM?}\label{cgmtoism} 
The higher metallicity of CGM$\rightarrow$ISM gas was a key factor in better enrichment of the RM. In principle, 
this is attributable to a) higher metallicity of freshly accreted gas for the RM, and b) higher metallicity of ISM$\rightarrow$CGM gas. Although it is not straightforward to investigate the origin of the former, and is outside
the scope of this work, we can provide a short explanation for the latter below.

After arriving at the galaxy from the circumgalactic space, some clouds are transformed into stars, while the others -- along with those already present in the galaxy -- are enriched by stars due to supernovae or stellar winds. The supernovae also heat up the gas around the stars through energy feedback, which can drive them out of the ISM. A similar transport can be induced by feedback from actively-accreting BHs in the galaxy, mostly affecting the gas near the galactic centre, which tends to be the most enriched gas in the galaxy. Therefore, feedbacks tend to drive out relatively enriched gas, and ISM$\rightarrow$CGM gas is always more enriched than CGM$\rightarrow$ISM gas [as evident in panel (e)]. Since the gas around star forming regions was
enriched more in the RM (due to the reasons delineated in Section~\ref{metflow}), and this tends to be the gas
affected by feedbacks, the ISM$\rightarrow$CGM flows also exhibit higher metallicities for
the RM.

\subsubsection{What do the gas flows suggest about the feedback mechanism?}\label{feedback}
Fig.~\ref{cenex} shows that there was a moment when the ISM escape rate for the RM started to rise relative to the BA, and for $M_\star\lesssim 10^{11.1}\,{\rm M}_\odot$, this was also accompanied by a simultaneous
decline in the efficiency with which the CGM cooled and flowed towards the galaxy (green-yellow). These trends are rather 
expected, considering that feedbacks are known to eject mass from the galaxy, and increase
the cooling time of the surrounding diffuse gas by heating it (the so-called ejective and preventative nature). We note that ISM escape did not follow star formation for the RM at most masses. One would expect otherwise if stellar feedback was indeed the main modulating factor for the escape rate. This also applies to ISM$\rightarrow$CGM flows. We believe this is due to the contribution from AGN feedback, at least for the centrals at $M_\star\approx 10^{9.9}\,{\rm M}_\odot$ and $M_\star\approx 10^{10.7}\,{\rm M}_\odot$ (see Section~\ref{feedhist}). Altogether, these trends complement the results in Section~\ref{agn}, Section~\ref{hotgas} and Section~\ref{feedhist} in highlighting the importance of \textit{both} feedback mechanisms in the formation of RM.


\subsection{Satellites}\label{satflow}
\begin{figure*}
\centering
  \includegraphics[width=2.07\columnwidth]{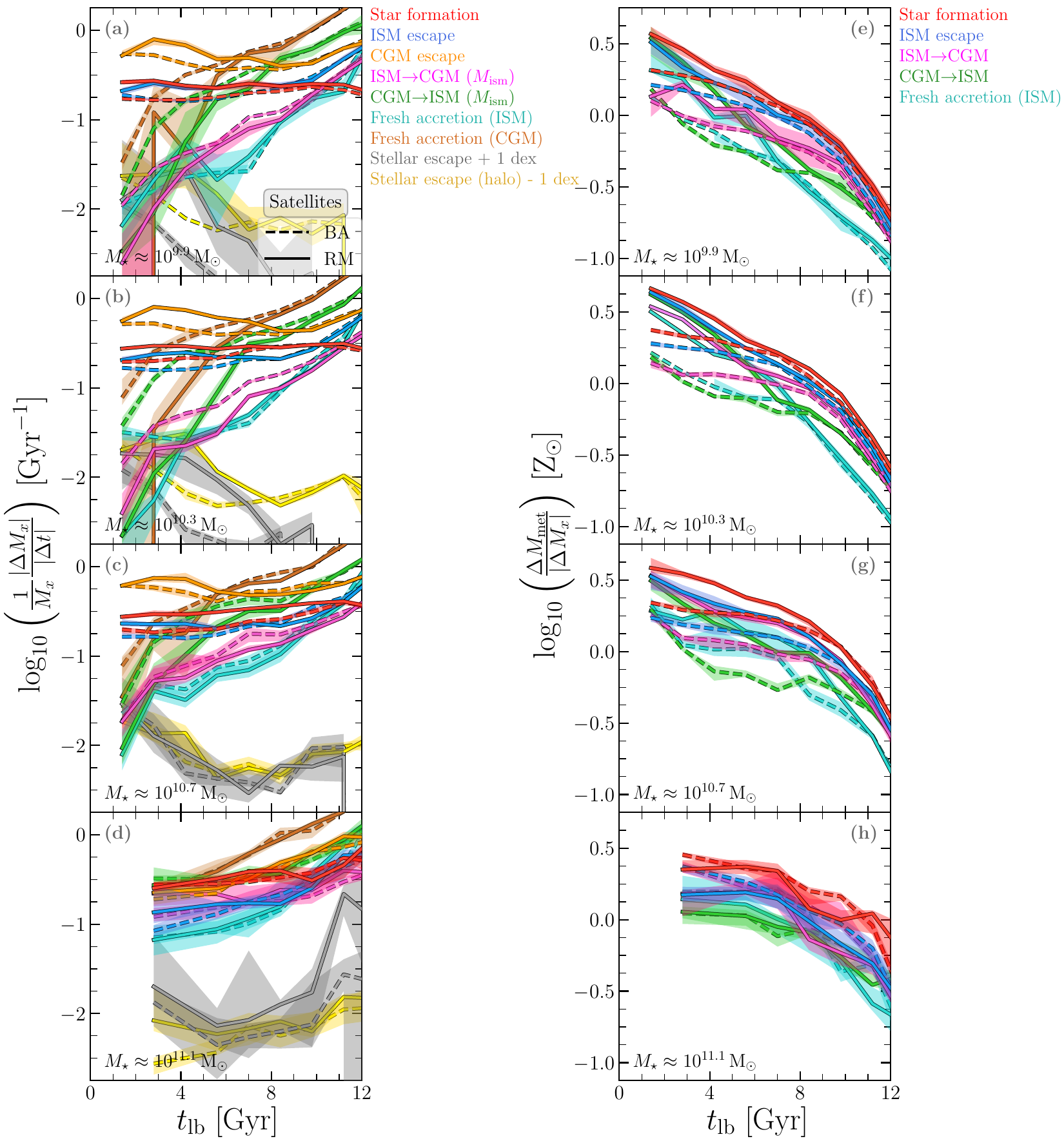}
   \caption{The mass-transfer rates and metallicities associated with various processes acting on the satellites. The
   plotting style is the same as Fig.~\ref{cenex}.}
   \label{satex}
\end{figure*}
\subsubsection{The growth of stellar mass}\label{sfhsat}
The rates associated with the ISM in panel (a) of Fig.~\ref{satex} convey that both the RM and the BA were undergoing a net growth of gas reservoir before $t_{\rm lb}\approx 8$~Gyr, driven by CGM$\rightarrow$ISM (green) and with equal efficiencies. The SFEs were slightly higher for the RM, resulting in higher SFRs (bottom-right panel, Fig.~\ref{bharnsfr}). Post $t_{\rm lb}\approx 8$~Gyr, the replenishment rate (CGM$\rightarrow$ISM) reduced and was lower than the total depletion rate due to star formation and ISM escape. Though the RM still had higher SFEs, they were also losing ISM at a faster rate due to smaller CGM$\rightarrow$ISM rates and higher ISM escape rates. Therefore, their SFRs incurred a steeper decline, and eventually reduced below those for the BA. This explains the age-offset at $M_\star\approx 10^{9.9}\,{\rm M}_\odot$ (Fig.~\ref{damvsmstar}). 

The curves at higher $M_\star$ exhibit qualitatively similar behaviours. For instance, SFEs were again higher for RM for
at least half of their history, and the replenishment of ISM (via CGM$\rightarrow$ISM) was less efficient. However, the discrepancy between the total depletion rates (star formation + ISM escape) reduced between the RM and the BA, which manifested
as smaller offsets in the median ages. The offset is absent at $M_\star\sim 10^{11}\,{\rm M}_\odot$ because the replenishment rates
were also equal between the classes.

\subsubsection{How do the processes relate to the final stellar metallicity?}\label{metflowsat}
Prior to $t_{\rm lb}\gtrsim 8$~Gyr, the SFE was mildly enhanced for the RM at $M_\star\approx 10^{9.9}\,{\rm M}_\odot$. As expected, this induced a better enrichment of the ISM (bottom panel, Fig.~\ref{dism}), which also reflected as higher metallicities for the processes involving the ISM [i.e. star formation, ISM escape, and ISM$\rightarrow$CGM; Fig.~\ref{satex}]. This disparity in SFE, and therefore the ISM metallicity (close to that of ISM escape), got even more pronounced for the evolution post $t_{\rm lb}\approx 8$~Gyr. Similar to the centrals (Section~\ref{metflow}), the cooling flow from the CGM promoted dilution (metallicity was less than that of the ISM). However, this was more effective for the BA due to a larger offset between the metallicity of the ISM and CGM$\rightarrow$ISM. The galaxies at higher masses exhibit similar trends, and hence, $Z_\star$-offsets (right column, Fig.~\ref{damvsmstar}).

\subsubsection{Indications of intragroup processing}\label{intragroup}
Recall that the degree of influence from the group environment on the satellites is expected to vary between the RM and the BA (Section~\ref{grouphist}). As it turns out, this is also supported by some of the trends presented in
Fig.~\ref{satex}, as explained below.

One clear indication of environment is discernible in the fresh-accretion rate for the CGM (brown), which has been specifically included in the figure for illustrating this effect. The curve shows that the rate was lower for the RM for the past 6 Gyr or so. Both the RM and the BA were orbiting other galaxies over this whole time span (bottom-right panel, Fig.~\ref{bharnsfr}). Interestingly, we find that $t_{\rm lb}\approx 6$~Gyr corresponds to the instant when the RM typically began to enter haloes more massive than the ones hosting the BA. As such, the drop in the accretion rate for the RM is well within the expected behaviour of accretion rate with halo mass \citep{van2017,Wright2022}. The primary source of \textit{fresh} accretion for these galaxies is the intragroup gas, which tends to get hotter with $M_{200}$ [see equation~(\ref{tvir})]. In fact, the CGM$\rightarrow$ISM rate also exhibits a stronger reduction for the RM for $t_{\rm lb}\lesssim 10$~Gyr. This discrepancy is not exhibited by the centrals (Fig.~\ref{cenex}) because they have been residing in similar-mass haloes (as alluded to by Fig.~\ref{m200vsmstar}). 

CGM escape rate (orange) has also been higher for the RM, and involved removal $\gtrsim 50$ per cent of their CGM since the last 6 Gyr. In fact, at its peak ($t_{\rm lb}\approx 2.8$~Gyr), nearly 80 per cent of the gas was escaping from the RM' CGM. Additionally, the rate had increased even when the SFE or the ISM escape rate remained constant. This indicates that a process in addition to feedbacks has possibly been driving this loss, particularly in light of the results in Section~\ref{grouphist}. 

Collisionless mass elements, like stars, are more apt for discerning tidal stripping, because gravitational force is the sole governing factor in their dynamics. Indeed, we find that the rates for stellar escape, from the galaxy (grey) as well as the stellar halo (yellow), were generally higher for the RM by $\gtrsim 0.3$~dex in the last 8 Gyr. We also find greater loss of dark matter for these galaxies (not shown). In addition, the curves for CGM escape and stellar escape (halo) exhibit similar shapes, hinting that the escape modes were \textit{predominantly} governed by the same mechanism.

\section{Gas angular momentum, accretion geometry, and star formation efficiency}\label{accnsfe}
\begin{figure*}
\centering
  \includegraphics[width=1.18\columnwidth]{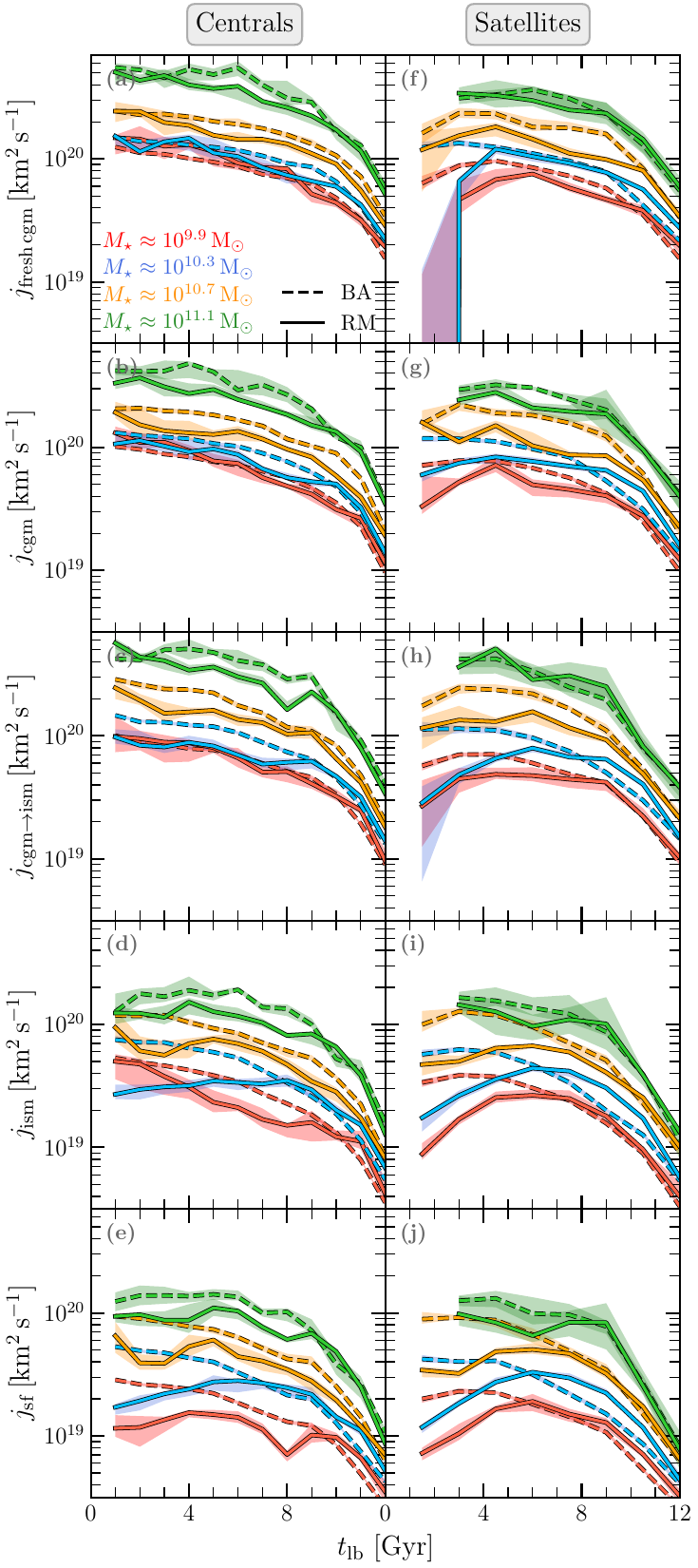}
   \caption{Specific angular momenta across time for gas flows, ISM and CGM in our centrals (left column) and satellites (right column). The plotting style is the same as Fig.~\ref{bharnsfr}. The gas corresponding to the quantity is mentioned in the subscript of the $y$-axis' label.}
   \label{gaskin}
\end{figure*}
The ability of incoming gas to fuel star formation is contingent on the magnitude of its spin, and the direction from which it is being accreted. If this gas possesses a high specific angular momentum (angular momentum normalised by the mass: $j$) 
well-aligned to that of the galaxy, it settles into rotationally-supported orbits at large radii. This is typical of cool gas accreting onto simulated galaxies \citep{Stewart2011,Stanic2016,Trapp2022}, but can sometimes be valid for gas that has originated in a hot halo \citep{Hafen2022}. However, it is the inner, low-$j$ gas that exhibits the conditions conducive for compression and collapse \citep{Obreschkow2016,Badry2018} -- implying that the gas is more likely to undergo star formation if it reaches inner parts of the galaxy. The prevailing understanding is that this occurs via inward advection due to angular-momentum loss, induced by non-axisymmetric torques from galactic structures formed as a result of gravitational instabilities \citep{Goldbaum2015,Grand2015,Krumholz2018,Trapp2022}. The SFE is, therefore, partially regulated by the pace of this migration, which is better facilitated if the gas has a low $j$ \citep[e.g.][]{Wang2022}. Additionally, misaligned accretion can induce a global loss of $j$ \citep{Sales2012}. In what follows, we explore this potential link between SFE and gas dynamics, as this could very well be fundamental to the higher SFEs of RM (shown in Figs~\ref{cenex} and~\ref{satex}).

We know that CGM kinematics has important implications for the growth of galaxy's $j$.
This has especially been true for our galaxies, as the cooling flows were the main source of their ISM's replenishment. Moreover, such flows can sometimes represent a stage in the fountain cycle, wherein the gas ejected from the galaxy acquires angular momentum by mixing with the high $j$ CGM, and comes back to the ISM with more $j$ than the initial value \citep{Ubler2014,DeFelippis2017,Grand2019}. Hence, we first examine the $j$ of our galaxies' CGM over time.

Panels (a) and (f) in Fig.~\ref{gaskin} show the $j$ of freshly accreted CGM for the centrals and the satellites, respectively. Each colour corresponds to a certain $z=0$ stellar mass, in accordance with the legend in the bottom-left corner of panel (a). Fresh accretion was the primary mode of the CGM's replenishment, and this generally carried slightly lower $j$ for RM. Consequently, the global $j$ of CGM ($j_{\rm cgm}$) was also smaller [panels (b) and (g)], albeit with a larger offset. The latter, for the centrals, was due to mass-, momentum- and energy-transfer linked to feedback events. For the satellites, there was additional influence from tidal forces and intrahalo-gas pressure during orbit traversal. 

Unsurprisingly, the disparity in $j_{\rm cgm}$ also reflected in the cooling flow from CGM to ISM, such that RM ended up accreting gas with lower $j$ than BA [panels (c) and (h)]. Note that this is the $j$ \textit{before} the gas was incorporated into the ISM. Since the ISM was mainly accumulated via this mode, $j_{\rm ism}$ has also been lower for RM [panels (d) and (i)]\footnote{Note that
feedback-driven expulsion of gas does not reduce $j_{\rm ism}$, because this mainly removes the low angular momentum gas, and facilitates the growth of $j_{\rm ism}$ instead.}. As expected, the lowest $j$ ISM gas that eventually underwent star formation had also carried lower $j$ for RM at the snapshot before its collapse. This suggests faster loss of angular momentum and advection of gas towards the central regions, which means faster gas$\rightarrow$star conversion or higher SFE. The most-massive centrals at $M_\star\approx 10^{11.1}\,{\rm M}_\odot$ are exceptions
to this, as the RM exhibit lower $j_{\rm ism}$ for the last $\approx 8$~Gyr, and yet, do not show strong differences in SFE (Fig.~\ref{cenex}). The
precise reason remains abstruse, and requires further investigation that we defer to future studies.

\begin{figure}
\centering
  \includegraphics[width=0.8\columnwidth]{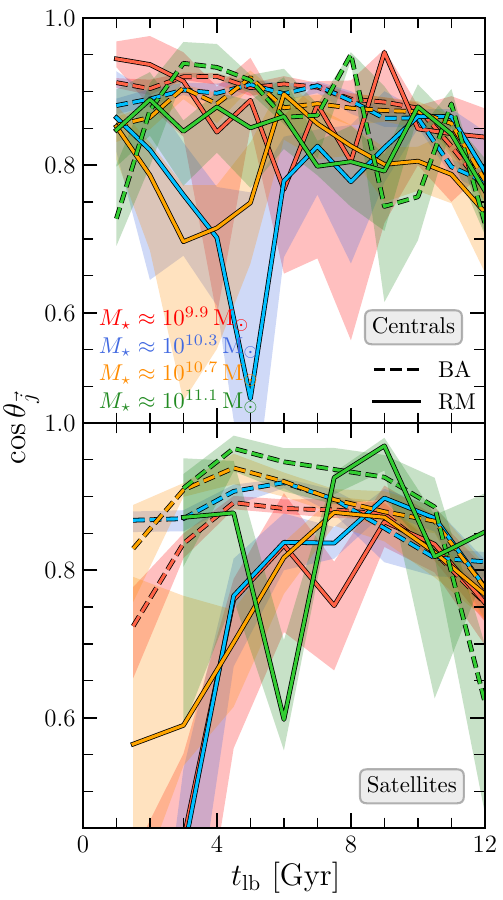}
   \caption{The alignment between the angular momenta of the ISM and the gas that is to be accreted from the CGM. The
   values increase with the degree of alignment, with $1$ being perfect, and $-1$ corresponding to diametrically
   opposite vectors. The top and bottom panels show the alignments across time for the centrals and the satellites, respectively. The accreting gas was generally less aligned with the ISM for RM, with a larger offset for the satellites.}
   \label{accprop}
\end{figure}

The $j_{\rm ism}$ has, on average, been less than $j$ of the accreted gas by $\gtrsim 0.3$~dex. One would indeed expect this, considering the influence from galactic structures and feedbacks. To add to this, accretion that is kinematically-misaligned would only partially contribute to the co-rotation within the ISM, and if it is counter-rotating, may even reduce it \citep[e.g][]{Taylor2018}. Fig.~\ref{accprop} presents the alignment between $\vec{j}_{\rm ism}$ and the flow's $\vec{j}$ \textit{prior} to its accretion. This has been computed as the cosine of the angle between the vectors, where a value of $1$ represents perfect alignment, and $-1$ represents opposing directions.

It shows that the cooling flows were, in general, well-aligned with the galaxy at $\gtrsim 50$ per cent, and more so for BA ($\gtrsim 80$ per cent); that is, they rarely carried a counter-rotating component. However, the central RM at $M_\star/{\rm M}_\odot = [10^{10.3},10^{10.7}]$ (in the top panel) exhibit poorer alignment for most of their history. As per the bottom panel, this also applies to the satellites, but the differences were more pronounced and rather drastic during the last 7 Gyr. Moreover, unlike the centrals, this discrepancy was experienced by the satellites at all masses. The larger deficits than the centrals are, at least in part, due to stronger environmental influence on the RM that is understood to be mostly from tidal effects \citep{Marasco2016} for their $M_{200}$ span (Fig.~\ref{m200vsmstar}). These results are in line with a scenario where, not only did the cooling flow arrive with smaller $j$ for RM (Fig.~\ref{gaskin}), less of its spin could couple to co-rotation within the ISM.

\section{Summary and Conclusions}\label{conc}
`Red misfits' (RM) refers to the group of star-forming (SF) galaxies that appear red in their $g-r$ colour, but are distinct from red spirals or S0s \citep{Evans2018}. Although the existence of this class has been known for about two decades \citep[e.g.][]{Wolf2005}, a direct theoretical investigation of their formation has not been undertaken. We conducted the first numerical study of this kind using the flagship run (Ref-L0100N1504; \citealt{Mcalpine2016}) from the \eagle suite of hydrodynamical simulations \citep{Crain2015,Schaye2015}.

We drew inferences about the physics of RM' formation by comparing the statistical behaviour of RM and BA in various bins of stellar mass. For this, we identified and segregated SF galaxies into the two classes by enforcing thresholds on $M_\star$, sSFR and $g-r$ in accordance with \citet{Chown2022}; and ensuring that the galaxies possess well-resolved dust distributions \citep{Camps2018}. We found that nearly all the SF galaxies below $M_\star\approx 10^{9.9}\,{\rm M}_\odot$ are BA and those above $M_\star\approx 10^{11.1}\,{\rm M}_\odot$ are RM (Fig.~\ref{damvsmstar}), and therefore focussed the analysis on the range bracketed by these limits to facilitate like-for-like comparisons. This was carried out separately for the centrals and the satellites.

The main outcomes are as follows:
\begin{enumerate}
    \item RM exhibit lower dust contents (by $\approx 0.5$~dex), older ages ($\lesssim 0.1$~dex), and greater metallicities ($\approx 0.05$~dex), such that the offset in age generally reduces with increasing $M_\star$ (Fig.~\ref{damvsmstar}). These trends are in agreement with observations \citep{Evans2018,Chown2022}, and provide the first theoretical support for the idea that the red colouration of RM is \textit{not} a product of
    dust-reddening, and rather stems from slightly older ages and higher metallicities of RM.
    \item Central RM do not differ from the BA in terms of \textit{in situ} fraction (top panel, Fig.~\ref{invsexfig}), mass of the host group ($M_{200}$; top panel, Fig.~\ref{m200vsmstar}), or its formation time (Fig.~\ref{tformhalo}). This means that their origin cannot be attributed to varied assembly histories of the host haloes of RM and BA.
    \item However, they do possess more-massive BHs ($\lesssim 0.45$~dex; Fig.~\ref{mbhvsmstar}) and smaller hot gas fractions ($\lesssim 0.4$; Fig.~\ref{hotgasvsmstar}), suggesting a significant role of feedbacks. In fact, compared to central
    BA, the RM typically had higher SFRs and/or BHARs (depending on the $z=0$ $M_\star$) for considerable portions of their histories (left column, Fig.~\ref{bharnsfr}).
    \item Satellite RM, on the other hand, clearly show signs of a higher degree of intragroup processing
    relative to their BA counterparts. This is evident, for example, in the mass of host group (bottom panel, Fig.~\ref{m200vsmstar}) and molecular-to-atomic gas ratio ($\Hmol/\HI$; bottom panel, Fig.~\ref{h2byhivsmstar}), both of which are greater for the RM by more than $0.5$~dex and $0.3$~dex, respectively. This is in addition to the impact of feedbacks, which usually sourced from stellar activity (right column, Fig.~\ref{bharnsfr}), but did not necessarily act at the same time as environmental mechanisms (refer Section~\ref{feedhist}).
    \item Tracking of baryons over time showed that the difference in star formation histories of RM and BA (bottom row, Fig.~\ref{bharnsfr}) -- which caused the offsets in age (Fig.~\ref{damvsmstar}) -- arose due to disparity in the efficiencies of gas flows and star formation (Section~\ref{sfh} and Section~\ref{sfhsat}). RM typically had higher SFEs than BA (left columns, Figs~\ref{cenex} and~\ref{satex}), which resulted in higher SFRs during early phases for a vast majority. However, since the accretion rates were declining over time, this also induced an earlier (and faster) depletion of their gas reservoir, and a steeper decline of their SFRs. The latter was also facilitated by a more efficient escape of ISM and, for the centrals at $M_\star\approx 10^{10.7}\,{\rm M}_\odot$, migration of gas from ISM to CGM. As stated in (iii) and (iv), the mechanisms responsible for the escape varied between the centrals and the satellites (Section~\ref{intragroup}).
    \item This interplay of gas flows and star formation, together with the corresponding metallicities, allowed us to elucidate the physical underpinnings of the higher metallicities of RM (Section~\ref{metflow} and Section~\ref{metflowsat}). The ISM in RM had enriched more efficiently than that in BA due to two reasons. First is the higher SFEs of RM that aided in faster transfer of metals from stars to ISM through supernovae and stellar winds (illustrated in Fig.~\ref{dism}; Appendix~\ref{sfenrich}). Second is the more efficient dilution of gas metallicity in BA via mixing with cooling flows from the CGM, partly due to lower metallicity of these flows in comparison to those interacting with RM' ISM (right columns, Figs~\ref{cenex} and~\ref{satex}).
    \item We found that the disparity in SFEs of RM and BA can be owed to the smaller $j$ of the freshly-accreted CGM around RM (Fig.~\ref{gaskin}). This was conducive for the lower $j$ of CGM around a typical RM, in addition to some other factors (Section~\ref{accnsfe}). Eventually, this also reflected in the kinematics of ISM due to incorporation of the cooling flows. 
    We believe that the lower $j$ aided in faster loss of angular momentum and advection of gas towards the inner regions of RM,
    thereby inducing a more rapid conversion of the cold gas reservoir to stars. 
    \item Additionally, we found that the spin of cooling flows was typically less aligned with that of the ISM of central RM at intermediate masses, and that of satellite RM at all masses (Fig.~\ref{accprop}). This suggests poorer dynamical-coupling of the incoming gas with the ISM,
    which was conducive for further reduction in $j_{\rm ism}$.
\end{enumerate}

Our findings highlight several important aspects of the formation of RM, but there is immense scope for further developments in this line of research. To begin with, the overwhelming absence of red misfits at $M_\star<10^{9.9}\,{\rm M}_\odot$ and their presence at $M_\star>10^{11.1}\,{\rm M}_\odot$ [see the PDFs in (d) panels in Fig.~\ref{damvsmstar}] are potential avenues for prospective works. One can utilise the full potential of \eagle model by re-simulating the volume at better mass resolutions, as it would enhance the statistics of galaxies with resolved dust, particularly in the bottom-left corner of Fig.~\ref{ssfrvsmstar}. Some of our results in Section~\ref{juxt} can serve as predictions that ought to be compared to observations: this includes $M_{200}$--$M_\star$ relations (Fig.~\ref{m200vsmstar}), $\Hmol/\HI$ ratios (Fig.~\ref{h2byhivsmstar}), the $M_{\rm bh}$--$M_{\star}$ relation (Fig.~\ref{mbhvsmstar}), and the hot-gas fractions (Fig.~\ref{hotgasvsmstar}).  Also, there are some alternative cosmological, hydrodynamical simulations based on $\Lambda$CDM cosmologies that can be employed. These simulations present a broad agreement on a wide variety of characteristics, but they do not really arrive at those via identical pathways \citep[see the review by][]{Crain2023}. They even differ significantly in some aspects due to varied implementation of feedback physics -- like gas flow rates and halo gas fractions, for instance, which has direct implications for the results in Section~\ref{hotgas} and Section~\ref{galhist}. Additional discrepancies could be introduced due to the absence of certain physics from the model altogether. For
example, magnetic fields were included in \textsc{\large illustristng} \citep{Nelson2018,Pillepich2018} and have notable influence on gas dynamics and morphology \citep{Kauffmann2019}, but they were not folded into \eagleNS. Hence, while one should not expect identical results, studies similar to ours using other simulations can help ensure that the conclusions presented are sufficiently robust against variations in numerical schemes, hydrodynamical prescriptions, and the simulation model. We hope that the aforementioned endeavours would help devise a more wholesome understanding of the physics of RM, and provide novel and crucial insights into galaxy evolution.

\section*{Acknowledgements}
We would like to thank the anonymous referee for a meticulous review that helped in the improvement of key parts of the paper. The simulation used in this work was performed using the Distributed Research utilising Advanced Computing (DiRAC-2) facility at Durham, managed by the Institute for Computational Cosmology, and the Partnership for Advanced Computing in Europe (PRACE) facility Curie based in France at Tr\`{e}s Grand Centre de Calcul, Commissariat \`{a} l'\'{e}nergie atomique et aux \'{e}nergies alternatives, Bruy\`{e}res-le-Ch\^{a}tel. We would 
like to thank the Virgo Consortium for making the simulation data available. The analysis 
was carried out on the OzSTAR supercomputer, funded by Astronomy Australia Limited (AAL) 
and the Australian Commonwealth Government, and managed through the Centre for 
Astrophysics \& Supercomputing at the Swinburne University of Technology. The key \textsc{\large python} packages employed for the analysis are \textsc{\large matplotlib} \citep{Matplotlib}, \textsc{\large numpy} \citep{Numpy}, \textsc{\large scipy} \citep{Scipy}, and \textsc{\large astropy} \citep{Astropy}. The paper has been typeset using Overleaf,\footnote{\url{https://www.overleaf.com/}} an online freemium academic writing environment.
\section*{Data Availability}
The halo catalogues and particle data for the \eagle simulation -- described in \citet{Mcalpine2016} and \citet{eaglepart}, respectively -- can be accessed at \url{http://icc.dur.ac.uk/Eagle/database.php}. The cold gas contents, kinematic assessments,
and the rates from the particle-tracking analysis can be obtained through a reasonable request to the corresponding author.


\bibliographystyle{mnras}
\bibliography{references} 

\begin{thebibliography}{}
\makeatletter
\relax
\def\mn@urlcharsother{\let\do\@makeother \do\$\do\&\do\#\do\^\do\_\do\%\do\~}
\def\mn@doi{\begingroup\mn@urlcharsother \@ifnextchar [ {\mn@doi@} {\mn@doi@[]}}
\def\mn@doi@[#1]#2{\def\@tempa{#1}\ifx\@tempa\@empty \href {http://dx.doi.org/#2} {doi:#2}\else \href {http://dx.doi.org/#2} {#1}\fi \endgroup}
\def\mn@eprint#1#2{\mn@eprint@#1:#2::\@nil}
\def\mn@eprint@arXiv#1{\href {http://arxiv.org/abs/#1} {{\tt arXiv:#1}}}
\def\mn@eprint@dblp#1{\href {http://dblp.uni-trier.de/rec/bibtex/#1.xml} {dblp:#1}}
\def\mn@eprint@#1:#2:#3:#4\@nil{\def\@tempa {#1}\def\@tempb {#2}\def\@tempc {#3}\ifx \@tempc \@empty \let \@tempc \@tempb \let \@tempb \@tempa \fi \ifx \@tempb \@empty \def\@tempb {arXiv}\fi \@ifundefined {mn@eprint@\@tempb}{\@tempb:\@tempc}{\expandafter \expandafter \csname mn@eprint@\@tempb\endcsname \expandafter{\@tempc}}}

\bibitem[\protect\citeauthoryear{{Astropy Collaboration} et~al.,}{{Astropy Collaboration} et~al.}{2018}]{Astropy}
{Astropy Collaboration} et~al., 2018, \mn@doi [\aj] {10.3847/1538-3881/aabc4f}, \href {https://ui.adsabs.harvard.edu/abs/2018AJ....156..123A} {156, 123}

\bibitem[\protect\citeauthoryear{{Babyk} \& {McNamara}}{{Babyk} \& {McNamara}}{2023}]{Babyk2023}
{Babyk} I.~V.,  {McNamara} B.~R.,  2023, \mn@doi [\apj] {10.3847/1538-4357/acbf4b}, \href {https://ui.adsabs.harvard.edu/abs/2023ApJ...946...54B} {946, 54}

\bibitem[\protect\citeauthoryear{{Baes} et~al.,}{{Baes} et~al.}{2003}]{Baes2003}
{Baes} M.,  et~al., 2003, \mn@doi [\mnras] {10.1046/j.1365-8711.2003.06770.x}, \href {https://ui.adsabs.harvard.edu/abs/2003MNRAS.343.1081B} {343, 1081}

\bibitem[\protect\citeauthoryear{{Baes}, {Verstappen}, {De Looze}, {Fritz}, {Saftly}, {Vidal P{\'e}rez}, {Stalevski}  \& {Valcke}}{{Baes} et~al.}{2011}]{Baes2011}
{Baes} M.,  {Verstappen} J.,  {De Looze} I.,  {Fritz} J.,  {Saftly} W.,  {Vidal P{\'e}rez} E.,  {Stalevski} M.,   {Valcke} S.,  2011, \mn@doi [\apjs] {10.1088/0067-0049/196/2/22}, \href {https://ui.adsabs.harvard.edu/abs/2011ApJS..196...22B} {196, 22}

\bibitem[\protect\citeauthoryear{{Baes} et~al.,}{{Baes} et~al.}{2020}]{Baes2020}
{Baes} M.,  et~al., 2020, \mn@doi [\mnras] {10.1093/mnras/staa990}, \href {https://ui.adsabs.harvard.edu/abs/2020MNRAS.494.2912B} {494, 2912}

\bibitem[\protect\citeauthoryear{{Bah{\'e}} \& {McCarthy}}{{Bah{\'e}} \& {McCarthy}}{2015}]{Bahe2015}
{Bah{\'e}} Y.~M.,  {McCarthy} I.~G.,  2015, \mn@doi [\mnras] {10.1093/mnras/stu2293}, \href {https://ui.adsabs.harvard.edu/abs/2015MNRAS.447..969B} {447, 969}

\bibitem[\protect\citeauthoryear{{Baldry}, {Glazebrook}, {Brinkmann}, {Ivezi{\'c}}, {Lupton}, {Nichol}  \& {Szalay}}{{Baldry} et~al.}{2004}]{Baldry2004}
{Baldry} I.~K.,  {Glazebrook} K.,  {Brinkmann} J.,  {Ivezi{\'c}} {\v{Z}}.,  {Lupton} R.~H.,  {Nichol} R.~C.,   {Szalay} A.~S.,  2004, \mn@doi [\apj] {10.1086/380092}, \href {https://ui.adsabs.harvard.edu/abs/2004ApJ...600..681B} {600, 681}

\bibitem[\protect\citeauthoryear{{Baldry}, {Balogh}, {Bower}, {Glazebrook}, {Nichol}, {Bamford}  \& {Budavari}}{{Baldry} et~al.}{2006}]{Baldry2006}
{Baldry} I.~K.,  {Balogh} M.~L.,  {Bower} R.~G.,  {Glazebrook} K.,  {Nichol} R.~C.,  {Bamford} S.~P.,   {Budavari} T.,  2006, \mn@doi [\mnras] {10.1111/j.1365-2966.2006.11081.x}, \href {https://ui.adsabs.harvard.edu/abs/2006MNRAS.373..469B} {373, 469}

\bibitem[\protect\citeauthoryear{{Balogh}, {Baldry}, {Nichol}, {Miller}, {Bower}  \& {Glazebrook}}{{Balogh} et~al.}{2004}]{Balogh2004}
{Balogh} M.~L.,  {Baldry} I.~K.,  {Nichol} R.,  {Miller} C.,  {Bower} R.,   {Glazebrook} K.,  2004, \mn@doi [\apjl] {10.1086/426079}, \href {https://ui.adsabs.harvard.edu/abs/2004ApJ...615L.101B} {615, L101}

\bibitem[\protect\citeauthoryear{{Blanton} et~al.,}{{Blanton} et~al.}{2003}]{Blanton2003}
{Blanton} M.~R.,  et~al., 2003, \mn@doi [\apj] {10.1086/375528}, \href {https://ui.adsabs.harvard.edu/abs/2003ApJ...594..186B} {594, 186}

\bibitem[\protect\citeauthoryear{{Bluck}, {Piotrowska}  \& {Maiolino}}{{Bluck} et~al.}{2023}]{Bluck2023}
{Bluck} A. F.~L.,  {Piotrowska} J.~M.,   {Maiolino} R.,  2023, \mn@doi [\apj] {10.3847/1538-4357/acac7c}, \href {https://ui.adsabs.harvard.edu/abs/2023ApJ...944..108B} {944, 108}

\bibitem[\protect\citeauthoryear{{Booth} \& {Schaye}}{{Booth} \& {Schaye}}{2009}]{Booth2009}
{Booth} C.~M.,  {Schaye} J.,  2009, \mn@doi [\mnras] {10.1111/j.1365-2966.2009.15043.x}, \href {https://ui.adsabs.harvard.edu/abs/2009MNRAS.398...53B} {398, 53}

\bibitem[\protect\citeauthoryear{{Boroson} \& {Lauer}}{{Boroson} \& {Lauer}}{2009}]{Boroson2009}
{Boroson} T.~A.,  {Lauer} T.~R.,  2009, \mn@doi [\nat] {10.1038/nature07779}, \href {https://ui.adsabs.harvard.edu/abs/2009Natur.458...53B} {458, 53}

\bibitem[\protect\citeauthoryear{{Borrow}, {Vogelsberger}, {O'Neil}, {McDonald}  \& {Smith}}{{Borrow} et~al.}{2023}]{Borrow2023}
{Borrow} J.,  {Vogelsberger} M.,  {O'Neil} S.,  {McDonald} M.~A.,   {Smith} A.,  2023, \mn@doi [\mnras] {10.1093/mnras/stad045}, \href {https://ui.adsabs.harvard.edu/abs/2023MNRAS.520..649B} {520, 649}

\bibitem[\protect\citeauthoryear{{Boselli}, {Fossati}  \& {Sun}}{{Boselli} et~al.}{2022}]{Boselli2022}
{Boselli} A.,  {Fossati} M.,   {Sun} M.,  2022, \mn@doi [\aapr] {10.1007/s00159-022-00140-3}, \href {https://ui.adsabs.harvard.edu/abs/2022A&ARv..30....3B} {30, 3}

\bibitem[\protect\citeauthoryear{{Brammer} et~al.,}{{Brammer} et~al.}{2009}]{Brammer2009}
{Brammer} G.~B.,  et~al., 2009, \mn@doi [\apjl] {10.1088/0004-637X/706/1/L173}, \href {https://ui.adsabs.harvard.edu/abs/2009ApJ...706L.173B} {706, L173}

\bibitem[\protect\citeauthoryear{{Brand} et~al.,}{{Brand} et~al.}{2009}]{Brand2009}
{Brand} K.,  et~al., 2009, \mn@doi [\apj] {10.1088/0004-637X/693/1/340}, \href {https://ui.adsabs.harvard.edu/abs/2009ApJ...693..340B} {693, 340}

\bibitem[\protect\citeauthoryear{{Bravo}, {Robotham}, {Lagos}, {Davies}, {Bellstedt}  \& {Thorne}}{{Bravo} et~al.}{2022}]{Bravo2022}
{Bravo} M.,  {Robotham} A. S.~G.,  {Lagos} C. d.~P.,  {Davies} L. J.~M.,  {Bellstedt} S.,   {Thorne} J.~E.,  2022, \mn@doi [\mnras] {10.1093/mnras/stac321}, \href {https://ui.adsabs.harvard.edu/abs/2022MNRAS.511.5405B} {511, 5405}

\bibitem[\protect\citeauthoryear{{Brown} et~al.,}{{Brown} et~al.}{2017}]{Brown2017}
{Brown} T.,  et~al., 2017, \mn@doi [\mnras] {10.1093/mnras/stw2991}, \href {https://ui.adsabs.harvard.edu/abs/2017MNRAS.466.1275B} {466, 1275}

\bibitem[\protect\citeauthoryear{{Bruzual} \& {Charlot}}{{Bruzual} \& {Charlot}}{2003}]{Bruzual2003}
{Bruzual} G.,  {Charlot} S.,  2003, \mn@doi [\mnras] {10.1046/j.1365-8711.2003.06897.x}, \href {https://ui.adsabs.harvard.edu/abs/2003MNRAS.344.1000B} {344, 1000}

\bibitem[\protect\citeauthoryear{{Camps} \& {Baes}}{{Camps} \& {Baes}}{2015}]{Camps2015}
{Camps} P.,  {Baes} M.,  2015, \mn@doi [Astronomy and Computing] {10.1016/j.ascom.2014.10.004}, \href {https://ui.adsabs.harvard.edu/abs/2015A&C.....9...20C} {9, 20}

\bibitem[\protect\citeauthoryear{{Camps}, {Trayford}, {Baes}, {Theuns}, {Schaller}  \& {Schaye}}{{Camps} et~al.}{2016}]{Camps2016}
{Camps} P.,  {Trayford} J.~W.,  {Baes} M.,  {Theuns} T.,  {Schaller} M.,   {Schaye} J.,  2016, \mn@doi [\mnras] {10.1093/mnras/stw1735}, \href {https://ui.adsabs.harvard.edu/abs/2016MNRAS.462.1057C} {462, 1057}

\bibitem[\protect\citeauthoryear{{Camps} et~al.,}{{Camps} et~al.}{2018}]{Camps2018}
{Camps} P.,  et~al., 2018, \mn@doi [\apjs] {10.3847/1538-4365/aaa24c}, \href {https://ui.adsabs.harvard.edu/abs/2018ApJS..234...20C} {234, 20}

\bibitem[\protect\citeauthoryear{{Carleton}, {Errani}, {Cooper}, {Kaplinghat}, {Pe{\~n}arrubia}  \& {Guo}}{{Carleton} et~al.}{2019}]{Carleton2019}
{Carleton} T.,  {Errani} R.,  {Cooper} M.,  {Kaplinghat} M.,  {Pe{\~n}arrubia} J.,   {Guo} Y.,  2019, \mn@doi [\mnras] {10.1093/mnras/stz383}, \href {https://ui.adsabs.harvard.edu/abs/2019MNRAS.485..382C} {485, 382}

\bibitem[\protect\citeauthoryear{{Cassata} et~al.,}{{Cassata} et~al.}{2008}]{Cassata2008}
{Cassata} P.,  et~al., 2008, \mn@doi [\aap] {10.1051/0004-6361:200809881}, \href {https://ui.adsabs.harvard.edu/abs/2008A&A...483L..39C} {483, L39}

\bibitem[\protect\citeauthoryear{{Cattaneo} et~al.,}{{Cattaneo} et~al.}{2007}]{Cattaneo2007}
{Cattaneo} A.,  et~al., 2007, \mn@doi [\mnras] {10.1111/j.1365-2966.2007.11597.x}, \href {https://ui.adsabs.harvard.edu/abs/2007MNRAS.377...63C} {377, 63}

\bibitem[\protect\citeauthoryear{{Chown} et~al.,}{{Chown} et~al.}{2022}]{Chown2022}
{Chown} R.,  et~al., 2022, \mn@doi [\mnras] {10.1093/mnras/stac2193}, \href {https://ui.adsabs.harvard.edu/abs/2022MNRAS.516...84C} {516, 84}

\bibitem[\protect\citeauthoryear{{Clauwens}, {Schaye}, {Franx}  \& {Bower}}{{Clauwens} et~al.}{2018}]{Clauwens2018}
{Clauwens} B.,  {Schaye} J.,  {Franx} M.,   {Bower} R.~G.,  2018, \mn@doi [\mnras] {10.1093/mnras/sty1229}, \href {https://ui.adsabs.harvard.edu/abs/2018MNRAS.478.3994C} {478, 3994}

\bibitem[\protect\citeauthoryear{{Coenda}, {Mart{\'\i}nez}  \& {Muriel}}{{Coenda} et~al.}{2018}]{Coenda2018}
{Coenda} V.,  {Mart{\'\i}nez} H.~J.,   {Muriel} H.,  2018, \mn@doi [\mnras] {10.1093/mnras/stx2707}, \href {https://ui.adsabs.harvard.edu/abs/2018MNRAS.473.5617C} {473, 5617}

\bibitem[\protect\citeauthoryear{{Correa}, {Schaye}, {Wyithe}, {Duffy}, {Theuns}, {Crain}  \& {Bower}}{{Correa} et~al.}{2018}]{Correa2018}
{Correa} C.~A.,  {Schaye} J.,  {Wyithe} J. S.~B.,  {Duffy} A.~R.,  {Theuns} T.,  {Crain} R.~A.,   {Bower} R.~G.,  2018, \mn@doi [\mnras] {10.1093/mnras/stx2332}, \href {https://ui.adsabs.harvard.edu/abs/2018MNRAS.473..538C} {473, 538}

\bibitem[\protect\citeauthoryear{{Correa}, {Schaye}  \& {Trayford}}{{Correa} et~al.}{2019}]{Correa2019}
{Correa} C.~A.,  {Schaye} J.,   {Trayford} J.~W.,  2019, \mn@doi [\mnras] {10.1093/mnras/stz295}, \href {https://ui.adsabs.harvard.edu/abs/2019MNRAS.484.4401C} {484, 4401}

\bibitem[\protect\citeauthoryear{{Crain} \& {van de Voort}}{{Crain} \& {van de Voort}}{2023}]{Crain2023}
{Crain} R.~A.,  {van de Voort} F.,  2023, \mn@doi [\araa] {10.1146/annurev-astro-041923-043618}, \href {https://ui.adsabs.harvard.edu/abs/2023ARA&A..61..473C} {61, 473}

\bibitem[\protect\citeauthoryear{{Crain} et~al.,}{{Crain} et~al.}{2015}]{Crain2015}
{Crain} R.~A.,  et~al., 2015, \mn@doi [\mnras] {10.1093/mnras/stv725}, \href {https://ui.adsabs.harvard.edu/abs/2015MNRAS.450.1937C} {450, 1937}

\bibitem[\protect\citeauthoryear{{Crain} et~al.,}{{Crain} et~al.}{2017}]{Crain2017}
{Crain} R.~A.,  et~al., 2017, \mn@doi [\mnras] {10.1093/mnras/stw2586}, \href {https://ui.adsabs.harvard.edu/abs/2017MNRAS.464.4204C} {464, 4204}

\bibitem[\protect\citeauthoryear{{Cui}, {Dav{\'e}}, {Peacock}, {Angl{\'e}s-Alc{\'a}zar}  \& {Yang}}{{Cui} et~al.}{2021}]{Cui2021}
{Cui} W.,  {Dav{\'e}} R.,  {Peacock} J.~A.,  {Angl{\'e}s-Alc{\'a}zar} D.,   {Yang} X.,  2021, \mn@doi [Nature Astronomy] {10.1038/s41550-021-01404-1}, \href {https://ui.adsabs.harvard.edu/abs/2021NatAs...5.1069C} {5, 1069}

\bibitem[\protect\citeauthoryear{{Dav{\'e}}, {Angl{\'e}s-Alc{\'a}zar}, {Narayanan}, {Li}, {Rafieferantsoa}  \& {Appleby}}{{Dav{\'e}} et~al.}{2019}]{Dave2019}
{Dav{\'e}} R.,  {Angl{\'e}s-Alc{\'a}zar} D.,  {Narayanan} D.,  {Li} Q.,  {Rafieferantsoa} M.~H.,   {Appleby} S.,  2019, \mn@doi [\mnras] {10.1093/mnras/stz937}, \href {https://ui.adsabs.harvard.edu/abs/2019MNRAS.486.2827D} {486, 2827}

\bibitem[\protect\citeauthoryear{{Davis}, {Efstathiou}, {Frenk}  \& {White}}{{Davis} et~al.}{1985}]{Davis1985}
{Davis} M.,  {Efstathiou} G.,  {Frenk} C.~S.,   {White} S.~D.~M.,  1985, \mn@doi [\apj] {10.1086/163168}, \href {https://ui.adsabs.harvard.edu/abs/1985ApJ...292..371D} {292, 371}

\bibitem[\protect\citeauthoryear{{Davison}, {Norris}, {Pfeffer}, {Davies}  \& {Crain}}{{Davison} et~al.}{2020}]{Davidson2020}
{Davison} T.~A.,  {Norris} M.~A.,  {Pfeffer} J.~L.,  {Davies} J.~J.,   {Crain} R.~A.,  2020, \mn@doi [\mnras] {10.1093/mnras/staa1816}, \href {https://ui.adsabs.harvard.edu/abs/2020MNRAS.497...81D} {497, 81}

\bibitem[\protect\citeauthoryear{{Davoodi} et~al.,}{{Davoodi} et~al.}{2006}]{Davoodi2006}
{Davoodi} P.,  et~al., 2006, \mn@doi [\mnras] {10.1111/j.1365-2966.2006.10793.x}, \href {https://ui.adsabs.harvard.edu/abs/2006MNRAS.371.1113D} {371, 1113}

\bibitem[\protect\citeauthoryear{{DeFelippis}, {Genel}, {Bryan}  \& {Fall}}{{DeFelippis} et~al.}{2017}]{DeFelippis2017}
{DeFelippis} D.,  {Genel} S.,  {Bryan} G.~L.,   {Fall} S.~M.,  2017, \mn@doi [\apj] {10.3847/1538-4357/aa6dfc}, \href {https://ui.adsabs.harvard.edu/abs/2017ApJ...841...16D} {841, 16}

\bibitem[\protect\citeauthoryear{{Deane} et~al.,}{{Deane} et~al.}{2014}]{Deane2014}
{Deane} R.~P.,  et~al., 2014, \mn@doi [\nat] {10.1038/nature13454}, \href {https://ui.adsabs.harvard.edu/abs/2014Natur.511...57D} {511, 57}

\bibitem[\protect\citeauthoryear{{Dekel} \& {Birnboim}}{{Dekel} \& {Birnboim}}{2006}]{Dekel2006}
{Dekel} A.,  {Birnboim} Y.,  2006, \mn@doi [\mnras] {10.1111/j.1365-2966.2006.10145.x}, \href {https://ui.adsabs.harvard.edu/abs/2006MNRAS.368....2D} {368, 2}

\bibitem[\protect\citeauthoryear{{Di Cintio}, {Mostoghiu}, {Knebe}  \& {Navarro}}{{Di Cintio} et~al.}{2021}]{Cintio2021}
{Di Cintio} A.,  {Mostoghiu} R.,  {Knebe} A.,   {Navarro} J.~F.,  2021, \mn@doi [\mnras] {10.1093/mnras/stab1682}, \href {https://ui.adsabs.harvard.edu/abs/2021MNRAS.506..531D} {506, 531}

\bibitem[\protect\citeauthoryear{{Diamond-Stanic}, {Coil}, {Moustakas}, {Tremonti}, {Sell}, {Mendez}, {Hickox}  \& {Rudnick}}{{Diamond-Stanic} et~al.}{2016}]{Stanic2016}
{Diamond-Stanic} A.~M.,  {Coil} A.~L.,  {Moustakas} J.,  {Tremonti} C.~A.,  {Sell} P.~H.,  {Mendez} A.~J.,  {Hickox} R.~C.,   {Rudnick} G.~H.,  2016, \mn@doi [\apj] {10.3847/0004-637X/824/1/24}, \href {https://ui.adsabs.harvard.edu/abs/2016ApJ...824...24D} {824, 24}

\bibitem[\protect\citeauthoryear{{D{\'\i}az-Garc{\'\i}a} et~al.,}{{D{\'\i}az-Garc{\'\i}a} et~al.}{2019}]{Garcia2019}
{D{\'\i}az-Garc{\'\i}a} L.~A.,  et~al., 2019, \mn@doi [\aap] {10.1051/0004-6361/201832788}, \href {https://ui.adsabs.harvard.edu/abs/2019A&A...631A.156D} {631, A156}

\bibitem[\protect\citeauthoryear{{Diemer} et~al.,}{{Diemer} et~al.}{2018}]{Diemer2018}
{Diemer} B.,  et~al., 2018, \mn@doi [\apjs] {10.3847/1538-4365/aae387}, \href {https://ui.adsabs.harvard.edu/abs/2018ApJS..238...33D} {238, 33}

\bibitem[\protect\citeauthoryear{{Doi} et~al.,}{{Doi} et~al.}{2010}]{Doi2010}
{Doi} M.,  et~al., 2010, \mn@doi [\aj] {10.1088/0004-6256/139/4/1628}, \href {https://ui.adsabs.harvard.edu/abs/2010AJ....139.1628D} {139, 1628}

\bibitem[\protect\citeauthoryear{{Dolag}, {Borgani}, {Murante}  \& {Springel}}{{Dolag} et~al.}{2009}]{Dolag2009}
{Dolag} K.,  {Borgani} S.,  {Murante} G.,   {Springel} V.,  2009, \mn@doi [\mnras] {10.1111/j.1365-2966.2009.15034.x}, \href {https://ui.adsabs.harvard.edu/abs/2009MNRAS.399..497D} {399, 497}

\bibitem[\protect\citeauthoryear{{Driver} et~al.,}{{Driver} et~al.}{2006}]{Driver2006}
{Driver} S.~P.,  et~al., 2006, \mn@doi [\mnras] {10.1111/j.1365-2966.2006.10126.x}, \href {https://ui.adsabs.harvard.edu/abs/2006MNRAS.368..414D} {368, 414}

\bibitem[\protect\citeauthoryear{{Driver} et~al.,}{{Driver} et~al.}{2011}]{Driver2011}
{Driver} S.~P.,  et~al., 2011, \mn@doi [\mnras] {10.1111/j.1365-2966.2010.18188.x}, \href {https://ui.adsabs.harvard.edu/abs/2011MNRAS.413..971D} {413, 971}

\bibitem[\protect\citeauthoryear{{El-Badry} et~al.,}{{El-Badry} et~al.}{2018}]{Badry2018}
{El-Badry} K.,  et~al., 2018, \mn@doi [\mnras] {10.1093/mnras/stx2482}, \href {https://ui.adsabs.harvard.edu/abs/2018MNRAS.473.1930E} {473, 1930}

\bibitem[\protect\citeauthoryear{{Evans}, {Parker}  \& {Roberts}}{{Evans} et~al.}{2018}]{Evans2018}
{Evans} F.~A.,  {Parker} L.~C.,   {Roberts} I.~D.,  2018, \mn@doi [\mnras] {10.1093/mnras/sty581}, \href {https://ui.adsabs.harvard.edu/abs/2018MNRAS.476.5284E} {476, 5284}

\bibitem[\protect\citeauthoryear{{Fang}, {Faber}, {Salim}, {Graves}  \& {Rich}}{{Fang} et~al.}{2012}]{Fang2012}
{Fang} J.~J.,  {Faber} S.~M.,  {Salim} S.,  {Graves} G.~J.,   {Rich} R.~M.,  2012, \mn@doi [\apj] {10.1088/0004-637X/761/1/23}, \href {https://ui.adsabs.harvard.edu/abs/2012ApJ...761...23F} {761, 23}

\bibitem[\protect\citeauthoryear{{Feldmann}}{{Feldmann}}{2017}]{Feldmann2017}
{Feldmann} R.,  2017, \mn@doi [\mnras] {10.1093/mnrasl/slx073}, \href {https://ui.adsabs.harvard.edu/abs/2017MNRAS.470L..59F} {470, L59}

\bibitem[\protect\citeauthoryear{{Font} et~al.,}{{Font} et~al.}{2008}]{Font2008}
{Font} A.~S.,  et~al., 2008, \mn@doi [\mnras] {10.1111/j.1365-2966.2008.13698.x}, \href {https://ui.adsabs.harvard.edu/abs/2008MNRAS.389.1619F} {389, 1619}

\bibitem[\protect\citeauthoryear{{Forbes}, {Krumholz}  \& {Burkert}}{{Forbes} et~al.}{2012}]{Forbes2012}
{Forbes} J.,  {Krumholz} M.,   {Burkert} A.,  2012, \mn@doi [\apj] {10.1088/0004-637X/754/1/48}, \href {https://ui.adsabs.harvard.edu/abs/2012ApJ...754...48F} {754, 48}

\bibitem[\protect\citeauthoryear{{Fraser-McKelvie}, {Brown}, {Pimbblet}, {Dolley}, {Crossett}  \& {Bonne}}{{Fraser-McKelvie} et~al.}{2016}]{Fraser2016}
{Fraser-McKelvie} A.,  {Brown} M. J.~I.,  {Pimbblet} K.~A.,  {Dolley} T.,  {Crossett} J.~P.,   {Bonne} N.~J.,  2016, \mn@doi [\mnras] {10.1093/mnrasl/slw117}, \href {https://ui.adsabs.harvard.edu/abs/2016MNRAS.462L..11F} {462, L11}

\bibitem[\protect\citeauthoryear{{Fukugita}, {Ichikawa}, {Gunn}, {Doi}, {Shimasaku}  \& {Schneider}}{{Fukugita} et~al.}{1996}]{Fukugita1996}
{Fukugita} M.,  {Ichikawa} T.,  {Gunn} J.~E.,  {Doi} M.,  {Shimasaku} K.,   {Schneider} D.~P.,  1996, AJ, 111, 1748

\bibitem[\protect\citeauthoryear{{Furlong} et~al.,}{{Furlong} et~al.}{2015}]{Furlong2015}
{Furlong} M.,  et~al., 2015, \mn@doi [\mnras] {10.1093/mnras/stv852}, \href {https://ui.adsabs.harvard.edu/abs/2015MNRAS.450.4486F} {450, 4486}

\bibitem[\protect\citeauthoryear{{Gabor} \& {Dav{\'e}}}{{Gabor} \& {Dav{\'e}}}{2012}]{Gabor2012}
{Gabor} J.~M.,  {Dav{\'e}} R.,  2012, \mn@doi [\mnras] {10.1111/j.1365-2966.2012.21640.x}, \href {https://ui.adsabs.harvard.edu/abs/2012MNRAS.427.1816G} {427, 1816}

\bibitem[\protect\citeauthoryear{{Garratt-Smithson}, {Power}, {Lagos}, {Stevens}, {Allison}  \& {Sadler}}{{Garratt-Smithson} et~al.}{2021}]{Smithson2021}
{Garratt-Smithson} L.,  {Power} C.,  {Lagos} C. d.~P.,  {Stevens} A. R.~H.,  {Allison} J.~R.,   {Sadler} E.~M.,  2021, \mn@doi [\mnras] {10.1093/mnras/staa3870}, \href {https://ui.adsabs.harvard.edu/abs/2021MNRAS.501.4396G} {501, 4396}

\bibitem[\protect\citeauthoryear{{George} \& {Zingade}}{{George} \& {Zingade}}{2015}]{George2015}
{George} K.,  {Zingade} K.,  2015, \mn@doi [\aap] {10.1051/0004-6361/201424826}, \href {https://ui.adsabs.harvard.edu/abs/2015A&A...583A.103G} {583, A103}

\bibitem[\protect\citeauthoryear{{Gnedin} \& {Draine}}{{Gnedin} \& {Draine}}{2014}]{GD14}
{Gnedin} N.~Y.,  {Draine} B.~T.,  2014, \mn@doi [\apj] {10.1088/0004-637X/795/1/37}, \href {https://ui.adsabs.harvard.edu/abs/2014ApJ...795...37G} {795, 37}

\bibitem[\protect\citeauthoryear{{Goldbaum}, {Krumholz}  \& {Forbes}}{{Goldbaum} et~al.}{2015}]{Goldbaum2015}
{Goldbaum} N.~J.,  {Krumholz} M.~R.,   {Forbes} J.~C.,  2015, \mn@doi [\apj] {10.1088/0004-637X/814/2/131}, \href {https://ui.adsabs.harvard.edu/abs/2015ApJ...814..131G} {814, 131}

\bibitem[\protect\citeauthoryear{{Goto} et~al.,}{{Goto} et~al.}{2003a}]{Goto2003}
{Goto} T.,  et~al., 2003a, \mn@doi [\pasj] {10.1093/pasj/55.4.771}, \href {https://ui.adsabs.harvard.edu/abs/2003PASJ...55..771G} {55, 771}

\bibitem[\protect\citeauthoryear{{Goto} et~al.,}{{Goto} et~al.}{2003b}]{Goto2003b}
{Goto} T.,  et~al., 2003b, \mn@doi [\pasj] {10.1093/pasj/55.4.757}, \href {https://ui.adsabs.harvard.edu/abs/2003PASJ...55..757G} {55, 757}

\bibitem[\protect\citeauthoryear{{Grand}, {Kawata}  \& {Cropper}}{{Grand} et~al.}{2015}]{Grand2015}
{Grand} R. J.~J.,  {Kawata} D.,   {Cropper} M.,  2015, \mn@doi [\mnras] {10.1093/mnras/stv016}, \href {https://ui.adsabs.harvard.edu/abs/2015MNRAS.447.4018G} {447, 4018}

\bibitem[\protect\citeauthoryear{{Grand} et~al.,}{{Grand} et~al.}{2019}]{Grand2019}
{Grand} R. J.~J.,  et~al., 2019, \mn@doi [\mnras] {10.1093/mnras/stz2928}, \href {https://ui.adsabs.harvard.edu/abs/2019MNRAS.490.4786G} {490, 4786}

\bibitem[\protect\citeauthoryear{{Groves}, {Dopita}, {Sutherland}, {Kewley}, {Fischera}, {Leitherer}, {Brandl}  \& {van Breugel}}{{Groves} et~al.}{2008}]{Groves2008}
{Groves} B.,  {Dopita} M.~A.,  {Sutherland} R.~S.,  {Kewley} L.~J.,  {Fischera} J.,  {Leitherer} C.,  {Brandl} B.,   {van Breugel} W.,  2008, \mn@doi [\apjs] {10.1086/528711}, \href {https://ui.adsabs.harvard.edu/abs/2008ApJS..176..438G} {176, 438}

\bibitem[\protect\citeauthoryear{{Haardt} \& {Madau}}{{Haardt} \& {Madau}}{2001}]{Haardt2001}
{Haardt} F.,  {Madau} P.,  2001, in {Neumann} D.~M.,  {Tran} J.~T.~V.,  eds, Clusters of Galaxies and the High Redshift Universe Observed in X-rays. p.~64 (\mn@eprint {arXiv} {astro-ph/0106018})

\bibitem[\protect\citeauthoryear{{Haardt} \& {Madau}}{{Haardt} \& {Madau}}{2012}]{Haardt2012}
{Haardt} F.,  {Madau} P.,  2012, \mn@doi [\apj] {10.1088/0004-637X/746/2/125}, \href {https://ui.adsabs.harvard.edu/abs/2012ApJ...746..125H} {746, 125}

\bibitem[\protect\citeauthoryear{{Haas}, {Schaye}  \& {Jeeson-Daniel}}{{Haas} et~al.}{2012}]{Haas2012}
{Haas} M.~R.,  {Schaye} J.,   {Jeeson-Daniel} A.,  2012, \mn@doi [\mnras] {10.1111/j.1365-2966.2011.19863.x}, \href {https://ui.adsabs.harvard.edu/abs/2012MNRAS.419.2133H} {419, 2133}

\bibitem[\protect\citeauthoryear{{Hafen} et~al.,}{{Hafen} et~al.}{2022}]{Hafen2022}
{Hafen} Z.,  et~al., 2022, \mn@doi [\mnras] {10.1093/mnras/stac1603}, \href {https://ui.adsabs.harvard.edu/abs/2022MNRAS.514.5056H} {514, 5056}

\bibitem[\protect\citeauthoryear{{Haines}, {Gargiulo}  \& {Merluzzi}}{{Haines} et~al.}{2008}]{Haines2008}
{Haines} C.~P.,  {Gargiulo} A.,   {Merluzzi} P.,  2008, \mn@doi [\mnras] {10.1111/j.1365-2966.2008.12954.x}, \href {https://ui.adsabs.harvard.edu/abs/2008MNRAS.385.1201H} {385, 1201}

\bibitem[\protect\citeauthoryear{{Hewett}, {Warren}, {Leggett}  \& {Hodgkin}}{{Hewett} et~al.}{2006}]{Hewett2006}
{Hewett} P.~C.,  {Warren} S.~J.,  {Leggett} S.~K.,   {Hodgkin} S.~T.,  2006, \mn@doi [\mnras] {10.1111/j.1365-2966.2005.09969.x}, \href {https://ui.adsabs.harvard.edu/abs/2006MNRAS.367..454H} {367, 454}

\bibitem[\protect\citeauthoryear{{Hogg} et~al.,}{{Hogg} et~al.}{2002}]{Hogg2002}
{Hogg} D.~W.,  et~al., 2002, \mn@doi [\aj] {10.1086/341392}, \href {https://ui.adsabs.harvard.edu/abs/2002AJ....124..646H} {124, 646}

\bibitem[\protect\citeauthoryear{{Houghton}}{{Houghton}}{2015}]{Houghton2015}
{Houghton} R.~C.~W.,  2015, \mn@doi [\mnras] {10.1093/mnras/stv1113}, \href {https://ui.adsabs.harvard.edu/abs/2015MNRAS.451.3427H} {451, 3427}

\bibitem[\protect\citeauthoryear{{Huertas-Company}, {Aguerri}, {Tresse}, {Bolzonella}, {Koekemoer}  \& {Maier}}{{Huertas-Company} et~al.}{2010}]{Company2010}
{Huertas-Company} M.,  {Aguerri} J.~A.~L.,  {Tresse} L.,  {Bolzonella} M.,  {Koekemoer} A.~M.,   {Maier} C.,  2010, \mn@doi [\aap] {10.1051/0004-6361/200913188}, \href {https://ui.adsabs.harvard.edu/abs/2010A&A...515A...3H} {515, A3}

\bibitem[\protect\citeauthoryear{{Hunter}}{{Hunter}}{2007}]{Matplotlib}
{Hunter} J.~D.,  2007, \mn@doi [Computing in Science and Engineering] {10.1109/MCSE.2007.55}, \href {https://ui.adsabs.harvard.edu/abs/2007CSE.....9...90H} {9, 90}

\bibitem[\protect\citeauthoryear{{Ilbert} et~al.,}{{Ilbert} et~al.}{2010}]{Ilbert2010}
{Ilbert} O.,  et~al., 2010, \mn@doi [\apj] {10.1088/0004-637X/709/2/644}, \href {https://ui.adsabs.harvard.edu/abs/2010ApJ...709..644I} {709, 644}

\bibitem[\protect\citeauthoryear{{Jackson} et~al.,}{{Jackson} et~al.}{2021}]{Jackson2021}
{Jackson} R.~A.,  et~al., 2021, \mn@doi [\mnras] {10.1093/mnras/stab093}, \href {https://ui.adsabs.harvard.edu/abs/2021MNRAS.502.1785J} {502, 1785}

\bibitem[\protect\citeauthoryear{{Jaff{\'e}}, {Smith}, {Candlish}, {Poggianti}, {Sheen}  \& {Verheijen}}{{Jaff{\'e}} et~al.}{2015}]{Jaffe2015}
{Jaff{\'e}} Y.~L.,  {Smith} R.,  {Candlish} G.~N.,  {Poggianti} B.~M.,  {Sheen} Y.-K.,   {Verheijen} M. A.~W.,  2015, \mn@doi [\mnras] {10.1093/mnras/stv100}, \href {https://ui.adsabs.harvard.edu/abs/2015MNRAS.448.1715J} {448, 1715}

\bibitem[\protect\citeauthoryear{{Jian} et~al.,}{{Jian} et~al.}{2020}]{Jian2020}
{Jian} H.-Y.,  et~al., 2020, \mn@doi [\apj] {10.3847/1538-4357/ab86a8}, \href {https://ui.adsabs.harvard.edu/abs/2020ApJ...894..125J} {894, 125}

\bibitem[\protect\citeauthoryear{{Jiang}, {Dekel}, {Freundlich}, {Romanowsky}, {Dutton}, {Macci{\`o}}  \& {Di Cintio}}{{Jiang} et~al.}{2019}]{Jiang2019}
{Jiang} F.,  {Dekel} A.,  {Freundlich} J.,  {Romanowsky} A.~J.,  {Dutton} A.~A.,  {Macci{\`o}} A.~V.,   {Di Cintio} A.,  2019, \mn@doi [\mnras] {10.1093/mnras/stz1499}, \href {https://ui.adsabs.harvard.edu/abs/2019MNRAS.487.5272J} {487, 5272}

\bibitem[\protect\citeauthoryear{{Kang}, {Jing}, {Mo}  \& {B{\"o}rner}}{{Kang} et~al.}{2005}]{Kang2005}
{Kang} X.,  {Jing} Y.~P.,  {Mo} H.~J.,   {B{\"o}rner} G.,  2005, \mn@doi [\apj] {10.1086/432493}, \href {https://ui.adsabs.harvard.edu/abs/2005ApJ...631...21K} {631, 21}

\bibitem[\protect\citeauthoryear{{Katsianis} et~al.,}{{Katsianis} et~al.}{2017}]{Katsianis2017}
{Katsianis} A.,  et~al., 2017, \mn@doi [\mnras] {10.1093/mnras/stx2020}, \href {https://ui.adsabs.harvard.edu/abs/2017MNRAS.472..919K} {472, 919}

\bibitem[\protect\citeauthoryear{{Katsianis} et~al.,}{{Katsianis} et~al.}{2020}]{Katsianis2020}
{Katsianis} A.,  et~al., 2020, \mn@doi [\mnras] {10.1093/mnras/staa157}, \href {https://ui.adsabs.harvard.edu/abs/2020MNRAS.492.5592K} {492, 5592}

\bibitem[\protect\citeauthoryear{{Katsianis} et~al.,}{{Katsianis} et~al.}{2021}]{Katsianis2021}
{Katsianis} A.,  et~al., 2021, \mn@doi [\mnras] {10.1093/mnras/staa3236}, \href {https://ui.adsabs.harvard.edu/abs/2021MNRAS.500.2036K} {500, 2036}

\bibitem[\protect\citeauthoryear{{Kauffmann} et~al.,}{{Kauffmann} et~al.}{2003}]{Kauffmann2003}
{Kauffmann} G.,  et~al., 2003, \mn@doi [\mnras] {10.1046/j.1365-8711.2003.06292.x}, \href {https://ui.adsabs.harvard.edu/abs/2003MNRAS.341...54K} {341, 54}

\bibitem[\protect\citeauthoryear{{Kauffmann}, {Nelson}, {Borthakur}, {Heckman}, {Hernquist}, {Marinacci}, {Pakmor}  \& {Pillepich}}{{Kauffmann} et~al.}{2019}]{Kauffmann2019}
{Kauffmann} G.,  {Nelson} D.,  {Borthakur} S.,  {Heckman} T.,  {Hernquist} L.,  {Marinacci} F.,  {Pakmor} R.,   {Pillepich} A.,  2019, \mn@doi [\mnras] {10.1093/mnras/stz1029}, \href {https://ui.adsabs.harvard.edu/abs/2019MNRAS.486.4686K} {486, 4686}

\bibitem[\protect\citeauthoryear{{Kaviraj}, {Tan}, {Ellis}  \& {Silk}}{{Kaviraj} et~al.}{2011}]{Kaviraj2011}
{Kaviraj} S.,  {Tan} K.-M.,  {Ellis} R.~S.,   {Silk} J.,  2011, \mn@doi [\mnras] {10.1111/j.1365-2966.2010.17754.x}, \href {https://ui.adsabs.harvard.edu/abs/2011MNRAS.411.2148K} {411, 2148}

\bibitem[\protect\citeauthoryear{{Kelvin} et~al.,}{{Kelvin} et~al.}{2012}]{Kelvin2012}
{Kelvin} L.~S.,  et~al., 2012, \mn@doi [\mnras] {10.1111/j.1365-2966.2012.20355.x}, \href {https://ui.adsabs.harvard.edu/abs/2012MNRAS.421.1007K} {421, 1007}

\bibitem[\protect\citeauthoryear{{Krumholz}, {Burkhart}, {Forbes}  \& {Crocker}}{{Krumholz} et~al.}{2018}]{Krumholz2018}
{Krumholz} M.~R.,  {Burkhart} B.,  {Forbes} J.~C.,   {Crocker} R.~M.,  2018, \mn@doi [\mnras] {10.1093/mnras/sty852}, \href {https://ui.adsabs.harvard.edu/abs/2018MNRAS.477.2716K} {477, 2716}

\bibitem[\protect\citeauthoryear{{Lagos} et~al.,}{{Lagos} et~al.}{2015}]{Lagos2015}
{Lagos} C. d.~P.,  et~al., 2015, \mn@doi [\mnras] {10.1093/mnras/stv1488}, \href {https://ui.adsabs.harvard.edu/abs/2015MNRAS.452.3815L} {452, 3815}

\bibitem[\protect\citeauthoryear{{Lazar}, {Kaviraj}, {Martin}, {Laigle}, {Watkins}  \& {Jackson}}{{Lazar} et~al.}{2023}]{Lazar2023}
{Lazar} I.,  {Kaviraj} S.,  {Martin} G.,  {Laigle} C.,  {Watkins} A.,   {Jackson} R.~A.,  2023, \mn@doi [\mnras] {10.1093/mnras/stad224}, \href {https://ui.adsabs.harvard.edu/abs/2023MNRAS.520.2109L} {520, 2109}

\bibitem[\protect\citeauthoryear{{Liske} et~al.,}{{Liske} et~al.}{2015}]{Liske2015}
{Liske} J.,  et~al., 2015, \mn@doi [\mnras] {10.1093/mnras/stv1436}, \href {https://ui.adsabs.harvard.edu/abs/2015MNRAS.452.2087L} {452, 2087}

\bibitem[\protect\citeauthoryear{{Lopes}, {Rembold}, {Ribeiro}, {Nascimento}  \& {Vajgel}}{{Lopes} et~al.}{2016}]{Lopes2016}
{Lopes} P.~A.~A.,  {Rembold} S.~B.,  {Ribeiro} A.~L.~B.,  {Nascimento} R.~S.,   {Vajgel} B.,  2016, \mn@doi [\mnras] {10.1093/mnras/stw1497}, \href {https://ui.adsabs.harvard.edu/abs/2016MNRAS.461.2559L} {461, 2559}

\bibitem[\protect\citeauthoryear{{Mahajan} \& {Raychaudhury}}{{Mahajan} \& {Raychaudhury}}{2009}]{Mahajan2009}
{Mahajan} S.,  {Raychaudhury} S.,  2009, \mn@doi [\mnras] {10.1111/j.1365-2966.2009.15512.x}, \href {https://ui.adsabs.harvard.edu/abs/2009MNRAS.400..687M} {400, 687}

\bibitem[\protect\citeauthoryear{{Manuwal} \& {Stevens}}{{Manuwal} \& {Stevens}}{2023}]{Manuwal2023}
{Manuwal} A.,  {Stevens} A. R.~H.,  2023, \mn@doi [\mnras] {10.1093/mnras/stad1587}, \href {https://ui.adsabs.harvard.edu/abs/2023MNRAS.tmp.1537M} {}

\bibitem[\protect\citeauthoryear{{Manuwal}, {Ludlow}, {Stevens}, {Wright}  \& {Robotham}}{{Manuwal} et~al.}{2022}]{Manuwal2022}
{Manuwal} A.,  {Ludlow} A.~D.,  {Stevens} A. R.~H.,  {Wright} R.~J.,   {Robotham} A. S.~G.,  2022, \mn@doi [\mnras] {10.1093/mnras/stab3534}, \href {https://ui.adsabs.harvard.edu/abs/2022MNRAS.510.3408M} {510, 3408}

\bibitem[\protect\citeauthoryear{{Marasco}, {Crain}, {Schaye}, {Bah{\'e}}, {van der Hulst}, {Theuns}  \& {Bower}}{{Marasco} et~al.}{2016}]{Marasco2016}
{Marasco} A.,  {Crain} R.~A.,  {Schaye} J.,  {Bah{\'e}} Y.~M.,  {van der Hulst} T.,  {Theuns} T.,   {Bower} R.~G.,  2016, \mn@doi [\mnras] {10.1093/mnras/stw1498}, \href {https://ui.adsabs.harvard.edu/abs/2016MNRAS.461.2630M} {461, 2630}

\bibitem[\protect\citeauthoryear{{Marinacci}, {Binney}, {Fraternali}, {Nipoti}, {Ciotti}  \& {Londrillo}}{{Marinacci} et~al.}{2010}]{Marinacci2010}
{Marinacci} F.,  {Binney} J.,  {Fraternali} F.,  {Nipoti} C.,  {Ciotti} L.,   {Londrillo} P.,  2010, \mn@doi [\mnras] {10.1111/j.1365-2966.2010.16352.x}, \href {https://ui.adsabs.harvard.edu/abs/2010MNRAS.404.1464M} {404, 1464}

\bibitem[\protect\citeauthoryear{{Martin} et~al.,}{{Martin} et~al.}{2007}]{Martin2007}
{Martin} D.~C.,  et~al., 2007, \mn@doi [\apjs] {10.1086/516639}, \href {https://ui.adsabs.harvard.edu/abs/2007ApJS..173..342M} {173, 342}

\bibitem[\protect\citeauthoryear{{Masters} et~al.,}{{Masters} et~al.}{2010}]{Masters2010}
{Masters} K.~L.,  et~al., 2010, \mn@doi [\mnras] {10.1111/j.1365-2966.2010.16503.x}, \href {https://ui.adsabs.harvard.edu/abs/2010MNRAS.405..783M} {405, 783}

\bibitem[\protect\citeauthoryear{{Mateus}, {Sodr{\'e}}, {Cid Fernandes}, {Stasi{\'n}ska}, {Schoenell}  \& {Gomes}}{{Mateus} et~al.}{2006}]{Mateus2006}
{Mateus} A.,  {Sodr{\'e}} L.,  {Cid Fernandes} R.,  {Stasi{\'n}ska} G.,  {Schoenell} W.,   {Gomes} J.~M.,  2006, \mn@doi [\mnras] {10.1111/j.1365-2966.2006.10565.x}, \href {https://ui.adsabs.harvard.edu/abs/2006MNRAS.370..721M} {370, 721}

\bibitem[\protect\citeauthoryear{{McAlpine} et~al.,}{{McAlpine} et~al.}{2016}]{Mcalpine2016}
{McAlpine} S.,  et~al., 2016, \mn@doi [Astronomy and Computing] {10.1016/j.ascom.2016.02.004}, \href {https://ui.adsabs.harvard.edu/abs/2016A&C....15...72M} {15, 72}

\bibitem[\protect\citeauthoryear{{McGee}, {Balogh}, {Wilman}, {Bower}, {Mulchaey}, {Parker}  \& {Oemler}}{{McGee} et~al.}{2011}]{McGee2011}
{McGee} S.~L.,  {Balogh} M.~L.,  {Wilman} D.~J.,  {Bower} R.~G.,  {Mulchaey} J.~S.,  {Parker} L.~C.,   {Oemler} A.,  2011, \mn@doi [\mnras] {10.1111/j.1365-2966.2010.18189.x}, \href {https://ui.adsabs.harvard.edu/abs/2011MNRAS.413..996M} {413, 996}

\bibitem[\protect\citeauthoryear{{Mendez}, {Coil}, {Lotz}, {Salim}, {Moustakas}  \& {Simard}}{{Mendez} et~al.}{2011}]{Mendez2011}
{Mendez} A.~J.,  {Coil} A.~L.,  {Lotz} J.,  {Salim} S.,  {Moustakas} J.,   {Simard} L.,  2011, \mn@doi [\apj] {10.1088/0004-637X/736/2/110}, \href {https://ui.adsabs.harvard.edu/abs/2011ApJ...736..110M} {736, 110}

\bibitem[\protect\citeauthoryear{{Mesa}, {Duplancic}, {Alonso}, {Coldwell}  \& {Lambas}}{{Mesa} et~al.}{2014}]{Mesa2014}
{Mesa} V.,  {Duplancic} F.,  {Alonso} S.,  {Coldwell} G.,   {Lambas} D.~G.,  2014, \mn@doi [\mnras] {10.1093/mnras/stt2317}, \href {https://ui.adsabs.harvard.edu/abs/2014MNRAS.438.1784M} {438, 1784}

\bibitem[\protect\citeauthoryear{{Mishra}, {Wadadekar}  \& {Barway}}{{Mishra} et~al.}{2019}]{Mishra2019}
{Mishra} P.~K.,  {Wadadekar} Y.,   {Barway} S.,  2019, \mn@doi [\mnras] {10.1093/mnras/stz1621}, \href {https://ui.adsabs.harvard.edu/abs/2019MNRAS.487.5572M} {487, 5572}

\bibitem[\protect\citeauthoryear{{Moresco} et~al.,}{{Moresco} et~al.}{2013}]{Moresco2013}
{Moresco} M.,  et~al., 2013, \mn@doi [\aap] {10.1051/0004-6361/201321797}, \href {https://ui.adsabs.harvard.edu/abs/2013A&A...558A..61M} {558, A61}

\bibitem[\protect\citeauthoryear{{Nelson} et~al.,}{{Nelson} et~al.}{2018}]{Nelson2018}
{Nelson} D.,  et~al., 2018, \mn@doi [\mnras] {10.1093/mnras/stx3040}, \href {https://ui.adsabs.harvard.edu/abs/2018MNRAS.475..624N} {475, 624}

\bibitem[\protect\citeauthoryear{{Obreschkow}, {Glazebrook}, {Kilborn}  \& {Lutz}}{{Obreschkow} et~al.}{2016}]{Obreschkow2016}
{Obreschkow} D.,  {Glazebrook} K.,  {Kilborn} V.,   {Lutz} K.,  2016, \mn@doi [\apjl] {10.3847/2041-8205/824/2/L26}, \href {https://ui.adsabs.harvard.edu/abs/2016ApJ...824L..26O} {824, L26}

\bibitem[\protect\citeauthoryear{{Oh} et~al.,}{{Oh} et~al.}{2019}]{Oh2019}
{Oh} S.,  et~al., 2019, \mn@doi [\mnras] {10.1093/mnras/stz1920}, \href {https://ui.adsabs.harvard.edu/abs/2019MNRAS.488.4169O} {488, 4169}

\bibitem[\protect\citeauthoryear{{Oke}}{{Oke}}{1974}]{Oke1974}
{Oke} J.~B.,  1974, \mn@doi [\apjs] {10.1086/190287}, \href {https://ui.adsabs.harvard.edu/abs/1974ApJS...27...21O} {27, 21}

\bibitem[\protect\citeauthoryear{{Oser}, {Ostriker}, {Naab}, {Johansson}  \& {Burkert}}{{Oser} et~al.}{2010}]{Oser2010}
{Oser} L.,  {Ostriker} J.~P.,  {Naab} T.,  {Johansson} P.~H.,   {Burkert} A.,  2010, \mn@doi [\apj] {10.1088/0004-637X/725/2/2312}, \href {https://ui.adsabs.harvard.edu/abs/2010ApJ...725.2312O} {725, 2312}

\bibitem[\protect\citeauthoryear{{Pillepich} et~al.,}{{Pillepich} et~al.}{2018}]{Pillepich2018}
{Pillepich} A.,  et~al., 2018, \mn@doi [\mnras] {10.1093/mnras/stx2656}, \href {https://ui.adsabs.harvard.edu/abs/2018MNRAS.473.4077P} {473, 4077}

\bibitem[\protect\citeauthoryear{{Piotrowska}, {Bluck}, {Maiolino}  \& {Peng}}{{Piotrowska} et~al.}{2022}]{Piotrowska2022}
{Piotrowska} J.~M.,  {Bluck} A. F.~L.,  {Maiolino} R.,   {Peng} Y.,  2022, \mn@doi [\mnras] {10.1093/mnras/stab3673}, \href {https://ui.adsabs.harvard.edu/abs/2022MNRAS.512.1052P} {512, 1052}

\bibitem[\protect\citeauthoryear{{Planck Collaboration} et~al.,}{{Planck Collaboration} et~al.}{2014}]{Planck2014}
{Planck Collaboration} et~al., 2014, \mn@doi [\aap] {10.1051/0004-6361/201321591}, \href {https://ui.adsabs.harvard.edu/abs/2014A&A...571A..16P} {571, A16}

\bibitem[\protect\citeauthoryear{{Poggianti}, {Smail}, {Dressler}, {Couch}, {Barger}, {Butcher}, {Ellis}  \& {Oemler}}{{Poggianti} et~al.}{1999}]{Poggianti1999}
{Poggianti} B.~M.,  {Smail} I.,  {Dressler} A.,  {Couch} W.~J.,  {Barger} A.~J.,  {Butcher} H.,  {Ellis} R.~S.,   {Oemler} Augustus J.,  1999, \mn@doi [\apj] {10.1086/307322}, \href {https://ui.adsabs.harvard.edu/abs/1999ApJ...518..576P} {518, 576}

\bibitem[\protect\citeauthoryear{{Poggianti}, {Bridges}, {Komiyama}, {Yagi}, {Carter}, {Mobasher}, {Okamura}  \& {Kashikawa}}{{Poggianti} et~al.}{2004}]{Poggianti2004}
{Poggianti} B.~M.,  {Bridges} T.~J.,  {Komiyama} Y.,  {Yagi} M.,  {Carter} D.,  {Mobasher} B.,  {Okamura} S.,   {Kashikawa} N.,  2004, \mn@doi [\apj] {10.1086/380195}, \href {https://ui.adsabs.harvard.edu/abs/2004ApJ...601..197P} {601, 197}

\bibitem[\protect\citeauthoryear{{Qu} et~al.,}{{Qu} et~al.}{2017}]{Qu2017}
{Qu} Y.,  et~al., 2017, \mn@doi [\mnras] {10.1093/mnras/stw2437}, \href {https://ui.adsabs.harvard.edu/abs/2017MNRAS.464.1659Q} {464, 1659}

\bibitem[\protect\citeauthoryear{{Rahmati}, {Pawlik}, {Rai{\v{c}}evi{\'c}}  \& {Schaye}}{{Rahmati} et~al.}{2013}]{Rahmati2013}
{Rahmati} A.,  {Pawlik} A.~H.,  {Rai{\v{c}}evi{\'c}} M.,   {Schaye} J.,  2013, \mn@doi [\mnras] {10.1093/mnras/stt066}, \href {https://ui.adsabs.harvard.edu/abs/2013MNRAS.430.2427R} {430, 2427}

\bibitem[\protect\citeauthoryear{{Rodriguez-Gomez} et~al.,}{{Rodriguez-Gomez} et~al.}{2016}]{Gomez2016}
{Rodriguez-Gomez} V.,  et~al., 2016, \mn@doi [\mnras] {10.1093/mnras/stw456}, \href {https://ui.adsabs.harvard.edu/abs/2016MNRAS.458.2371R} {458, 2371}

\bibitem[\protect\citeauthoryear{{Sales}, {Navarro}, {Theuns}, {Schaye}, {White}, {Frenk}, {Crain}  \& {Dalla Vecchia}}{{Sales} et~al.}{2012}]{Sales2012}
{Sales} L.~V.,  {Navarro} J.~F.,  {Theuns} T.,  {Schaye} J.,  {White} S. D.~M.,  {Frenk} C.~S.,  {Crain} R.~A.,   {Dalla Vecchia} C.,  2012, \mn@doi [\mnras] {10.1111/j.1365-2966.2012.20975.x}, \href {https://ui.adsabs.harvard.edu/abs/2012MNRAS.423.1544S} {423, 1544}

\bibitem[\protect\citeauthoryear{{Salim} et~al.,}{{Salim} et~al.}{2007}]{Salim2007}
{Salim} S.,  et~al., 2007, \mn@doi [\apjs] {10.1086/519218}, \href {https://ui.adsabs.harvard.edu/abs/2007ApJS..173..267S} {173, 267}

\bibitem[\protect\citeauthoryear{{Santini} et~al.,}{{Santini} et~al.}{2009}]{Santini2009}
{Santini} P.,  et~al., 2009, \mn@doi [\aap] {10.1051/0004-6361/200811434}, \href {https://ui.adsabs.harvard.edu/abs/2009A&A...504..751S} {504, 751}

\bibitem[\protect\citeauthoryear{{Schawinski} et~al.,}{{Schawinski} et~al.}{2009}]{Schawinski2009}
{Schawinski} K.,  et~al., 2009, \mn@doi [\mnras] {10.1111/j.1365-2966.2009.14793.x}, \href {https://ui.adsabs.harvard.edu/abs/2009MNRAS.396..818S} {396, 818}

\bibitem[\protect\citeauthoryear{{Schawinski} et~al.,}{{Schawinski} et~al.}{2014}]{Schawinski2014}
{Schawinski} K.,  et~al., 2014, \mn@doi [\mnras] {10.1093/mnras/stu327}, \href {https://ui.adsabs.harvard.edu/abs/2014MNRAS.440..889S} {440, 889}

\bibitem[\protect\citeauthoryear{{Schaye} et~al.,}{{Schaye} et~al.}{2015}]{Schaye2015}
{Schaye} J.,  et~al., 2015, \mn@doi [\mnras] {10.1093/mnras/stu2058}, \href {https://ui.adsabs.harvard.edu/abs/2015MNRAS.446..521S} {446, 521}

\bibitem[\protect\citeauthoryear{{Simpson}, {Grand}, {G{\'o}mez}, {Marinacci}, {Pakmor}, {Springel}, {Campbell}  \& {Frenk}}{{Simpson} et~al.}{2018}]{Simpson2018}
{Simpson} C.~M.,  {Grand} R. J.~J.,  {G{\'o}mez} F.~A.,  {Marinacci} F.,  {Pakmor} R.,  {Springel} V.,  {Campbell} D. J.~R.,   {Frenk} C.~S.,  2018, \mn@doi [\mnras] {10.1093/mnras/sty774}, \href {https://ui.adsabs.harvard.edu/abs/2018MNRAS.478..548S} {478, 548}

\bibitem[\protect\citeauthoryear{{Smethurst} et~al.,}{{Smethurst} et~al.}{2015}]{Smethurst2015}
{Smethurst} R.~J.,  et~al., 2015, \mn@doi [\mnras] {10.1093/mnras/stv161}, \href {https://ui.adsabs.harvard.edu/abs/2015MNRAS.450..435S} {450, 435}

\bibitem[\protect\citeauthoryear{{Smith} et~al.,}{{Smith} et~al.}{2022}]{Smith2022}
{Smith} D.,  et~al., 2022, \mn@doi [\mnras] {10.1093/mnras/stac2258}, \href {https://ui.adsabs.harvard.edu/abs/2022MNRAS.517.4575S} {517, 4575}

\bibitem[\protect\citeauthoryear{{Springel}}{{Springel}}{2005}]{Springel2005}
{Springel} V.,  2005, \mn@doi [\mnras] {10.1111/j.1365-2966.2005.09655.x}, \href {https://ui.adsabs.harvard.edu/abs/2005MNRAS.364.1105S} {364, 1105}

\bibitem[\protect\citeauthoryear{{Springel}, {White}, {Tormen}  \& {Kauffmann}}{{Springel} et~al.}{2001}]{Springel2001}
{Springel} V.,  {White} S. D.~M.,  {Tormen} G.,   {Kauffmann} G.,  2001, \mn@doi [\mnras] {10.1046/j.1365-8711.2001.04912.x}, \href {https://ui.adsabs.harvard.edu/abs/2001MNRAS.328..726S} {328, 726}

\bibitem[\protect\citeauthoryear{{Springel}, {Di Matteo}  \& {Hernquist}}{{Springel} et~al.}{2005}]{Springel2005d}
{Springel} V.,  {Di Matteo} T.,   {Hernquist} L.,  2005, \mn@doi [\apjl] {10.1086/428772}, \href {https://ui.adsabs.harvard.edu/abs/2005ApJ...620L..79S} {620, L79}

\bibitem[\protect\citeauthoryear{{Stark}, {Kannappan}, {Wei}, {Baker}, {Leroy}, {Eckert}  \& {Vogel}}{{Stark} et~al.}{2013}]{Stark2013}
{Stark} D.~V.,  {Kannappan} S.~J.,  {Wei} L.~H.,  {Baker} A.~J.,  {Leroy} A.~K.,  {Eckert} K.~D.,   {Vogel} S.~N.,  2013, \mn@doi [\apj] {10.1088/0004-637X/769/1/82}, \href {https://ui.adsabs.harvard.edu/abs/2013ApJ...769...82S} {769, 82}

\bibitem[\protect\citeauthoryear{{Steinhauser}, {Schindler}  \& {Springel}}{{Steinhauser} et~al.}{2016}]{Steinhauser2016}
{Steinhauser} D.,  {Schindler} S.,   {Springel} V.,  2016, \mn@doi [\aap] {10.1051/0004-6361/201527705}, \href {https://ui.adsabs.harvard.edu/abs/2016A&A...591A..51S} {591, A51}

\bibitem[\protect\citeauthoryear{{Stevens}, {Martig}, {Croton}  \& {Feng}}{{Stevens} et~al.}{2014}]{Stevens2014}
{Stevens} A. R.~H.,  {Martig} M.,  {Croton} D.~J.,   {Feng} Y.,  2014, \mn@doi [\mnras] {10.1093/mnras/stu1724}, \href {https://ui.adsabs.harvard.edu/abs/2014MNRAS.445..239S} {445, 239}

\bibitem[\protect\citeauthoryear{{Stevens}, {Lagos}, {Contreras}, {Croton}, {Padilla}, {Schaller}, {Schaye}  \& {Theuns}}{{Stevens} et~al.}{2017}]{Stevens2017}
{Stevens} A. R.~H.,  {Lagos} C. d.~P.,  {Contreras} S.,  {Croton} D.~J.,  {Padilla} N.~D.,  {Schaller} M.,  {Schaye} J.,   {Theuns} T.,  2017, \mn@doi [\mnras] {10.1093/mnras/stx243}, \href {https://ui.adsabs.harvard.edu/abs/2017MNRAS.467.2066S} {467, 2066}

\bibitem[\protect\citeauthoryear{{Stevens} et~al.,}{{Stevens} et~al.}{2019}]{Stevens2019}
{Stevens} A. R.~H.,  et~al., 2019, \mn@doi [\mnras] {10.1093/mnras/sty3451}, \href {https://ui.adsabs.harvard.edu/abs/2019MNRAS.483.5334S} {483, 5334}

\bibitem[\protect\citeauthoryear{{Stevens} et~al.,}{{Stevens} et~al.}{2021}]{Stevens2021}
{Stevens} A. R.~H.,  et~al., 2021, \mn@doi [\mnras] {10.1093/mnras/staa3662}, \href {https://ui.adsabs.harvard.edu/abs/2021MNRAS.502.3158S} {502, 3158}

\bibitem[\protect\citeauthoryear{{Stewart}, {Kaufmann}, {Bullock}, {Barton}, {Maller}, {Diemand}  \& {Wadsley}}{{Stewart} et~al.}{2011}]{Stewart2011}
{Stewart} K.~R.,  {Kaufmann} T.,  {Bullock} J.~S.,  {Barton} E.~J.,  {Maller} A.~H.,  {Diemand} J.,   {Wadsley} J.,  2011, \mn@doi [\apj] {10.1088/0004-637X/738/1/39}, \href {https://ui.adsabs.harvard.edu/abs/2011ApJ...738...39S} {738, 39}

\bibitem[\protect\citeauthoryear{{Strateva} et~al.,}{{Strateva} et~al.}{2001}]{Strateva2001}
{Strateva} I.,  et~al., 2001, \mn@doi [\aj] {10.1086/323301}, \href {https://ui.adsabs.harvard.edu/abs/2001AJ....122.1861S} {122, 1861}

\bibitem[\protect\citeauthoryear{{Tacchella} et~al.,}{{Tacchella} et~al.}{2019}]{Tacchella2019}
{Tacchella} S.,  et~al., 2019, \mn@doi [\mnras] {10.1093/mnras/stz1657}, \href {https://ui.adsabs.harvard.edu/abs/2019MNRAS.487.5416T} {487, 5416}

\bibitem[\protect\citeauthoryear{{Taylor} \& {Kobayashi}}{{Taylor} \& {Kobayashi}}{2015}]{Taylor2015a}
{Taylor} P.,  {Kobayashi} C.,  2015, \mn@doi [\mnras] {10.1093/mnras/stv139}, \href {https://ui.adsabs.harvard.edu/abs/2015MNRAS.448.1835T} {448, 1835}

\bibitem[\protect\citeauthoryear{{Taylor} et~al.,}{{Taylor} et~al.}{2015}]{Taylor2015}
{Taylor} E.~N.,  et~al., 2015, \mn@doi [\mnras] {10.1093/mnras/stu1900}, \href {https://ui.adsabs.harvard.edu/abs/2015MNRAS.446.2144T} {446, 2144}

\bibitem[\protect\citeauthoryear{{Taylor}, {Federrath}  \& {Kobayashi}}{{Taylor} et~al.}{2018}]{Taylor2018}
{Taylor} P.,  {Federrath} C.,   {Kobayashi} C.,  2018, \mn@doi [\mnras] {10.1093/mnras/sty1439}, \href {https://ui.adsabs.harvard.edu/abs/2018MNRAS.479..141T} {479, 141}

\bibitem[\protect\citeauthoryear{{The EAGLE team}}{{The EAGLE team}}{2017}]{eaglepart}
{The EAGLE team} 2017, arXiv e-prints, \href {https://ui.adsabs.harvard.edu/abs/2017arXiv170609899T} {p. arXiv:1706.09899}

\bibitem[\protect\citeauthoryear{{Tollet}, {Cattaneo}, {Mamon}, {Moutard}  \& {van den Bosch}}{{Tollet} et~al.}{2017}]{Tollet2017}
{Tollet} {\'E}.,  {Cattaneo} A.,  {Mamon} G.~A.,  {Moutard} T.,   {van den Bosch} F.~C.,  2017, \mn@doi [\mnras] {10.1093/mnras/stx1840}, \href {https://ui.adsabs.harvard.edu/abs/2017MNRAS.471.4170T} {471, 4170}

\bibitem[\protect\citeauthoryear{{Trapp} et~al.,}{{Trapp} et~al.}{2022}]{Trapp2022}
{Trapp} C.~W.,  et~al., 2022, \mn@doi [\mnras] {10.1093/mnras/stab3251}, \href {https://ui.adsabs.harvard.edu/abs/2022MNRAS.509.4149T} {509, 4149}

\bibitem[\protect\citeauthoryear{{Trayford} et~al.,}{{Trayford} et~al.}{2015}]{Trayford2015}
{Trayford} J.~W.,  et~al., 2015, \mn@doi [\mnras] {10.1093/mnras/stv1461}, \href {https://ui.adsabs.harvard.edu/abs/2015MNRAS.452.2879T} {452, 2879}

\bibitem[\protect\citeauthoryear{{Trayford}, {Theuns}, {Bower}, {Crain}, {Lagos}, {Schaller}  \& {Schaye}}{{Trayford} et~al.}{2016}]{Trayford2016}
{Trayford} J.~W.,  {Theuns} T.,  {Bower} R.~G.,  {Crain} R.~A.,  {Lagos} C. d.~P.,  {Schaller} M.,   {Schaye} J.,  2016, \mn@doi [\mnras] {10.1093/mnras/stw1230}, \href {https://ui.adsabs.harvard.edu/abs/2016MNRAS.460.3925T} {460, 3925}

\bibitem[\protect\citeauthoryear{{Trayford} et~al.,}{{Trayford} et~al.}{2017}]{Trayford2017}
{Trayford} J.~W.,  et~al., 2017, \mn@doi [\mnras] {10.1093/mnras/stx1051}, \href {https://ui.adsabs.harvard.edu/abs/2017MNRAS.470..771T} {470, 771}

\bibitem[\protect\citeauthoryear{{Tremmel}, {Governato}, {Volonteri}  \& {Quinn}}{{Tremmel} et~al.}{2015}]{Tremmel2015}
{Tremmel} M.,  {Governato} F.,  {Volonteri} M.,   {Quinn} T.~R.,  2015, \mn@doi [\mnras] {10.1093/mnras/stv1060}, \href {https://ui.adsabs.harvard.edu/abs/2015MNRAS.451.1868T} {451, 1868}

\bibitem[\protect\citeauthoryear{{Tr{\v{c}}ka} et~al.,}{{Tr{\v{c}}ka} et~al.}{2020}]{Trcka2020}
{Tr{\v{c}}ka} A.,  et~al., 2020, \mn@doi [\mnras] {10.1093/mnras/staa857}, \href {https://ui.adsabs.harvard.edu/abs/2020MNRAS.494.2823T} {494, 2823}

\bibitem[\protect\citeauthoryear{{{\"U}bler}, {Naab}, {Oser}, {Aumer}, {Sales}  \& {White}}{{{\"U}bler} et~al.}{2014}]{Ubler2014}
{{\"U}bler} H.,  {Naab} T.,  {Oser} L.,  {Aumer} M.,  {Sales} L.~V.,   {White} S. D.~M.,  2014, \mn@doi [\mnras] {10.1093/mnras/stu1275}, \href {https://ui.adsabs.harvard.edu/abs/2014MNRAS.443.2092U} {443, 2092}

\bibitem[\protect\citeauthoryear{{Virtanen} et~al.,}{{Virtanen} et~al.}{2020}]{Scipy}
{Virtanen} P.,  et~al., 2020, \mn@doi [Nature Methods] {10.1038/s41592-019-0686-2}, \href {https://ui.adsabs.harvard.edu/abs/2020NatMe..17..261V} {17, 261}

\bibitem[\protect\citeauthoryear{{Wang}, {Xu}, {Lu}, {Cai}, {Xiang}, {Mao}, {Springel}  \& {Hernquist}}{{Wang} et~al.}{2022}]{Wang2022}
{Wang} S.,  {Xu} D.,  {Lu} S.,  {Cai} Z.,  {Xiang} M.,  {Mao} S.,  {Springel} V.,   {Hernquist} L.,  2022, \mn@doi [\mnras] {10.1093/mnras/stab3167}, \href {https://ui.adsabs.harvard.edu/abs/2022MNRAS.509.3148W} {509, 3148}

\bibitem[\protect\citeauthoryear{{Watts}, {Power}, {Catinella}, {Cortese}  \& {Stevens}}{{Watts} et~al.}{2020}]{Watts2020}
{Watts} A.~B.,  {Power} C.,  {Catinella} B.,  {Cortese} L.,   {Stevens} A. R.~H.,  2020, \mn@doi [\mnras] {10.1093/mnras/staa3200}, \href {https://ui.adsabs.harvard.edu/abs/2020MNRAS.499.5205W} {499, 5205}

\bibitem[\protect\citeauthoryear{{Weinmann}, {van den Bosch}, {Yang}  \& {Mo}}{{Weinmann} et~al.}{2006}]{Weinmann2006}
{Weinmann} S.~M.,  {van den Bosch} F.~C.,  {Yang} X.,   {Mo} H.~J.,  2006, \mn@doi [\mnras] {10.1111/j.1365-2966.2005.09865.x}, \href {https://ui.adsabs.harvard.edu/abs/2006MNRAS.366....2W} {366, 2}

\bibitem[\protect\citeauthoryear{{Wetzel}}{{Wetzel}}{2011}]{Wetzel2011}
{Wetzel} A.~R.,  2011, \mn@doi [\mnras] {10.1111/j.1365-2966.2010.17877.x}, \href {https://ui.adsabs.harvard.edu/abs/2011MNRAS.412...49W} {412, 49}

\bibitem[\protect\citeauthoryear{{Whitaker} et~al.,}{{Whitaker} et~al.}{2011}]{Whitaker2011}
{Whitaker} K.~E.,  et~al., 2011, \mn@doi [\apj] {10.1088/0004-637X/735/2/86}, \href {https://ui.adsabs.harvard.edu/abs/2011ApJ...735...86W} {735, 86}

\bibitem[\protect\citeauthoryear{{Wijers} \& {Schaye}}{{Wijers} \& {Schaye}}{2022}]{Wijers2022}
{Wijers} N.~A.,  {Schaye} J.,  2022, \mn@doi [\mnras] {10.1093/mnras/stac1580}, \href {https://ui.adsabs.harvard.edu/abs/2022MNRAS.514.5214W} {514, 5214}

\bibitem[\protect\citeauthoryear{{Wijers}, {Schaye}  \& {Oppenheimer}}{{Wijers} et~al.}{2020}]{Wijers2020}
{Wijers} N.~A.,  {Schaye} J.,   {Oppenheimer} B.~D.,  2020, \mn@doi [\mnras] {10.1093/mnras/staa2456}, \href {https://ui.adsabs.harvard.edu/abs/2020MNRAS.498..574W} {498, 574}

\bibitem[\protect\citeauthoryear{{Wild}, {Walcher}, {Johansson}, {Tresse}, {Charlot}, {Pollo}, {Le F{\`e}vre}  \& {de Ravel}}{{Wild} et~al.}{2009}]{Wild2009}
{Wild} V.,  {Walcher} C.~J.,  {Johansson} P.~H.,  {Tresse} L.,  {Charlot} S.,  {Pollo} A.,  {Le F{\`e}vre} O.,   {de Ravel} L.,  2009, \mn@doi [\mnras] {10.1111/j.1365-2966.2009.14537.x}, \href {https://ui.adsabs.harvard.edu/abs/2009MNRAS.395..144W} {395, 144}

\bibitem[\protect\citeauthoryear{{Williams}, {Quadri}, {Franx}, {van Dokkum}  \& {Labb{\'e}}}{{Williams} et~al.}{2009}]{Williams2009}
{Williams} R.~J.,  {Quadri} R.~F.,  {Franx} M.,  {van Dokkum} P.,   {Labb{\'e}} I.,  2009, \mn@doi [\apj] {10.1088/0004-637X/691/2/1879}, \href {https://ui.adsabs.harvard.edu/abs/2009ApJ...691.1879W} {691, 1879}

\bibitem[\protect\citeauthoryear{{Wilman} \& {Erwin}}{{Wilman} \& {Erwin}}{2012}]{Wilman2012}
{Wilman} D.~J.,  {Erwin} P.,  2012, \mn@doi [\apj] {10.1088/0004-637X/746/2/160}, \href {https://ui.adsabs.harvard.edu/abs/2012ApJ...746..160W} {746, 160}

\bibitem[\protect\citeauthoryear{{Wolf}, {Gray}  \& {Meisenheimer}}{{Wolf} et~al.}{2005}]{Wolf2005}
{Wolf} C.,  {Gray} M.~E.,   {Meisenheimer} K.,  2005, \mn@doi [\aap] {10.1051/0004-6361:20053585}, \href {https://ui.adsabs.harvard.edu/abs/2005A&A...443..435W} {443, 435}

\bibitem[\protect\citeauthoryear{{Wolf} et~al.,}{{Wolf} et~al.}{2009}]{Wolf2009}
{Wolf} C.,  et~al., 2009, \mn@doi [\mnras] {10.1111/j.1365-2966.2008.14204.x}, \href {https://ui.adsabs.harvard.edu/abs/2009MNRAS.393.1302W} {393, 1302}

\bibitem[\protect\citeauthoryear{{Wright}, {Lagos}, {Power}, {Stevens}, {Cortese}  \& {Poulton}}{{Wright} et~al.}{2022}]{Wright2022}
{Wright} R.~J.,  {Lagos} C. d.~P.,  {Power} C.,  {Stevens} A. R.~H.,  {Cortese} L.,   {Poulton} R. J.~J.,  2022, \mn@doi [\mnras] {10.1093/mnras/stac2042}, \href {https://ui.adsabs.harvard.edu/abs/2022MNRAS.516.2891W} {516, 2891}

\bibitem[\protect\citeauthoryear{{York} et~al.,}{{York} et~al.}{2000}]{York2000}
{York} D.~G.,  et~al., 2000, \mn@doi [\aj] {10.1086/301513}, \href {https://ui.adsabs.harvard.edu/abs/2000AJ....120.1579Y} {120, 1579}

\bibitem[\protect\citeauthoryear{{Young} et~al.,}{{Young} et~al.}{2014}]{Young2014}
{Young} L.~M.,  et~al., 2014, \mn@doi [\mnras] {10.1093/mnras/stt2474}, \href {https://ui.adsabs.harvard.edu/abs/2014MNRAS.444.3408Y} {444, 3408}

\bibitem[\protect\citeauthoryear{{Zamojski} et~al.,}{{Zamojski} et~al.}{2007}]{Zamojski2007}
{Zamojski} M.~A.,  et~al., 2007, \mn@doi [\apjs] {10.1086/516593}, \href {https://ui.adsabs.harvard.edu/abs/2007ApJS..172..468Z} {172, 468}

\bibitem[\protect\citeauthoryear{{Zinger} et~al.,}{{Zinger} et~al.}{2020}]{Zinger2020}
{Zinger} E.,  et~al., 2020, \mn@doi [\mnras] {10.1093/mnras/staa2607}, \href {https://ui.adsabs.harvard.edu/abs/2020MNRAS.499..768Z} {499, 768}

\bibitem[\protect\citeauthoryear{{de Graaff}, {Trayford}, {Franx}, {Schaller}, {Schaye}  \& {van der Wel}}{{de Graaff} et~al.}{2022}]{deGraaff2022}
{de Graaff} A.,  {Trayford} J.,  {Franx} M.,  {Schaller} M.,  {Schaye} J.,   {van der Wel} A.,  2022, \mn@doi [\mnras] {10.1093/mnras/stab3510}, \href {https://ui.adsabs.harvard.edu/abs/2022MNRAS.511.2544D} {511, 2544}

\bibitem[\protect\citeauthoryear{{van de Voort}, {Bah{\'e}}, {Bower}, {Correa}, {Crain}, {Schaye}  \& {Theuns}}{{van de Voort} et~al.}{2017}]{van2017}
{van de Voort} F.,  {Bah{\'e}} Y.~M.,  {Bower} R.~G.,  {Correa} C.~A.,  {Crain} R.~A.,  {Schaye} J.,   {Theuns} T.,  2017, \mn@doi [\mnras] {10.1093/mnras/stw3356}, \href {https://ui.adsabs.harvard.edu/abs/2017MNRAS.466.3460V} {466, 3460}

\bibitem[\protect\citeauthoryear{{van der Walt}, {Colbert}  \& {Varoquaux}}{{van der Walt} et~al.}{2011}]{Numpy}
{van der Walt} S.,  {Colbert} S.~C.,   {Varoquaux} G.,  2011, \mn@doi [Computing in Science and Engineering] {10.1109/MCSE.2011.37}, \href {https://ui.adsabs.harvard.edu/abs/2011CSE....13b..22V} {13, 22}

\makeatother
\end{thebibliography}



\appendix
\section{Enrichment of ISM due to the transfer of metals from stars}\label{sfenrich}
\begin{figure}
\centering
  \includegraphics[width=0.8\columnwidth]{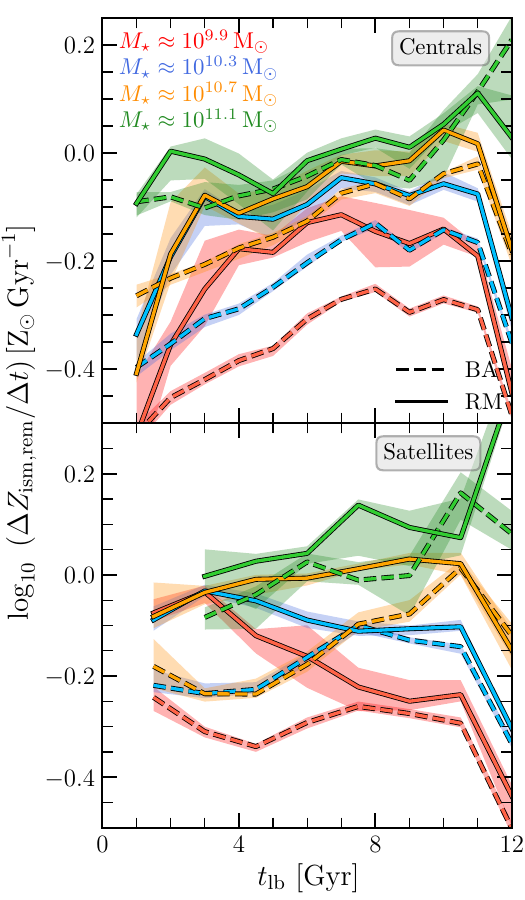}
   \caption{The enrichment rates for the ISM that remained between successive snapshots for our sample of SF galaxies. 
   The top and bottom panels show the results for the centrals and the satellites, respectively. In each panel, the solid and dashed curves correspond to the BA and the RM with $z=0$ stellar masses within $0.1$~dex of the $M_\star$ displayed in the top-left corner of the top panel; the $M_\star$ can be inferred through the colour correspondence. The curves show that, for virtually the whole history, the ISM in RM was typically enriched more by stellar winds and supernovae than that in BA.}
   \label{dism}
\end{figure}

The gas around stars in the simulation gets enriched via the mass-transfers driven by stellar winds and supernovae \citep{Schaye2015}.
We attempt to probe this effect in our galaxies by examining the change in metallicity of the gas particles that remain
in the ISM between successive snapshots ($\Delta Z_{\rm ism,rem}$). Fig.~\ref{dism} shows the rate of this change throughout
history for the centrals (top panel) and the satellites (bottom panel). Each curve shows the median rate for the galaxies whose $z=0$ stellar masses lie within $\pm 0.1$~dex of the value displayed in the top-left corner, and has the same colour
as the text. The solid and dashed curves correspond to BA and RM, respectively.

Regardless of the panel that we choose to focus on, we find that the enrichment rate for the ISM was generally higher
for the RM, similar to the SFEs (Figs~\ref{cenex} and~\ref{satex}). In fact, a qualitative comparison of the enrichment rates against the SFEs reveals that the offsets between the RM and the BA for the former have been somewhat correlated to those in the latter. This suggests a causal connection between the two processes, and is rather
reasonable. A higher SFE essentially means a higher SFR for a fixed gas mass. Each star enriches various gas particles around it over the course of its lifetime. More the stars, more the cumulative metallic mass transferred to gas, which increases the overall metallicity of the part of the ISM that remains in the galaxy during this whole process.


\bsp	
\label{lastpage}
\end{document}